\documentclass[a4paper,10pt]{article}
\pdfoutput=1 

\usepackage{jheppub} 
\usepackage[export]{adjustbox}
\usepackage[utf8]{inputenc}
\usepackage{multirow}
\usepackage{bbold}
\usepackage[table]{xcolor}
\usepackage{slashed}
\usepackage{bm}
\usepackage{subfig}
\usepackage{float}
\usepackage{fancyvrb}
\usepackage{fvextra}
\usepackage{soul}
\usepackage{comment}
\usepackage{cancel} 
\usepackage{csquotes} 
\usepackage[title]{appendix}
\usepackage{amsmath,amssymb,amsthm}
\usepackage{mathtools}
\usepackage{enumitem}
\usepackage{hyperref}

\usepackage{tabularx}

\usepackage{subfig}
\usepackage{color,graphicx,slashed,hyperref}
\usepackage[utf8]{inputenc}
 \usepackage{verbatim}
\usepackage{setspace}

\usepackage{booktabs}
\usepackage{amstext} 
\usepackage{array}   
\newcolumntype{C}{>{$}c<{$}} 
\newcolumntype{L}{>{$}l<{$}} 

\newcommand{\Mpl}{M_{\rm Pl}}

\theoremstyle{plain}

\theoremstyle{definition}

\title{Adiabatic Deformations of Black Hole Moduli:\\
II. The Magnetic GHS Family}

\author[a]{Karim Benakli,}
\emailAdd{kbenakli@lpthe.jussieu.fr}

\author[a]{Anna Chrysostomou}
\emailAdd{chrysostomou@lpthe.jussieu.fr}

\affiliation[a]{Laboratoire de Physique Th\'eorique et Hautes \'Energies - LPTHE, Sorbonne Universit\'e, CNRS, 4 Place Jussieu, 75005 Paris, France}

\abstract{
We study the leading adiabatic response of magnetic GHS black holes in massless Einstein--Maxwell--dilaton theory to a prescribed exterior modulus varying slowly in advanced time. In ingoing Eddington--Finkelstein coordinates with the exact areal radius, we derive the time-dependent spherical equations and expand about a slowly evolving instantaneous static GHS representative, keeping its horizon distinct from the exact dynamical marginal horizon. The magnetic GHS family realizes the canonical-slice and Fredholm framework developed in the companion paper. Its integrability yields in closed form the canonical tangent, the finite-rank source profile, a horizon-regular solution of the auxiliary inhomogeneous equation, and a globally regular generator of the sliced-operator kernel. For the leading Robin functional arising in the massless large-radius overlap, we evaluate the Fredholm denominator analytically and prove that it is strictly positive on the charged non-extremal branch. This remains true where the conventional normalization of the kernel generator degenerates, showing that these loci are normalization singularities rather than degeneracies of the kernel mode. Within the controlled large-radius overlap regime, this positivity proves existence and uniqueness of the completed leading-order quasi-static near-zone problem. We then solve the forced radial equation explicitly. The scalar lag is the sum of a universal forced profile and a sliced homogeneous mode whose amplitude is fixed by the exterior matching datum. The mass lag follows algebraically from the radial constraint, and the lapse lag from a single elementary quadrature. The exact marginal horizon is recovered from the lag-corrected solution. The result gives the complete leading quasi-static near-zone response, exact in its radial dependence, for every admissible exterior datum within the specified massless overlap model. }


\begin{document}
\maketitle
\flushbottom

\newpage


\section{Introduction}
\label{Introduction}

Light scalar fields with slowly evolving expectation values arise naturally
in extensions of General Relativity and of the Standard Model. String moduli,
dilatons, axion-like fields, and quintessence candidates may vary on
cosmological time and length scales while remaining approximately uniform
across the much smaller near zone of a compact gravitating object. It is
therefore natural to ask how a black hole responds when the exterior scalar
data evolve slowly in time.

Charged black holes provide a particularly sharp setting for this question.
When the gauge coupling depends on a scalar modulus, the static black hole
solution already carries a nontrivial radial scalar profile. Slow evolution
of the exterior modulus then affects not only the asymptotic scalar value but
the entire near-zone configuration. Black hole solution spaces have also
been used more broadly as probes of moduli-space geometry (see e.g. Refs.
\cite{Delgado:2022dkz,Sen:2025ljz,Benakli:2026zbm,Heidenreich:2026ztw}), while in the extremal setting the attractor mechanism
identifies distinguished critical points in moduli space
\cite{Ferrara:1997tw}. The problem addressed here is different: we consider
a non-extremal static solution family and determine its leading response to
slowly varying exterior scalar data.

The magnetic Garfinkle--Horowitz--Strominger (GHS) family
\cite{Gibbons:1987ps,Garfinkle:1990qj} provides an especially useful example.
In massless Einstein--Maxwell--dilaton (EMD) theory, its metric and scalar profiles
are known analytically for arbitrary values of the static modulus label.
This exact integrability makes it possible to carry the complete
leading-order near-zone construction through in closed form. The dynamical
problem is nevertheless nontrivial: replacing the static modulus parameter
by a function of time does not produce a solution of the time-dependent field
equations. The promoted static configuration develops an adiabatic defect,
and the exact solution differs from it by scalar and metric-sector lag fields.

A vanishing scalar potential does not imply that the exterior modulus must be
constant. A homogeneous massless scalar may evolve through its kinetic
energy, as in a kination regime, or may form part of a more general slowly
varying exterior solution. Such configurations motivate the massless problem
studied here. We do not, however, solve a specific global cosmological
background. Instead, the exterior is represented locally by slowly varying
data supplied at a finite matching radius. The detailed exterior solution
determines the matching functional and its inhomogeneous datum.

The finite near-zone formulation is essential. We work on an interval
bounded internally by the horizon of an instantaneous static representative
and externally by a matching surface \(r=r_{\rm m}\), with
\begin{equation}
r_{\rm H}^{\rm inst}
<
r_{\rm m}
\ll
L_{\rm cos}.
\label{eq:introduction_near_zone_hierarchy}
\end{equation}
No limit to spatial infinity is part of the near-zone boundary-value problem.
The static GHS solution may be extended to infinity and used to obtain
analytical background profiles, but the dynamical exterior information enters
through matching at finite radius.

The explicit massless matching model considered later requires, in addition,
that the matching surface lie in the asymptotic end of the GHS exterior.
Writing
\begin{equation}
b_{\rm m}
\equiv
\frac{\rho(r_{\rm m})-r_+}{r_+},
\end{equation}
the controlled large-radius overlap regime is
\begin{equation}
b_{\rm m}\gg1,
\qquad
r_{\rm m}\ll L_{\rm cos}.
\label{eq:introduction_controlled_overlap}
\end{equation}
Equivalently,
\begin{equation}
\frac{r_+}{\rho(r_{\rm m})}
=
\frac{1}{1+b_{\rm m}}
\ll1.
\end{equation}
This condition is stronger than merely requiring the matching surface to be
many areal-horizon radii away. Indeed,
\begin{equation}
\frac{r_{\rm m}}{r_{{\rm H},0}}
=
\sqrt{
\frac{(1+b_{\rm m})(a+b_{\rm m})}{a}
},
\end{equation}
which can become large near extremality even when
\(b_{\rm m}={\mathcal O}(1)\). The condition \(b_{\rm m}\gg1\) is the one that
controls the asymptotic expansion used to derive the leading massless
matching functional.

A second important distinction concerns the horizon. The promoted static
family possesses an instantaneous horizon
\(r_{\rm H}^{\rm inst}(v)\), determined by the collective coordinates of the
chosen static representative. The full dynamical solution instead possesses
an exact marginal horizon \(r_{\rm H}^{\rm exact}(v)\). These two radii may
differ at first adiabatic order once the lag fields are included, even though
their time derivatives begin only at second order. Maintaining this
distinction is necessary for a consistent decomposition of the exact
solution into an instantaneous static configuration and a lag field.

The response of black holes to slowly varying scalar backgrounds has been
studied from several complementary perspectives. Jacobson constructed a
horizon-regular scalar configuration around Schwarzschild for a linearly
rolling exterior scalar and showed the absence of gravitational memory
\cite{Jacobson:1999vr}. Subsequently, a
self-consistent slow-roll expansion for black holes embedded in
scalar-field dark-energy backgrounds was developed, incorporating scalar accretion and
its backreaction on the geometry \cite{Chadburn:2013mta}. Jacobson's
construction was also used to study phenomenological consequences,
including scalar dipole radiation and constraints from the binary quasar
OJ287 \cite{Horbatsch:2011ye}.

A separate literature established no-scalar-hair results under assumptions
of staticity, regularity, and specific scalar dynamics
\cite{Bekenstein:1995un,Hui:2012qt}, together with examples that evade
those assumptions through nonminimal scalar--tensor couplings or explicit
time dependence
\cite{Sotiriou:2013qea,Babichev:2013cya}; see
Ref.~\cite{Volkov:2016ehx} for a broader review. These works primarily ask
whether particular static or time-dependent scalar configurations can
exist, or how a prescribed cosmological scalar drives the evolution of an
individual black hole. The present problem is different: it concerns the
existence and uniqueness of the slowly forced deformation associated with
an entire manifold of static solutions, after the instantaneous
representative, the residual tangential ambiguity, and the exterior
matching data have been specified.

A related tradition is the moduli-space approximation for solitons
\cite{Manton:1981mp}, later applied to extremal and BPS black holes
\cite{Ferrell:1987gf,Gibbons:1986cq}. Such approaches project slow dynamics
onto the tangent space of a static solution manifold. The construction used
here shares this geometric starting point but differs in two respects.
First, the distinguished tangent direction is selected intrinsically from
the linearized gravitational constraints rather than by assuming an
orthogonality condition on the solution space. Second, the remaining
ambiguity is removed by a local slice at finite radius, which induces a
finite-rank modification of the reduced radial operator.

The general framework was developed in the companion paper
\emph{Adiabatic Deformations of Black Hole Moduli:
I.~Canonical Slices and Fredholm Solvability}, hereafter Paper~I
\cite{Benakli:2026PaperI}. There, the static solution manifold is
parametrized locally by a modulus label and a static horizon-mass parameter.
Differentiating the family defines an intrinsic covector on its tangent
space. Its kernel selects the canonical modulus direction at fixed static
horizon mass, whose scalar component is a zero mode of the unsliced reduced
operator.

After promotion of the collective coordinates, the exact solution is written
as an instantaneous static representative plus a lag field. A representation
convention separates the exact dynamical horizon from the horizon of the
instantaneous representative, while a local mass slice removes the remaining
tangential ambiguity. The slice converts the unsliced radial operator \(L\)
into the finite-rank perturbation
\begin{equation}
L_{\rm sl}
=
L
+
f_\varphi(r_{\rm m})
\,|K\rangle
\langle{\rm ev}_{r_{\rm m}}|.
\end{equation}
The canonical zero mode is generally displaced to a new one-dimensional
kernel generator \(\psi_\star\). Once an exterior matching functional
\(\mathcal B_{\rm m}\) is specified, existence and uniqueness of the
completed leading-order radial problem reduce to the single condition
\begin{equation}
\mathcal B_{\rm m}\psi_\star\neq0.
\end{equation}
The numerical value of this expression depends on the normalization chosen
for \(\psi_\star\), whereas its vanishing or non-vanishing does not.

The purpose of the present paper is to realize this construction explicitly
for the magnetic GHS family. We first derive the exact spherically symmetric
EMD system in advanced
Eddington--Finkelstein coordinates using the exact areal radius and the
Misner--Sharp mass function \cite{Misner:1964je}. We then specialize the
geometric and Fredholm framework of Paper~I to the GHS family.

The exact GHS profiles allow all intrinsic radial ingredients to be evaluated
analytically. We determine the canonical tangent
\(\varphi_\varphi\), the rank-one coefficient \(f_\varphi\), the source kernel
\(K\), a horizon-regular solution of \(L\chi_0=K\), and a globally regular
generator \(\widehat\psi_\star\) of the sliced kernel. These constructions are
exact on every nontrivial finite radial interval in the charged non-extremal
exterior and do not rely on a large-radius matching approximation.

For the model functional associated with the leading massless large-radius
overlap,
\begin{equation}
\mathcal B_{\rm m}
=
r_{\rm m}\partial_r+1,
\end{equation}
we compute the Fredholm denominator in closed form and prove the exact
algebraic inequality
\begin{equation}
\mathfrak D
\equiv
\mathcal B_{\rm m}\widehat\psi_\star
>0,
\qquad
0<a<1,
\qquad
b_{\rm m}\geq0.
\end{equation}
This includes every point at which the conventional normalization factor
\(\mathcal N_{\rm m}\) vanishes. The globally regular representative
\(\widehat\psi_\star\) shows that this locus is a normalization degeneracy
only and does not correspond to a discontinuity or bifurcation of the sliced
kernel.

The inequality above is an exact algebraic statement about the chosen model
functional acting on the exact GHS radial profiles. Its interpretation as
the non-degeneracy condition of a physical massless exterior matching problem
is controlled only in the large-radius overlap regime given in Eq.~\eqref{eq:introduction_controlled_overlap}, in which
\(\mathcal B_{\rm m}\) was derived. The normalization-degeneracy locus may
lie inside or outside this controlled regime depending on \(a\); outside the
overlap domain, the positivity remains mathematically valid but is not
interpreted as a physical exterior-matching statement.

The exact integrability of the GHS background also permits the forced radial
problem to be solved explicitly. Within the specified massless overlap model,
the scalar lag takes the form
\begin{equation}
\eta(v,r)
=
\dot\varphi_{\rm ext}(v)\,q_0(r)
+
\frac{
J_{\rm ext}(v)
-
\dot\varphi_{\rm ext}(v)\mathcal R_{\rm m}
}{
\mathfrak D
}
\widehat\psi_\star(r),
\end{equation}
where \(q_0\), \(\mathcal R_{\rm m}\), and
\(\mathfrak D\) are known analytically. The mass lag then follows
algebraically from the radial constraint, while the lapse lag is
reconstructed by one explicit elementary quadrature. The result is exact in
its radial dependence at leading adiabatic order. Its remaining exterior
inputs are the prescribed rolling rate
\(\dot\varphi_{\rm ext}(v)\) and the matching datum \(J_{\rm ext}(v)\).

The matching functional used for this explicit test is not claimed to be
universal. It is the leading Robin functional associated with the specified
massless large-radius overlap and with the convention that the exterior
modulus fixes the constant scalar mode. Subleading overlap corrections, a
different cosmological exterior, or a nonzero scalar potential may modify
both the matching functional and its inhomogeneous datum. Within the
controlled overlap regime and for admissible slowly varying matching data,
the strict positivity of \(\mathfrak D\) establishes existence and uniqueness
of the completed leading-order near-zone response. The construction of a
global exterior spacetime realizing a prescribed matching datum remains a
separate problem and is not established here.

We restrict the intrinsic GHS construction throughout to the charged
non-extremal branch,
\begin{equation}
0<a<1.
\end{equation}
The extremal endpoint lies outside the uniformly non-degenerate horizon
framework of Paper~I. The strict Schwarzschild endpoint is also degenerate
for the local mass slice used here, since the source kernel vanishes and the
slice ceases to be transverse. Both limits provide formal consistency checks
but do not belong to the domain of the sliced Fredholm construction. The
additional condition \(b_{\rm m}\gg1\) restricts the physical validity of the
particular massless matching model, not the intrinsic finite-interval GHS
operator construction.

The paper is organized as follows. Section~\ref{sec:EMD-background} introduces the
EMD theory, reviews the static magnetic GHS family,
derives the exact time-dependent near-zone system, and establishes the
relevant horizon identities. Section~\ref{sec:canonical_adiabatic_deformation} applies the canonical-tangent,
local-slice, and Fredholm construction of Paper~I while maintaining the
distinction between instantaneous and exact horizon quantities.
Section~\ref{sec:GHS_application} evaluates the canonical tangent, source kernel, displaced kernel
generator, and Fredholm denominator explicitly for the magnetic GHS family
and specifies the leading massless overlap matching model. Section~\ref{sec:physical_consequences}
constructs the forced response, reconstructs the scalar and metric-sector lag
fields, and discusses the role of the exterior data and possible extensions.
Section~\ref{sec:conclusions} summarizes the conclusions. Appendix~\ref{app:notation_dictionary}
collects the notation used in Paper~I, in the static GHS description, and in
the present analysis.

\section{Einstein--Maxwell--Dilaton theory and the magnetic GHS solution}
\label{sec:EMD-background}

In this section, we introduce the EMD theory that
underlies the present work. After presenting the general action and field
equations, we review the static magnetically-charged GHS black hole solution in its original
Schwarzschild-like coordinates. We then reformulate the solution in terms of
the exact areal radius, establishing the geometric framework used in the exact time-dependent Eddington--Finkelstein formulation developed in  Sec.~\ref{subsec:time_dependent_ansatz}.

\subsection{Einstein--Maxwell--dilaton theory}
\label{subsec:EMD}

We consider four-dimensional EMD theory in the
Einstein frame with metric signature \( (-,+,+,+) \). We define
\begin{equation}
F^2\equiv F_{\mu\nu}F^{\mu\nu},
\end{equation}
and relate the reduced Planck mass to Newton's constant through
\begin{equation}
\Mpl^{-2}=8\pi G.
\end{equation}

The action is
\begin{equation}
S=
\int d^4x\,\sqrt{-g}\,
\left[
\frac{\Mpl^2}{2}R
-\frac12\nabla_\mu\varphi\nabla^\mu\varphi
-V(\varphi)
-\frac14 B(\varphi)F_{\mu\nu}F^{\mu\nu}
\right].
\label{eq:EMD_action}
\end{equation}
Here, \(R\) is the Ricci scalar, \(\varphi\) is the dilaton (or modulus)
field, \(F_{\mu\nu}\) is the Maxwell field strength, \(V(\varphi)\) is the
scalar potential, and \(B(\varphi)\) is the gauge kinetic function.

The field equations are first derived for arbitrary
\(V(\varphi)\) and \(B(\varphi)\). The remainder of this paper specializes
to the massless magnetic GHS theory,
\begin{equation}
V(\varphi)=0,
\qquad
B(\varphi)=e^{-2\alpha\varphi/\Mpl},
\label{eq:GHS_coupling}
\end{equation}
with the canonical GHS coupling
\begin{equation}
\alpha=\frac{1}{\sqrt2},
\label{eq:GHS_alpha}
\end{equation}
in the normalization of Eq.~\eqref{eq:EMD_action}.

Varying the action, Eq.~\eqref{eq:EMD_action}, with respect to the metric, gauge
field, and scalar field yields the Einstein, Maxwell, and scalar equations.
The Einstein equations are
\begin{equation}
\Mpl^2 G_{\mu\nu}
=
T_{\mu\nu}^{(\varphi)}
+
T_{\mu\nu}^{(F)},
\label{eq:Einstein_general}
\end{equation}
where
\begin{equation}
T_{\mu\nu}^{(\varphi)}
=
\nabla_\mu\varphi\nabla_\nu\varphi
-
g_{\mu\nu}
\left[
\frac12(\nabla\varphi)^2
+
V(\varphi)
\right],
\label{eq:Tphi}
\end{equation}
and
\begin{equation}
T_{\mu\nu}^{(F)}
=
B(\varphi)
\left(
F_{\mu\rho}F_{\nu}{}^{\rho}
-
\frac14
g_{\mu\nu}
F_{\rho\sigma}F^{\rho\sigma}
\right).
\label{eq:TF}
\end{equation}

The Maxwell equation,
\begin{equation}
\nabla_\mu
\!\left(
B(\varphi)F^{\mu\nu}
\right)
=
0,
\label{eq:Maxwell_general}
\end{equation}
is supplemented by the Bianchi identity
\begin{equation}
\nabla_{[\mu}F_{\nu\rho]}=0.
\label{eq:Bianchi}
\end{equation}

The scalar equation is
\begin{equation}
\Box\varphi
=
V'(\varphi)
+
\frac14
B'(\varphi)
F_{\mu\nu}F^{\mu\nu},
\label{eq:scalar_general}
\end{equation}
where a prime denotes differentiation with respect to \(\varphi\).

These covariant equations will be rewritten exactly in advanced
Eddington--Finkelstein coordinates in
Sec.~\ref{subsec:time_dependent_ansatz}, where they reduce to the minimal
evolution system underlying the adiabatic construction.

We now specialize them to the static magnetic GHS family, which provides the reference background for the remainder of this work.


\subsection{Static magnetic GHS family}
\label{subsec:static_GHS}

We now specialize the EMD equations derived above to the canonical magnetic
GHS branch introduced in Sec.~\ref{subsec:EMD}. This family provides the
static reference background whose slowly time-dependent deformations will be
studied throughout the remainder of the paper.

For the canonical coupling in Eq.~\eqref{eq:GHS_alpha}, the static,
spherically symmetric magnetic black hole solution may be written in the
conventional GHS radial coordinate \(\rho\) as
\begin{equation}
ds^2
=
-f(\rho)\,dt^2
+
\frac{d\rho^2}{f(\rho)}
+
R^2(\rho)\,d\Omega_2^2,
\label{eq:ghs_metric}
\end{equation}
with
\begin{align}
f(\rho)
&=
1-\frac{r_+}{\rho},
\\
R^2(\rho)
&=
\rho(\rho-r_-).
\end{align}

The magnetic field strength is
\begin{equation}
F
=
Q_m\sin\theta\,d\theta\wedge d\phi,
\label{eq:magnetic_field}
\end{equation}
where \(Q_m\) is constant by virtue of the Bianchi identity,
Eq.~\eqref{eq:Bianchi}. We choose the orientation of the magnetic field so
that \(Q_m>0\); reversing its sign leaves the geometry and dilaton profile
unchanged.

The static dilaton profile is
\begin{equation}
e^{-2\alpha\varphi_{\rm GHS}(\rho)/\Mpl}
=
e^{-2\alpha\varphi/\Mpl}
\left(
1-\frac{r_-}{\rho}
\right),
\label{eq:ghs_dilaton}
\end{equation}
where the undecorated quantity \(\varphi\) denotes the asymptotic modulus
label,
\begin{equation}
\lim_{\rho\to\infty}
\varphi_{\rm GHS}(\rho)
=
\varphi.
\end{equation}
This convention distinguishes the constant label \(\varphi\) from the
radially varying profile \(\varphi_{\rm GHS}(\rho)\).

We henceforth restrict to a sector of fixed magnetic charge \(Q_m\). In this
sector, the GHS parameters obey the charge relation
\begin{equation}
r_+r_-
=
\frac{Q_m^2}{\Mpl^2}
e^{-2\alpha\varphi/\Mpl}.
\label{eq:GHS_charge_relation}
\end{equation}
The outer event horizon is located at
\begin{equation}
\rho=r_+,
\end{equation}
while \(r_-\) controls the dilatonic deformation away from the Schwarzschild
solution.

It is convenient to introduce the dimensionless non-extremality parameter
\begin{equation}
a
\equiv
\frac{r_+-r_-}{r_+},
\label{eq:a_parameter}
\end{equation}
with
\begin{equation}
0\leq a\leq1.
\end{equation}
The endpoint \(a=1\) corresponds to \(r_-=0\) and hence, for finite
asymptotic modulus, to the uncharged Schwarzschild solution. The extremal GHS
limit is obtained as \(a\to0\). Throughout the Fredholm construction below,
we restrict to the charged non-extremal branch
\begin{equation}
0<a<1.
\end{equation}

It is important to distinguish the conventional GHS coordinate \(\rho\)
from the exact areal radius used throughout Paper~I and in the
Eddington--Finkelstein formulation adopted below. The areal radius is
\begin{equation}
r
=
R(\rho)
=
\sqrt{\rho(\rho-r_-)},
\label{eq:areal_radius_dictionary}
\end{equation}
so that the static horizon has areal radius
\begin{equation}
r_{{\rm H},0}
=
R(r_+)
=
\sqrt{r_+(r_+-r_-)}
=
r_+\sqrt a.
\label{eq:rH-static}
\end{equation}

The corresponding static horizon-mass parameter is
\begin{equation}
M_{{\rm H},0}
=
\frac{\Mpl^2}{2}
r_{{\rm H},0}.
\label{eq:MH-static}
\end{equation}
Anticipating the dynamical notation used below, the static and exact horizon
quantities coincide on the strictly static solution:
\begin{equation}
r_{\rm H}^{\rm exact}
=
r_{{\rm H},0},
\qquad
M_{\rm H}^{\rm exact}
=
M_{{\rm H},0}.
\label{eq:static-exact-horizon-coincidence}
\end{equation}
After adiabatic promotion, \(r_{{\rm H},0}\) and \(M_{{\rm H},0}\) become
the horizon radius and horizon-mass parameter of the instantaneous static
representative, respectively, whereas \(r_{\rm H}^{\rm exact}\) and
\(M_{\rm H}^{\rm exact}\) refer to the full dynamical solution. These
quantities will be distinguished explicitly in
Sec.~\ref{sec:canonical_adiabatic_deformation}.

Using Eqs.~\eqref{eq:GHS_charge_relation} and
\eqref{eq:a_parameter}, the static horizon-mass parameter can be expressed
in terms of the asymptotic modulus label and the non-extremality parameter as
\begin{equation}
M_{{\rm H},0}(\varphi,a)
=
\frac{\Mpl Q_m}{2}
e^{-\alpha\varphi/\Mpl}
\sqrt{\frac{a}{1-a}},
\qquad
0<a<1.
\label{eq:MH0-Phi-a}
\end{equation}
At fixed \(\varphi\) and fixed magnetic charge \(Q_m>0\), this function is
strictly increasing:
\begin{equation}
\left.
\frac{\partial M_{{\rm H},0}}{\partial a}
\right|_{\varphi,Q_m}
=
\frac{\Mpl Q_m}{4}
e^{-\alpha\varphi/\Mpl}
\frac{1}{\sqrt a\,(1-a)^{3/2}}
>0.
\label{eq:dMH0-da-positive}
\end{equation}
Moreover,
\begin{equation}
M_{{\rm H},0}\longrightarrow0
\quad\text{as}\quad a\to0^+,
\qquad
M_{{\rm H},0}\longrightarrow\infty
\quad\text{as}\quad a\to1^-.
\end{equation}
The relation may therefore be inverted uniquely:
\begin{equation}
a
=
\frac{
M_{{\rm H},0}^{\,2}
}{
M_{{\rm H},0}^{\,2}
+
\displaystyle
\frac{\Mpl^2Q_m^2}{4}
e^{-2\alpha\varphi/\Mpl}
}.
\label{eq:a-Phi-MH0}
\end{equation}
Consequently, in the fixed-charge sector, the map
\begin{equation}
(\varphi,a)
\longmapsto
(\varphi,M_{{\rm H},0})
\end{equation}
is a smooth one-to-one change of coordinates on the charged non-extremal
static solution manifold. Thus
\((\varphi,a)\) and
\((\varphi,M_{{\rm H},0})\)
provide equivalent smooth coordinate systems on this branch.

Following our previous static analysis \cite{Benakli:2026zbm}, we introduce
the dimensionless throat coordinate
\begin{equation}
b
\equiv
\frac{\rho-r_+}{r_+},
\label{eq:b_coordinate}
\end{equation}
so that the event horizon is located at \(b=0\), while spatial infinity
corresponds to \(b\to\infty\).

In terms of the throat coordinate, the static dilaton profile becomes
\begin{equation}
\varphi_{\rm GHS}(b)
=
\varphi
+
\frac{\Mpl}{2\alpha}
\ln \!
\left[
\frac{1+b}{a+b}
\right],
\label{eq:phi_GHS_b}
\end{equation}
which will serve as the reference profile throughout this work.

In the following section, we rewrite the GHS solution in advanced
Eddington--Finkelstein coordinates using the exact areal radius \(r\),
thereby bringing it into the form employed by the general adiabatic
framework developed in Paper~I.

\subsection{Choice of radial coordinate and time-dependent ansatz}
\label{subsec:time_dependent_ansatz}

The static GHS solution reviewed in Sec.~\ref{subsec:static_GHS} is expressed
in terms of the conventional Schwarzschild-like radial coordinate \(\rho\),
for which the area of the symmetry spheres is
\begin{equation}
4\pi R^2(\rho),
\qquad
R^2(\rho)=\rho(\rho-r_-).
\end{equation}
Thus, \(\rho\) coincides with the areal radius only in the Schwarzschild limit
\(r_-=0\).

For the time-dependent formulation, we instead use the exact areal radius as
the radial coordinate. On the static GHS family, the two radial coordinates
are related by
\begin{equation}
r
=
R(\rho)
=
\sqrt{\rho(\rho-r_-)},
\label{eq:areal_radius_from_GHS}
\end{equation}
with inverse
\begin{equation}
\rho(r)
=
\frac12
\left(
r_-
+
\sqrt{r_-^2+4r^2}
\right),
\label{eq:GHS_coordinate_inverse}
\end{equation}
where the positive branch is selected by the requirement
\(\rho(r)\sim r\) as \(r\to\infty\).

Eq.~\eqref{eq:GHS_coordinate_inverse} is only the coordinate dictionary
relating the two descriptions of a static GHS solution. It is not promoted
to a time-dependent coordinate transformation. The exact dynamical fields
introduced below are independent fields and are not obtained by replacing
the static GHS parameters by functions of time.

For the static solution, the horizon is located at the areal radius
\(r_{{\rm H},0}\) defined in Eq.~\eqref{eq:rH-static}, and
\begin{equation}
r_{\rm H}^{\rm exact}
=
r_{{\rm H},0} \;,
\end{equation}
as stated in Eq.~\eqref{eq:static-exact-horizon-coincidence}. Once time
dependence is introduced, the exact dynamical horizon will in general differ
from the horizon of the instantaneous static representative, as discussed in
Sec.~\ref{sec:canonical_adiabatic_deformation}. From this point onward,
\(r\) always denotes the exact areal-radius coordinate, whereas \(\rho\) is
reserved for the conventional static GHS coordinate.

We now allow the geometry and the scalar field to depend on an advanced null
coordinate \(v\). In spherical symmetry, we adopt the ingoing
Eddington--Finkelstein gauge
\begin{equation}
ds^2
=
-e^{2\delta(v,r)}\mathcal F(v,r)\,dv^2
+
2e^{\delta(v,r)}\,dv\,dr
+
r^2d\Omega_2^2,
\label{eq:EF_metric_ansatz}
\end{equation}
where
\begin{equation}
\mathcal F(v,r)
=
1-\frac{2m(v,r)}{\Mpl^2r}.
\label{eq:EF_mass_function}
\end{equation}
Here, \(\delta(v,r)\) is a dimensionless lapse function and \(m(v,r)\) is
the Misner--Sharp quasi-local mass function. The scalar field is
\begin{equation}
\varphi=\varphi(v,r).
\label{eq:scalar_time_dependent_ansatz}
\end{equation}
No boundary relation between \(\varphi(v,r_{\rm m})\), evaluated at the matching surface \(r_{\rm m}\), and the prescribed
slowly varying exterior modulus \(\varphi_{\rm ext}(v)\) is imposed at this
stage. Their relation will be supplied later by the exterior matching
problem.

The exact future marginal horizon is defined by the vanishing of the outgoing
null expansion. In the gauge, Eq.~\eqref{eq:EF_metric_ansatz}, this condition is
equivalent to
\begin{equation}
\mathcal F
\bigl(
v,r_{\rm H}^{\rm exact}(v)
\bigr)
=
0.
\label{eq:marginal_horizon_definition}
\end{equation}
Provided that \(\delta\), \(m\), and the matter fields remain finite, the
ingoing chart is regular across a smooth future horizon even when
\(\mathcal F=0\). The vanishing of \(\mathcal F\) therefore identifies the
marginal horizon rather than a coordinate singularity.

The ansatz retains the residual freedom to reparametrize the advanced time,
\begin{equation}
v\longrightarrow\widetilde v(v).
\end{equation}
Under this transformation,
\begin{equation}
e^{\widetilde\delta(\widetilde v,r)}
=
e^{\delta(v,r)}
\frac{dv}{d\widetilde v}.
\label{eq:residual_time_reparametrization}
\end{equation}
For the finite near-zone problem, this freedom is fixed by imposing
\begin{equation}
\delta(v,r_{\rm m})=0,
\label{eq:matching_gauge_condition}
\end{equation}
where \(r_{\rm m}\) serves as a fixed areal matching radius chosen in the overlap
region between the near-zone geometry and the slowly varying exterior
background.

In the magnetic sector, spherical symmetry admits the exact ansatz, as in
Eq.~\eqref{eq:magnetic_field},
\begin{equation}
F
=
Q_m\sin\theta\,d\theta\wedge d\phi,
\label{eq:time_dependent_magnetic_ansatz}
\end{equation}
equivalently, the only independent non-zero coordinate components are
\begin{equation}
F_{\theta\phi} =-F_{\phi \theta}
=
Q_m\sin\theta.
\end{equation}
In the absence of magnetic sources, the Bianchi identity implies
\begin{equation}
\partial_vQ_m
=
\partial_rQ_m
=
0,
\label{eq:magnetic_charge_constant}
\end{equation}
so that \(Q_m\) is an exact constant. 
%
%
For this magnetic ansatz, the Maxwell equation is identically satisfied for
an arbitrary spherically symmetric scalar field \(\varphi(v,r)\) and an
arbitrary gauge kinetic function \(B(\varphi)\).

Equations~\eqref{eq:EF_metric_ansatz},
\eqref{eq:EF_mass_function},
\eqref{eq:scalar_time_dependent_ansatz}, and
\eqref{eq:time_dependent_magnetic_ansatz}
define the exact time-dependent near-zone ansatz used throughout the
remainder of this work.

In the next subsection, we evaluate the geometric identities implied by this
ansatz and reduce the covariant EMD equations to the minimal exact system of
evolution and constraint equations governing the dynamical fields.


\subsection{Geometric identities}
\label{subsec:verified_identities}

Before reducing the field equations, we collect several identities that
follow directly from the Eddington--Finkelstein ansatz introduced in
Sec.~\ref{subsec:time_dependent_ansatz}. The metric and matter identities
derived first require no use of the field equations. The dependence of the
angular Einstein equation is then established using the contracted Bianchi
identity together with the matter equations.

For the metric introduced in Eq.~\eqref{eq:EF_metric_ansatz},
\begin{equation}
ds^2
=
-e^{2\delta}\mathcal F\,dv^2
+
2e^\delta\,dv\,dr
+
r^2d\Omega_2^2,
\end{equation}
the nonvanishing components of the inverse metric are
\begin{equation}
g^{vr}
=
g^{rv}
=
e^{-\delta},
\qquad
g^{rr}
=
\mathcal F,
\qquad
g^{\theta\theta}
=
\frac{1}{r^2},
\qquad
g^{\phi\phi}
=
\frac{1}{r^2\sin^2\theta},
\end{equation}
while
\begin{equation}
g^{vv}=0,
\qquad
\sqrt{-g}
=
e^\delta r^2\sin\theta.
\end{equation}

Since the scalar field depends only on \(v\) and \(r\), its kinetic term is
\begin{equation}
(\nabla\varphi)^2
=
2e^{-\delta}\varphi_{,v}\varphi_{,r}
+
\mathcal F\,\varphi_{,r}^{\,2}.
\label{eq:scalar_kinetic_EF}
\end{equation}

For the purely magnetic ansatz,  the electromagnetic invariant is
\begin{equation}
F_{\mu\nu}F^{\mu\nu}
=
\frac{2Q_m^2}{r^4}.
\label{eq:F2_exact}
\end{equation}
Moreover,
\begin{equation}
F_{\mu\rho}F_\nu{}^\rho
=
0,
\qquad
\mu,\nu\in\{v,r\}.
\end{equation}
Consequently, the \(v\)-\(r\) components of the electromagnetic stress
tensor reduce to
\begin{equation}
T^{(F)}_{\mu\nu}
=
-\frac14 B(\varphi)g_{\mu\nu}F_{\rho\sigma}F^{\rho\sigma},
\qquad
\mu,\nu\in\{v,r\},
\end{equation}
and therefore satisfy the exact algebraic identity
\begin{equation}
T^{(F)}_{vv}
+
e^\delta\mathcal F\,
T^{(F)}_{vr}
=
0.
\label{eq:TF_identity}
\end{equation}

A useful purely geometric combination of Einstein-tensor components is
\begin{equation}
G_{vv}
+
e^\delta\mathcal F\,
G_{vr}
=
-
\frac{e^\delta}{r}\,
\partial_v\mathcal F.
\label{eq:Einstein_identity}
\end{equation}
Together with Eq.~\eqref{eq:TF_identity}, this identity yields the exact
evolution equation for the Misner--Sharp mass in the following subsection.

The angular Einstein equation is not independent. We may define the total
Einstein equation residual by
\begin{equation}
E_{\mu\nu}
\equiv
\Mpl^2G_{\mu\nu}
-
T_{\mu\nu},
\qquad
T_{\mu\nu}
\equiv
T_{\mu\nu}^{(\varphi)}
+
T_{\mu\nu}^{(F)}.
\end{equation}
The contracted Bianchi identity, together with the scalar and Maxwell
equations, implies
\begin{equation}
\nabla^\mu E_{\mu\nu}=0.
\end{equation}
For the magnetic ansatz, the Maxwell equation is satisfied identically. In
spherical symmetry,
\begin{equation}
E_{\phi\phi}
=
\sin^2\theta\,E_{\theta\theta}.
\end{equation}
Once the \((rr)\), \((vr)\), and \((vv)\) Einstein equations and the scalar
equation are satisfied, the \(\nu=r\) component of
\(\nabla^\mu E_{\mu\nu}=0\) reduces to the angular Einstein equation and
implies
\begin{equation}
E_{\theta\theta}=0.
\end{equation}
Thus, no independent angular Einstein equation remains.

These identities provide the geometric and algebraic input required to
reduce the covariant EMD equations to the exact minimal system derived in
the next subsection.


\subsection{Exact minimal system}
\label{subsec:exact_minimal_system}

Combining the covariant EMD equations derived in
Sec.~\ref{subsec:EMD} with the geometric identities established in
Sec.~\ref{subsec:verified_identities} yields a closed exact system for the
three dynamical fields,
\begin{equation}
\varphi(v,r),
\qquad
\delta(v,r),
\qquad
m(v,r).
\end{equation}
No adiabatic expansion or instantaneous-static ansatz is used in this
reduction.

The radial \((rr)\) Einstein equation determines the radial variation of the
lapse function,
\begin{equation}
\partial_r\delta
=
\frac{r}{2\Mpl^2}
\left(\partial_r\varphi\right)^2.
\label{eq:delta_exact}
\end{equation}
The \((vr)\) Einstein equation gives the radial constraint for the
Misner--Sharp mass,
\begin{equation}
\partial_r m
=
\frac{r^2}{4}
\mathcal F
\left(\partial_r\varphi\right)^2
+
\frac{r^2}{2}
V(\varphi)
+
\frac{Q_m^2}{4r^2}
B(\varphi).
\label{eq:mr_exact}
\end{equation}
Using the combination
\(G_{vv}+e^\delta\mathcal F G_{vr}\)
derived in Sec.~\ref{subsec:verified_identities}, the remaining independent
Einstein equation becomes the exact mass-balance relation,
\begin{equation}
\partial_v m
=
\frac{r^2}{2}
\left[
e^{-\delta}
\left(\partial_v\varphi\right)^2
+
\mathcal F
\left(\partial_v\varphi\right)
\left(\partial_r\varphi\right)
\right].
\label{eq:mv_exact}
\end{equation}

For later convenience, we define the scalar source
\begin{equation}
\mathcal S(\varphi,r)
\equiv
V'(\varphi)
+
\frac{Q_m^2}{2r^4}
B'(\varphi).
\label{eq:scalar_source_exact}
\end{equation}
The scalar equation then takes the conservative form
\begin{equation}
0
=
\partial_v\!\left(r^2\partial_r\varphi\right)
+
\partial_r\!\left[
r^2\partial_v\varphi
+
e^\delta r^2\mathcal F\,\partial_r\varphi
\right]
-
e^\delta r^2\mathcal S(\varphi,r).
\label{eq:scalar_exact_conservative}
\end{equation}
Equivalently, after expanding the derivatives,
\begin{align}
2e^{-\delta}\partial_v\partial_r\varphi
+
\frac{2e^{-\delta}}{r}\partial_v\varphi
&+
\left(
\frac{2\mathcal F}{r}
+
\partial_r\mathcal F
+
\mathcal F\,\partial_r\delta
\right)
\partial_r\varphi
+
\mathcal F\,\partial_r^2\varphi
\nonumber\\
&=
\mathcal S(\varphi,r).
\label{eq:scalar_exact}
\end{align}

Together with
\begin{equation}
\mathcal F(v,r)
=
1-
\frac{2m(v,r)}{\Mpl^2r},
\label{eq:F_definition_again}
\end{equation}
Eqs.~\eqref{eq:delta_exact},
\eqref{eq:mr_exact},
\eqref{eq:mv_exact}, and
\eqref{eq:scalar_exact_conservative}
form the exact minimal system governing the spherically symmetric magnetic
sector.

The magnetic charge \(Q_m\) is the exact constant established in
Eq.~\eqref{eq:magnetic_charge_constant}. For the purely magnetic ansatz, the
Maxwell equation is satisfied identically, while the angular Einstein
equation follows from the equations displayed above by the contracted
Bianchi identity, as discussed in
Sec.~\ref{subsec:verified_identities}.

 
\subsection{Static limit}
\label{subsec:static_limit}

We now verify that the exact Eddington--Finkelstein system of
Sec.~\ref{subsec:exact_minimal_system} reproduces the static magnetic EMD
equations in the exact areal-radius coordinate and, in particular, the
magnetic GHS solution reviewed in Sec.~\ref{subsec:static_GHS}.

We impose
\begin{equation}
\partial_v\varphi
=
\partial_v m
=
\partial_v\delta
=
0
\label{eq:static_limit_conditions}
\end{equation}
and write
\begin{equation}
\varphi(v,r)=\varphi_0(r),
\qquad
m(v,r)=m_0(r),
\qquad
\delta(v,r)=\delta_0(r),
\qquad
\mathcal F(v,r)=\mathcal F_0(r),
\end{equation}
with
\begin{equation}
\mathcal F_0(r)
=
1-\frac{2m_0(r)}{\Mpl^2r}.
\label{eq:F0_definition}
\end{equation}

The lapse equation, Eq.~\eqref{eq:delta_exact}, reduces to
\begin{equation}
\delta_0'
=
\frac{r}{2\Mpl^2}
\left(\varphi_0'\right)^2.
\label{eq:static_delta_equation}
\end{equation}
The radial mass constraint, Eq.~\eqref{eq:mr_exact}, becomes
\begin{equation}
m_0'
=
\frac{r^2}{4}\,
\mathcal F_0
\left(\varphi_0'\right)^2
+
\frac{r^2}{2}\,
V(\varphi_0)
+
\frac{Q_m^2}{4r^2}\,
B(\varphi_0).
\label{eq:static_mass_equation}
\end{equation}
The mass-balance equation, Eq.~\eqref{eq:mv_exact}, is then satisfied
identically.

The scalar equation reduces to
\begin{equation}
\mathcal F_0\varphi_0''
+
\left(
\frac{2\mathcal F_0}{r}
+
\mathcal F_0'
+
\mathcal F_0\delta_0'
\right)
\varphi_0'
=
V'(\varphi_0)
+
\frac{Q_m^2}{2r^4}\,
B'(\varphi_0).
\label{eq:static_scalar_equation_expanded}
\end{equation}
Equivalently, in conservative form,
\begin{equation}
\frac{e^{-\delta_0}}{r^2}
\frac{d}{dr}
\left(
e^{\delta_0}r^2\mathcal F_0\varphi_0'
\right)
=
V'(\varphi_0)
+
\frac{Q_m^2}{2r^4}\,
B'(\varphi_0).
\label{eq:static_scalar_equation}
\end{equation}
Eqs.~\eqref{eq:static_delta_equation},
\eqref{eq:static_mass_equation}, and
\eqref{eq:static_scalar_equation}
are the static magnetic EMD equations in the exact areal-radius coordinate
\(r\).

We now specialize to the massless magnetic GHS theory,
\begin{equation}
V(\varphi)=0,
\qquad
B(\varphi)=e^{-2\alpha\varphi/\Mpl},
\qquad
\alpha=\frac{1}{\sqrt2}.
\label{eq:GHS_theory_again}
\end{equation}
In the conventional Schwarzschild-like coordinate \(\rho\), the static
solution is
\begin{align}
ds^2
&=
-
\left(1-\frac{r_+}{\rho}\right)dt^2
+
\left(1-\frac{r_+}{\rho}\right)^{-1}d\rho^2
+
\rho(\rho-r_-)\,d\Omega_2^2,
\label{eq:GHS_static_metric_again}
\\
\varphi_{\rm GHS}(\rho)
&=
\varphi
+
\frac{\Mpl}{2\alpha}
\ln\!\left[
\frac{\rho}{\rho-r_-}
\right],
\label{eq:GHS_scalar_again}
\end{align}
with the charge relation
\begin{equation}
Q_m^2e^{-2\alpha\varphi/\Mpl}
=
\Mpl^2r_+r_-.
\label{eq:GHS_charge_relation_again}
\end{equation}

We introduce an advanced coordinate \(v_\infty\) adapted to the standard
asymptotic normalization,
\begin{equation}
dv_\infty
=
dt
+
\frac{d\rho}{1-r_+/\rho}.
\label{eq:GHS_advanced_time}
\end{equation}
The metric then becomes
\begin{equation}
ds^2
=
-
\left(1-\frac{r_+}{\rho}\right)dv_\infty^2
+
2\,dv_\infty\,d\rho
+
\rho(\rho-r_-)\,d\Omega_2^2.
\label{eq:GHS_metric_EF_rho}
\end{equation}
Passing to the exact areal radius,
\begin{equation}
r^2=\rho(\rho-r_-),
\label{eq:GHS_areal_relation_again}
\end{equation}
gives
\begin{equation}
ds^2
=
-
\left(1-\frac{r_+}{\rho(r)}\right)dv_\infty^2
+
2\frac{d\rho}{dr}\,dv_\infty\,dr
+
r^2d\Omega_2^2,
\label{eq:GHS_metric_EF_areal}
\end{equation}
where, as in Eq.~\eqref{eq:GHS_coordinate_inverse},
\begin{equation}
\rho(r)
=
\frac12
\left(
r_-+\sqrt{r_-^2+4r^2}
\right).
\label{eq:GHS_rho_of_r_again}
\end{equation}

Comparison with the Eddington--Finkelstein ansatz of
Sec.~\ref{subsec:time_dependent_ansatz} yields the static fields in the
asymptotic normalization:
\begin{align}
e^{\delta_0^{(\infty)}(r)}
&=
\frac{d\rho}{dr}
=
\frac{2r}{2\rho(r)-r_-},
\label{eq:GHS_delta_areal_infinity}
\\
\mathcal F_0(r)
&=
\left(
1-\frac{r_+}{\rho(r)}
\right)
\left(
\frac{dr}{d\rho}
\right)^2
\nonumber\\
&=
\left(
1-\frac{r_+}{\rho(r)}
\right)
\frac{\bigl(2\rho(r)-r_-\bigr)^2}{4r^2},
\label{eq:GHS_F_areal}
\\
\varphi_0(r)
&=
\varphi
+
\frac{\Mpl}{2\alpha}
\ln\!\left[
\frac{\rho(r)}{\rho(r)-r_-}
\right].
\label{eq:GHS_phi_areal}
\end{align}
The corresponding Misner--Sharp mass function is
\begin{equation}
m_0(r)
=
\frac{\Mpl^2r}{2}
\left[
1-
\left(
1-\frac{r_+}{\rho(r)}
\right)
\frac{\bigl(2\rho(r)-r_-\bigr)^2}{4r^2}
\right].
\label{eq:GHS_mass_areal}
\end{equation}

In this normalization,
\begin{equation}
\delta_0^{(\infty)}(\infty)=0.
\end{equation}
The finite near-zone construction instead uses
\begin{equation}
\delta_0(r_{\rm m})=0,
\end{equation}
as imposed in Eq.~\eqref{eq:matching_gauge_condition}. The required
normalization is obtained by the constant reparametrization
\begin{equation}
dv
=
e^{\delta_0^{(\infty)}(r_{\rm m})}\,dv_\infty,
\end{equation}
under which
\begin{equation}
\delta_0(r)
=
\delta_0^{(\infty)}(r)
-
\delta_0^{(\infty)}(r_{\rm m}).
\label{eq:static_matching_normalization}
\end{equation}
Defining
\begin{equation}
D(r)
\equiv
2\rho(r)-r_-
=
\sqrt{r_-^2+4r^2},
\qquad
D_{\rm m}
\equiv
D(r_{\rm m}),
\label{eq:D_definition_static}
\end{equation}
one obtains
\begin{equation}
e^{\delta_0(r)}
=
\frac{
e^{\delta_0^{(\infty)}(r)}
}{
e^{\delta_0^{(\infty)}(r_{\rm m})}
}
=
\frac{rD_{\rm m}}{r_{\rm m}D(r)}.
\label{eq:GHS_delta_matching}
\end{equation}
This constant rescaling of the advanced-time coordinate shifts
\(\delta_0\) by a constant but leaves
\(\mathcal F_0(r)\), \(m_0(r)\), and \(\varphi_0(r)\) unchanged.

Direct substitution of
Eqs.~\eqref{eq:GHS_F_areal},
\eqref{eq:GHS_phi_areal},
\eqref{eq:GHS_mass_areal}, and
\eqref{eq:GHS_delta_matching}
into Eqs.~\eqref{eq:static_delta_equation}--
\eqref{eq:static_scalar_equation}, together with
Eqs.~\eqref{eq:GHS_theory_again} and
\eqref{eq:GHS_charge_relation_again}, verifies all three static equations
identically.

The static horizon is located at
\begin{equation}
\mathcal F_0(r_{{\rm H},0})=0,
\qquad
r_{{\rm H},0}
=
\sqrt{r_+(r_+-r_-)}.
\label{eq:static_GHS_horizon_areal}
\end{equation}
For the strictly static configuration, this radius coincides with the exact
marginal-horizon radius introduced in
Sec.~\ref{subsec:time_dependent_ansatz}. After promotion of the static
family, \(r_{{\rm H},0}\) becomes the horizon
\(r_{\rm H}^{\rm inst}(v)\) of the instantaneous static representative and
must be distinguished from the exact dynamical horizon
\(r_{\rm H}^{\rm exact}(v)\).

The advanced Eddington--Finkelstein formulation is therefore an exact
rewriting of the static magnetic GHS solution in the exact areal-radius
coordinate. No additional approximation has been introduced.

\subsection{Horizon dynamics}
\label{subsec:horizon_dynamics}

The Eddington--Finkelstein formulation gives direct access to the evolution
of the future outer marginal horizon. Since \(r\) is the exact areal radius,
the horizon is the marginally trapped tube
\(r=r_{\rm H}^{\rm exact}(v)\) defined by
\begin{equation}
\mathcal F\!\left(
v,r_{\rm H}^{\rm exact}(v)
\right)
=
0.
\label{eq:apparent_horizon}
\end{equation}
Using
\begin{equation}
\mathcal F(v,r)
=
1-
\frac{2m(v,r)}{\Mpl^2r},
\end{equation}
this condition gives the exact kinematic identity
\begin{equation}
M_{\rm H}^{\rm exact}(v)
\equiv
m\!\left(
v,r_{\rm H}^{\rm exact}(v)
\right)
=
\frac{\Mpl^2}{2}\,
r_{\rm H}^{\rm exact}(v).
\label{eq:horizon_mass_relation}
\end{equation}
This relation holds for any spherically symmetric configuration admitting a
marginal horizon of the form~\eqref{eq:apparent_horizon}. It relies neither
on staticity nor on an adiabatic approximation.

Differentiating the horizon condition along the marginal-horizon tube gives
\begin{equation}
\left.
\partial_v\mathcal F
\right|_{\rm H}
+
\dot r_{\rm H}^{\rm exact}
\left.
\partial_r\mathcal F
\right|_{\rm H}
=
0,
\label{eq:horizon_evolution_general}
\end{equation}
where a subscript \({\rm H}\) denotes evaluation at
\(r=r_{\rm H}^{\rm exact}(v)\), and
\begin{equation}
\dot r_{\rm H}^{\rm exact}
\equiv
\frac{dr_{\rm H}^{\rm exact}}{dv}.
\end{equation}
Provided the horizon is non-degenerate,
\begin{equation}
\left.
\partial_r\mathcal F
\right|_{\rm H}
\neq0,
\end{equation}
one obtains
\begin{equation}
\dot r_{\rm H}^{\rm exact}
=
-
\left.
\frac{\partial_v\mathcal F}
{\partial_r\mathcal F}
\right|_{\rm H}.
\label{eq:horizon_velocity}
\end{equation}

From the definition of \(\mathcal F\),
\begin{equation}
\partial_v\mathcal F
=
-
\frac{2}{\Mpl^2r}\,
\partial_vm.
\label{eq:Fv_mass}
\end{equation}
Evaluating the exact mass-balance equation
\eqref{eq:mv_exact} on the horizon, where
\(\mathcal F|_{\rm H}=0\), gives
\begin{equation}
\left.
\partial_vm
\right|_{\rm H}
=
\frac{
\left(r_{\rm H}^{\rm exact}\right)^2
}{2}
e^{-\delta_{\rm H}}
\left(
\left.\partial_v\varphi\right|_{\rm H}
\right)^2.
\label{eq:horizon_mass_flux}
\end{equation}
Substitution into Eq.~\eqref{eq:horizon_velocity} yields the exact
horizon-growth law
\begin{equation}
\dot r_{\rm H}^{\rm exact}
=
\frac{
r_{\rm H}^{\rm exact}
e^{-\delta_{\rm H}}
\left(
\left.\partial_v\varphi\right|_{\rm H}
\right)^2
}{
\Mpl^2
\left.\partial_r\mathcal F\right|_{\rm H}
}.
\label{eq:exact_horizon_growth}
\end{equation}

With the advanced-time normalization fixed at \(r=r_{\rm m}\), we define the
normalization-dependent instantaneous surface gravity parameter
\begin{equation}
\kappa_{\rm H}^{\rm exact}
\equiv
\frac12
e^{\delta_{\rm H}}
\left.
\partial_r\mathcal F
\right|_{\rm H}.
\label{eq:exact_surface_gravity}
\end{equation}
For a static solution, this reduces to the Killing surface gravity in the
same normalization of the advanced-time coordinate. For a dynamical
marginal horizon, it is used only as a convenient instantaneous parameter;
no unique notion of dynamical surface gravity is assumed.

For a non-degenerate future outer marginal horizon,
\begin{equation}
\left.
\partial_r\mathcal F
\right|_{\rm H}>0,
\qquad
\kappa_{\rm H}^{\rm exact}>0,
\end{equation}
and Eq.~\eqref{eq:exact_horizon_growth} becomes
\begin{equation}
\dot r_{\rm H}^{\rm exact}
=
\frac{
r_{\rm H}^{\rm exact}
}{
2\Mpl^2\kappa_{\rm H}^{\rm exact}
}
\left(
\left.\partial_v\varphi\right|_{\rm H}
\right)^2
\geq0.
\label{eq:horizon_monotonicity}
\end{equation}
Thus, the exact marginal-horizon radius is non-decreasing, with equality
precisely when
\(\left.\partial_v\varphi\right|_{\rm H}=0\).

Using the kinematic identity, Eq.~\eqref{eq:horizon_mass_relation}, the total
derivative of the exact horizon mass is
\begin{equation}
\dot M_{\rm H}^{\rm exact}
\equiv
\frac{d}{dv}
m\!\left(
v,r_{\rm H}^{\rm exact}(v)
\right)
=
\frac{\Mpl^2}{2}\,
\dot r_{\rm H}^{\rm exact}
=
\frac{
r_{\rm H}^{\rm exact}
}{
4\kappa_{\rm H}^{\rm exact}
}
\left(
\left.\partial_v\varphi\right|_{\rm H}
\right)^2
\geq0.
\label{eq:exact_horizon_mass_growth}
\end{equation}
Note carefully that the partial derivative
\(\left.\partial_vm\right|_{\rm H}\) appearing in
Eq.~\eqref{eq:horizon_mass_flux} should not be confused with this total
derivative, since the latter also accounts for the motion of the horizon.
Both Eqs.~\eqref{eq:horizon_monotonicity} and
\eqref{eq:exact_horizon_mass_growth} are exact and require no adiabatic
approximation.

In the slowly varying regime,
\begin{equation}
\left.\partial_v\varphi\right|_{\rm H}
=
O_{\rm ad}(\epsilon),
\end{equation}
so that
\begin{equation}
\dot r_{\rm H}^{\rm exact}
=
O_{\rm ad}(\epsilon^2),
\qquad
\dot M_{\rm H}^{\rm exact}
=
O_{\rm ad}(\epsilon^2),
\label{eq:exact_horizon_adiabatic_order}
\end{equation}
provided the inverse surface gravity scale is not enhanced in the adiabatic
limit. Equivalently, the horizon must remain uniformly non-degenerate, with
\begin{equation}
\kappa_{\rm H}^{\rm exact}
r_{\rm H}^{\rm exact}
\gg
\epsilon.
\label{eq:exact_horizon_non_degeneracy}
\end{equation}
The ultra-near-extremal regime in which
\(\kappa_{\rm H}^{\rm exact}r_{\rm H}^{\rm exact}\) becomes comparable to
\(\epsilon\) lies outside the present expansion.

The exact quantities
\(r_{\rm H}^{\rm exact}(v)\) and
\(M_{\rm H}^{\rm exact}(v)\) introduced here must be distinguished from the
horizon radius and horizon-mass parameter of the instantaneous static
representative that shall be introduced in
Sec.~\ref{sec:canonical_adiabatic_deformation}. The two sets of quantities
may differ at first adiabatic order even though their time derivatives begin
only at second order.

\section{Canonical adiabatic deformation}
\label{sec:canonical_adiabatic_deformation}

We now apply the general adiabatic framework developed in Paper~I to the
time-dependent EMD system derived in Sec.~\ref{sec:EMD-background}. At this stage, the construction
uses only the existence of a smooth family of non-extremal static solutions
and the reduced radial structure inherited from the exact field equations.
The explicit specialization to the magnetic GHS family is deferred to
Sec.~\ref{sec:GHS_application}.


\subsection{Notation and dictionary with Paper~I}
\label{subsec:short_notation_dictionary}

From this point onward, \(r\) denotes the exact areal-radius coordinate
introduced in Sec.~\ref{subsec:time_dependent_ansatz}. The conventional GHS
coordinate \(\rho\) is used only when translating explicit static formulas,
and the coordinate \(b\) introduced in Eq.~\eqref{eq:b_coordinate},
\begin{equation}
b
\equiv
\frac{\rho-r_+}{r_+} \;,
\end{equation}
denotes the corresponding dimensionless throat coordinate. A complete
notation dictionary is collected in Appendix~\ref{app:notation_dictionary}.

The slowly varying exterior is characterized by a prescribed modulus
\(\varphi_{\rm ext}(v)\). This quantity is not an additional dynamical field
of the finite near-zone problem. Instead, it labels the exterior solution to which the
near-zone fields are matched at \(r=r_{\rm m}\), and its relation to the
near-zone scalar field is supplied by the matching condition. In particular,
we do not identify it a priori with either
\(\varphi(v,r_{\rm m})\) or a formal limit \(r\to\infty\) taken within the
finite near-zone construction.

The finite near-zone construction assumes that the prescribed exterior
modulus varies negligibly across the entire matching region. This follows
from the overlap hierarchy,
\begin{equation}
r_{\rm H}^{\rm inst}
<
r_{\rm m}
\ll
L_{\rm cos},
\end{equation}
where \(L_{\rm cos}\) is the shortest characteristic length scale of the
exterior solution. In particular, since
\(L_{\rm cos}\leq T_{\dot\varphi}\), with
\(T_{\dot\varphi}\equiv
\Mpl/|\dot\varphi_{\rm ext}|\), one has
\begin{equation}
\frac{
r_{\rm m}
\left|\dot\varphi_{\rm ext}\right|
}{\Mpl}
=
\frac{r_{\rm m}}{T_{\dot\varphi}}
\ll1.
\label{eq:finite_near_zone_modulus_variation}
\end{equation}
The exterior may therefore be represented, throughout the finite near zone,
by a single slowly varying modulus \(\varphi_{\rm ext}(v)\), while its
relation to the near-zone scalar field at \(r=r_{\rm m}\) is supplied by the
matching condition.

The static reference solutions form a smooth two-parameter family
\begin{equation}
\Psi_0(r;\lambda)
\equiv
\bigl(
\varphi_0(r;\lambda),
\delta_0(r;\lambda),
m_0(r;\lambda)
\bigr),
\end{equation}
with local coordinates
\begin{equation}
\lambda^A
=
\bigl(
\varphi,M_{{\rm H},0}
\bigr).
\label{eq:static_family_coordinates}
\end{equation}
Here, the undecorated quantity \(\varphi\) denotes the static modulus label,
whereas \(\varphi_0(r;\lambda)\) is the corresponding radial scalar profile.
The second coordinate is the static horizon-mass parameter,
\begin{equation}
M_{{\rm H},0}
=
m_0\!\left(r_{{\rm H},0};\lambda\right)
=
\frac{\Mpl^2}{2}\,
r_{{\rm H},0},
\label{eq:static_horizon_mass_parameter}
\end{equation}
where
\begin{equation}
\mathcal F_0\!\left(r_{{\rm H},0};\lambda\right)=0.
\label{eq:static_horizon_definition}
\end{equation}

Time dependence is introduced by promoting the collective coordinates,
\begin{equation}
\lambda^A
\longrightarrow
\lambda^A(v).
\end{equation}
The associated static horizon quantities then become those of the
instantaneous static representative,
\begin{equation}
r_{\rm H}^{\rm inst}(v)
\equiv
r_{{\rm H},0}\!\left(\lambda(v)\right),
\qquad
M_{\rm H}^{\rm inst}(v)
\equiv
M_{{\rm H},0}\!\left(\lambda(v)\right)
=
\frac{\Mpl^2}{2}\,
r_{\rm H}^{\rm inst}(v).
\label{eq:instantaneous_horizon_quantities}
\end{equation}

We reiterate that these quantities must be distinguished from the exact dynamical horizon
quantities introduced in Sec.~\ref{subsec:horizon_dynamics},
\begin{equation}
\mathcal F\!\left(
v,r_{\rm H}^{\rm exact}(v)
\right)
=
0,
\qquad
M_{\rm H}^{\rm exact}(v)
=
m\!\left(
v,r_{\rm H}^{\rm exact}(v)
\right)
=
\frac{\Mpl^2}{2}\,
r_{\rm H}^{\rm exact}(v).
\label{eq:exact_horizon_quantities_again}
\end{equation}
The instantaneous and exact horizon quantities coincide for an exactly
static configuration, but may differ at first adiabatic order once the lag
fields are included:
\begin{equation}
r_{\rm H}^{\rm exact}
-
r_{\rm H}^{\rm inst}
=
\mathcal O_{\rm ad}
\!\left(
\epsilon_\varphi r_{\rm H}^{\rm inst}
\right),
\qquad
M_{\rm H}^{\rm exact}
-
M_{\rm H}^{\rm inst}
=
\mathcal O_{\rm ad}
\!\left(
\epsilon_\varphi M_{\rm H}^{\rm inst}
\right).
\label{eq:exact_instantaneous_horizon_difference}
\end{equation}
Their explicit first-order relation is derived in
Sec.~\ref{subsec:exact_horizon_displacement}. Although the two horizons may
differ at first order, their time derivatives begin only at second
adiabatic order.

We emphasize that the two modulus quantities introduced above have distinct
roles. The collective coordinate \(\lambda^\varphi(v)\) labels the modulus
of the chosen instantaneous static representative, whereas
\(\varphi_{\rm ext}(v)\) is prescribed by the exterior solution. Their
relation is therefore not a pointwise identification imposed at the outset:
it depends on how the exact configuration is decomposed into an
instantaneous representative and a lag field.

At leading adiabatic order, the representative is chosen to track the
exterior evolution. A residual displacement along the static modulus
direction may nevertheless be redistributed between the collective
coordinate and the lag field without changing the physical configuration.
The allowed size and time dependence of this displacement, and the resulting
relation between
\(\dot\lambda^\varphi\) and
\(\dot\varphi_{\rm ext}\), are specified after the adiabatic counting and
the promotion of the static family have been introduced in the following
subsection.

\subsection{Adiabatic expansion}
\label{subsec:adiabatic_expansion}

The exact EMD system derived in
Sec.~\ref{subsec:exact_minimal_system} admits generic time-dependent
solutions, including freely excited perturbations evolving on the intrinsic
black hole timescale. Here, we isolate a different sector: the branch driven
by prescribed exterior scalar data that vary parametrically more slowly than
the local response time of the black hole. On this slowly forced branch, the
near-zone solution remains perturbatively close to the static magnetic GHS
family while adjusting to the data supplied at the matching surface
\(r=r_{\rm m}\).

The prescribed exterior modulus is denoted by
\(\varphi_{\rm ext}(v)\). Its rolling rate defines the dimensionless
black hole scale adiabatic parameter
\begin{equation}
\epsilon_\varphi
\equiv
\frac{
r_{\rm H}^{\rm inst}
\left|\dot\varphi_{\rm ext}\right|
}{
\Mpl
}
=
\frac{
r_{\rm H}^{\rm inst}
}{
T_{\dot\varphi}
},
\qquad
T_{\dot\varphi}
\equiv
\frac{\Mpl}{\left|\dot\varphi_{\rm ext}\right|}.
\label{eq:adiabatic_parameter}
\end{equation}
Replacing \(r_{\rm H}^{\rm inst}\) in this estimate by
\(r_{\rm H}^{\rm exact}\) changes \(\epsilon_\varphi\) only beyond the
leading adiabatic order. The overlap hierarchy,
\begin{equation}
r_{\rm H}^{\rm inst}
<
r_{\rm m}
\ll
L_{\rm cos}
\leq
T_{\dot\varphi},
\end{equation}
also ensures that the exterior modulus varies negligibly across the full
finite near zone:
\begin{equation}
\frac{
r_{\rm m}
\left|\dot\varphi_{\rm ext}\right|
}{\Mpl}
=
\frac{r_{\rm m}}{T_{\dot\varphi}}
\ll1.
\label{eq:matching_region_adiabaticity}
\end{equation}
This condition is stronger than
\(\epsilon_\varphi\ll1\) whenever
\(r_{\rm m}\gg r_{\rm H}^{\rm inst}\). The slowly forced black hole response
therefore requires
\begin{equation}
\epsilon_\varphi\ll1,
\label{eq:adiabatic_regime}
\end{equation}
together with uniform non-degeneracy of the instantaneous static horizon,
\begin{equation}
\kappa_{\rm H}^{\rm inst}r_{\rm H}^{\rm inst}
\gg
\epsilon_\varphi.
\label{eq:adiabatic_nonextremality}
\end{equation}
This is the instantaneous-static counterpart of the exact-horizon condition
in Eq.~\eqref{eq:exact_horizon_non_degeneracy}. It ensures that the inverse
surface-gravity scale is not parametrically enhanced relative to the
adiabatic expansion.
Here \(\kappa_{\rm H}^{\rm inst}\) is the surface gravity parameter of the
instantaneous static representative in the matching normalization. The last
condition ensures that the intrinsic response scale
\((\kappa_{\rm H}^{\rm inst})^{-1}\) remains parametrically shorter than the
exterior evolution timescale. The ultra-near-extremal regime, in which
\(\kappa_{\rm H}^{\rm inst}r_{\rm H}^{\rm inst}\) becomes comparable to
\(\epsilon_\varphi\), lies outside the present expansion.

On the slowly forced branch, every advanced-time derivative inherited from
the prescribed exterior evolution raises the adiabatic order by one, whereas
radial derivatives are retained exactly. This counting does not apply to a
generic freely excited homogeneous perturbation, whose time dependence may
instead occur on the black hole scale. The notation
\(\mathcal O_{\rm ad}(\epsilon_\varphi^n)\) records adiabatic order only; the
physical dimension of each quantity is fixed separately by the field or
equation in which it appears.

\subsubsection{Promotion of the static family}

Let
\begin{equation}
\lambda^A
=
\left(
\varphi,M_{{\rm H},0}
\right)
\label{eq:static_collective_coordinates}
\end{equation}
be local coordinates on the static solution manifold \(\mathcal S\), and
write
\begin{equation}
\Psi_0(r;\lambda)
\equiv
\Bigl(
\varphi_0(r;\lambda),
\delta_0(r;\lambda),
m_0(r;\lambda)
\Bigr)
\label{eq:GHS_static_family_lambda}
\end{equation}
for the corresponding static configuration in the exact areal-radius
coordinate. The static horizon radius and horizon-mass parameter satisfy
\begin{equation}
\mathcal F_0\!\left(
r_{{\rm H},0}(\lambda);\lambda
\right)
=
0,
\qquad
M_{{\rm H},0}
=
\frac{\Mpl^2}{2}\,
r_{{\rm H},0}.
\label{eq:static_horizon_coordinates_again}
\end{equation}

Following Paper~I, time dependence is introduced by promoting the collective
coordinates to slowly varying functions of the advanced time,
\begin{equation}
\lambda^A
\longrightarrow
\lambda^A(v).
\label{eq:promoted_collective_coordinates}
\end{equation}
The corresponding instantaneous static representative is
\begin{equation}
\Psi_{\rm inst}(v,r)
\equiv
\Psi_0\!\left(r;\lambda(v)\right).
\label{eq:GHS_instantaneous_tracking}
\end{equation}
Its horizon radius and horizon-mass parameter are
\begin{equation}
r_{\rm H}^{\rm inst}(v)
\equiv
r_{{\rm H},0}\!\left(\lambda(v)\right),
\qquad
M_{\rm H}^{\rm inst}(v)
\equiv
M_{{\rm H},0}\!\left(\lambda(v)\right)
=
\frac{\Mpl^2}{2}\,
r_{\rm H}^{\rm inst}(v).
\label{eq:promoted_instantaneous_horizon}
\end{equation}
The metric function of this representative is
\begin{equation}
\mathcal F_{\rm inst}(v,r)
=
1-
\frac{
2m_0(r;\lambda(v))
}{
\Mpl^2r
}
\equiv
\mathcal F_0(r;\lambda(v)),
\end{equation}
and hence
\begin{equation}
\mathcal F_{\rm inst}\!\left(
v,r_{\rm H}^{\rm inst}(v)
\right)
=
0.
\end{equation}
These relations define the chosen instantaneous representative. They do not
assert that the exact dynamical solution remains on \(\mathcal S\), nor do
they imply
\begin{equation}
r_{\rm H}^{\rm exact}
=
r_{\rm H}^{\rm inst}.
\end{equation}
The instantaneous quantities
\(r_{\rm H}^{\rm inst}\) and \(M_{\rm H}^{\rm inst}\) must therefore be
distinguished from the exact dynamical quantities
\(r_{\rm H}^{\rm exact}\) and \(M_{\rm H}^{\rm exact}\).

The exterior solution supplies the prescribed modulus
\(\varphi_{\rm ext}(v)\), whereas
\(\lambda^\varphi(v)\) labels the modulus of the chosen instantaneous static
representative. At leading adiabatic order, the representative is chosen to
track the exterior evolution. Their instantaneous values need not coincide,
however, because a first-order displacement along the static modulus
direction may be redistributed between the instantaneous representative and
the lag field without changing the total physical configuration. Accordingly,
\begin{equation}
\frac{
\lambda^\varphi(v)-\varphi_{\rm ext}(v)
}{
\Mpl
}
=
\mathcal O_{\rm ad}(\epsilon_\varphi).
\label{eq:Phi_collective_external_relation}
\end{equation}
Admissible changes of representative are restricted to the slowly varying
adiabatic sector. The time derivative of the difference in
Eq.~\eqref{eq:Phi_collective_external_relation} therefore begins only at
second adiabatic order. In dimensionless form,
\begin{equation}
\frac{r_{\rm H}^{\rm inst}}{\Mpl}
\left(
\dot\lambda^\varphi(v)
-
\dot\varphi_{\rm ext}(v)
\right)
=
\mathcal O_{\rm ad}(\epsilon_\varphi^2).
\label{eq:collective_modulus_velocity}
\end{equation}
Equivalently,
\begin{equation}
\dot\lambda^\varphi(v)
=
\dot\varphi_{\rm ext}(v)
+
\mathcal O_{\rm ad}
\left(
\epsilon_\varphi^2
\frac{\Mpl}{r_{\rm H}^{\rm inst}}
\right).
\end{equation}
Consequently, the residual freedom in the instantaneous value of
\(\lambda^\varphi\) does not modify the leading forcing, which depends only
on the common first-order rolling rate
\(\dot\varphi_{\rm ext}\). This modulus-tangential freedom is described
explicitly below and is ultimately removed by the local slice.

The promoted horizon-mass coordinate is chosen within the slowly varying
sector. Dimensionally, the corresponding counting is
\begin{equation}
\frac{
\dot M_{\rm H}^{\rm inst}
}{
M_{\rm H}^{\rm inst}/r_{\rm H}^{\rm inst}
}
=
\mathcal O_{\rm ad}(\epsilon_\varphi^2),
\qquad
\dot r_{\rm H}^{\rm inst}
=
\mathcal O_{\rm ad}(\epsilon_\varphi^2).
\label{eq:instantaneous_horizon_mass_second_order}
\end{equation}
Using
\begin{equation}
\frac{M_{\rm H}^{\rm inst}}{r_{\rm H}^{\rm inst}}
=
\frac{\Mpl^2}{2},
\end{equation}
the first relation may equivalently be written as
\begin{equation}
\frac{\dot M_{\rm H}^{\rm inst}}{\Mpl^2}
=
\mathcal O_{\rm ad}(\epsilon_\varphi^2).
\end{equation}
This counting is motivated by the exact quadratic horizon-flux law derived
in Sec.~\ref{subsec:horizon_dynamics}. It is nevertheless a condition on the
chosen instantaneous representative, rather than a pointwise identification
of \(M_{\rm H}^{\rm inst}\) with \(M_{\rm H}^{\rm exact}\). Moreover, the
second-order evolution condition does not by itself fix a possible
first-order displacement of the representative in the static horizon-mass
direction. That independent representation freedom is fixed separately by
the convention introduced in Sec.~\ref{subsec:local_slice_recap}.

\subsubsection{Time evolution of the instantaneous configuration}

Tangent vectors to \(\mathcal S\) are obtained by differentiating the static
fields with respect to the collective coordinates at fixed areal radius. We
write
\begin{equation}
\partial_A\Psi_0
=
\left(
\varphi_A,
\delta_A,
m_A
\right),
\qquad
\varphi_A
\equiv
\left.
\frac{\partial\varphi_0}{\partial\lambda^A}
\right|_r,
\quad
\delta_A
\equiv
\left.
\frac{\partial\delta_0}{\partial\lambda^A}
\right|_r,
\quad
m_A
\equiv
\left.
\frac{\partial m_0}{\partial\lambda^A}
\right|_r.
\end{equation}
In particular, the tangent generated by varying the static modulus label at
fixed static horizon-mass parameter is
\begin{equation}
\varphi_\varphi
\equiv
\left.
\frac{\partial\varphi_0}{\partial\varphi}
\right|_{M_{{\rm H},0},\,r},
\qquad
\delta_\varphi
\equiv
\left.
\frac{\partial\delta_0}{\partial\varphi}
\right|_{M_{{\rm H},0},\,r},
\qquad
m_\varphi
\equiv
\left.
\frac{\partial m_0}{\partial\varphi}
\right|_{M_{{\rm H},0},\,r}.
\label{eq:GHS_tangent_fields}
\end{equation}

At fixed areal radius, promotion gives the exact chain rule
\begin{equation}
\partial_v\Psi_{\rm inst}
=
\dot\lambda^A\,\partial_A\Psi_0.
\label{eq:instantaneous_chain_rule_compact}
\end{equation}
For the scalar component,
\begin{equation}
\partial_v\varphi_{\rm inst}
=
\dot\lambda^\varphi\,\varphi_\varphi
+
\dot M_{\rm H}^{\rm inst}\,
\varphi_{M_{\rm H}},
\label{eq:GHS_chain_rule_scalar}
\end{equation}
where
\begin{equation}
\varphi_{M_{\rm H}}
\equiv
\left.
\frac{\partial\varphi_0}{\partial M_{{\rm H},0}}
\right|_{\varphi,\,r}.
\end{equation}
Analogous expressions hold for \(\delta_{\rm inst}\) and
\(m_{\rm inst}\).

Using Eqs.~\eqref{eq:collective_modulus_velocity} and
\eqref{eq:instantaneous_horizon_mass_second_order}, the first-order chain
rules reduce to
\begin{align}
\partial_v\varphi_{\rm inst}
&=
\dot\varphi_{\rm ext}\,\varphi_\varphi
+
\mathcal O_{\rm ad}(\epsilon_\varphi^2),
\label{eq:leading_chain_varphi}
\\
\partial_v\delta_{\rm inst}
&=
\dot\varphi_{\rm ext}\,\delta_\varphi
+
\mathcal O_{\rm ad}(\epsilon_\varphi^2),
\label{eq:leading_chain_delta}
\\
\partial_vm_{\rm inst}
&=
\dot\varphi_{\rm ext}\,m_\varphi
+
\mathcal O_{\rm ad}(\epsilon_\varphi^2).
\label{eq:leading_chain_m}
\end{align}
The explicit tangent profiles of the magnetic GHS family are evaluated in
Sec.~\ref{subsec:explicit_GHS_tangent}.

\subsubsection{Adiabatic defect and lag fields}

For each fixed value of \(v\), the instantaneous configuration
\(\Psi_{\rm inst}\) is an exact member of the static family. Consequently,
parameter promotion preserves the purely radial constraint equations
pointwise. The instantaneous configuration does not, however, generally
satisfy those equations in the exact system that contain explicit
advanced-time derivatives. Evaluating the exact equations on
\(\Psi_{\rm inst}\) therefore produces a first-order adiabatic defect.

The exact dynamical solution is decomposed as
\begin{align}
\varphi(v,r)
&=
\varphi_{\rm inst}(v,r)
+
\eta(v,r)
+
\mathcal O_{\rm ad}(\epsilon_\varphi^2),
\label{eq:lag_decomposition_varphi}
\\
\delta(v,r)
&=
\delta_{\rm inst}(v,r)
+
\xi(v,r)
+
\mathcal O_{\rm ad}(\epsilon_\varphi^2),
\label{eq:lag_decomposition_delta}
\\
m(v,r)
&=
m_{\rm inst}(v,r)
+
\mu(v,r)
+
\mathcal O_{\rm ad}(\epsilon_\varphi^2),
\label{eq:lag_decomposition_m}
\end{align}
where the dimensionless relative orders are
\begin{equation}
\frac{\eta}{\Mpl}
=
\mathcal O_{\rm ad}(\epsilon_\varphi),
\qquad
\xi
=
\mathcal O_{\rm ad}(\epsilon_\varphi),
\qquad
\frac{\mu}{M_{\rm H}^{\rm inst}}
=
\mathcal O_{\rm ad}(\epsilon_\varphi).
\label{eq:GHS_lag_order}
\end{equation}
The triplet \((\eta,\xi,\mu)\) measures the first-order departure of the
exact solution from the chosen instantaneous representative. It must not be
confused with the tangent triplet
\((\varphi_\varphi,\delta_\varphi,m_\varphi)\), which instead describes an
infinitesimal displacement along the static solution manifold.

The decomposition is not unique. A slowly varying displacement of the
instantaneous representative along a tangent direction can be compensated by
the opposite displacement of the lag fields, leaving the total physical
solution unchanged to the order retained. The representation convention in
the static horizon-mass direction and the local slice that removes the
remaining modulus-tangential ambiguity are introduced in the following
subsections.

The radial constraints obeyed by \((\xi,\mu)\) have the same homogeneous
linearized differential structure as those obeyed by the corresponding
static tangent fields, since both arise by linearization about the same
instantaneous background. Their roles are nevertheless distinct. The tangent
fields encode the geometry of the static solution manifold, whereas the lag
fields solve the inhomogeneous problem generated by the adiabatic defect.


\subsection{Canonical tangent}
\label{subsec:canonical_tangent_recap}

The instantaneous configuration introduced in
Sec.~\ref{subsec:adiabatic_expansion} evolves on the static solution manifold
\(\mathcal S\). According to the Canonical Tangent Theorem of Paper~I, the
distinguished tangent direction relevant to the leading adiabatic response
is obtained by varying the static modulus label while holding the static
horizon-mass parameter fixed. Its field components are
\begin{equation}
\varphi_\varphi
\equiv
\left.
\frac{\partial\varphi_0}{\partial\varphi}
\right|_{M_{{\rm H},0},\,r},
\qquad
\delta_\varphi
\equiv
\left.
\frac{\partial\delta_0}{\partial\varphi}
\right|_{M_{{\rm H},0},\,r},
\qquad
m_\varphi
\equiv
\left.
\frac{\partial m_0}{\partial\varphi}
\right|_{M_{{\rm H},0},\,r}.
\label{eq:canonical_tangent_definition}
\end{equation}
All parameter derivatives are taken at fixed exact areal radius \(r\).
Because these fields are obtained by differentiating a family of exact
static solutions, they satisfy the frozen-background linearized equations.

The linearized lapse constraint gives
\begin{equation}
\delta_\varphi'
=
\frac{r}{\Mpl^2}\,
\varphi_0'\varphi_\varphi'.
\label{eq:deltaPhi_constraint}
\end{equation}
The linearized mass constraints imply, for a general tangent direction
\(\partial_A\),
\begin{equation}
m_A(r)
=
f_\varphi(r)\varphi_A(r)
+
C_{\mu,A}e^{-\delta_0(r)},
\qquad
f_\varphi(r)
\equiv
\frac{r^2}{2}\,
\mathcal F_0(r)\varphi_0'(r).
\label{eq:general_tangent_mass_decomposition}
\end{equation}
Here \(C_{\mu,A}\) is the residual mass-integration constant associated with
the tangent direction \(\partial_A\). Its linear dependence on the tangent
vector defines the intrinsic covector
\begin{equation}
C_\mu\in T_\lambda^\ast\mathcal S.
\end{equation}

Paper~I gives the horizon representation
\begin{equation}
C_{\mu,A}
=
e^{\delta_{0{\rm H}}}
\left[
\partial_A M_{{\rm H},0}
-
m_0'(r_{{\rm H},0})
\partial_A r_{{\rm H},0}
\right],
\label{eq:Cmu_horizon_formula}
\end{equation}
where the parameter derivatives on the right-hand side are evaluated along
the chosen tangent direction, and
\begin{equation}
\delta_{0{\rm H}}
\equiv
\delta_0(r_{{\rm H},0}),
\qquad
M_{{\rm H},0}
=
m_0\!\left(r_{{\rm H},0};\lambda\right)
=
\frac{\Mpl^2}{2}\,
r_{{\rm H},0},
\qquad
\mathcal F_0(r_{{\rm H},0};\lambda)=0.
\label{eq:static_horizon_mass_again}
\end{equation}

Along the adapted coordinate direction
\(\left.\partial_\varphi\right|_{M_{{\rm H},0}}\), the static
horizon-mass parameter is constant by definition. Since
\begin{equation}
r_{{\rm H},0}
=
\frac{2M_{{\rm H},0}}{\Mpl^2},
\end{equation}
the static horizon radius is constant along the same direction:
\begin{equation}
\left.
\frac{\partial r_{{\rm H},0}}
{\partial\varphi}
\right|_{M_{{\rm H},0}}
=
0.
\label{eq:fixed_horizon_radius}
\end{equation}
Both terms in Eq.~\eqref{eq:Cmu_horizon_formula} therefore vanish, and
\begin{equation}
C_{\mu,\varphi}=0.
\label{eq:CmuPhi_zero}
\end{equation}

The tangent mass decomposition consequently reduces to the algebraic
identity
\begin{equation}
m_\varphi(r)
=
f_\varphi(r)\varphi_\varphi(r).
\label{eq:mPhi_reduction}
\end{equation}
At the static horizon,
\begin{equation}
\mathcal F_0(r_{{\rm H},0})=0,
\qquad
f_\varphi(r_{{\rm H},0})=0,
\end{equation}
and hence
\begin{equation}
m_\varphi(r_{{\rm H},0})=0.
\label{eq:mPhi_horizon}
\end{equation}
These identities refer to the horizon of the static reference solution.
After promotion, \(r_{{\rm H},0}\) becomes
\(r_{\rm H}^{\rm inst}(v)\) and must be distinguished from the exact
dynamical horizon \(r_{\rm H}^{\rm exact}(v)\).

The metric components of the canonical tangent are therefore determined by
its scalar component. The mass tangent follows algebraically from
Eq.~\eqref{eq:mPhi_reduction}, while the lapse tangent is reconstructed from
Eq.~\eqref{eq:deltaPhi_constraint} together with
\begin{equation}
\delta_\varphi(r_{\rm m})=0.
\label{eq:deltaPhi_matching_normalization}
\end{equation}
This boundary value follows by differentiating the matching normalization
\begin{equation}
\delta_0(r_{\rm m};\lambda)=0
\end{equation}
at fixed \(r_{\rm m}\) throughout the static family.

The frozen-background reduced scalar equation acts on a general tangent
profile according to
\begin{equation}
L\varphi_A
=
C_{\mu,A}K(r),
\qquad
K(r)
\equiv
\frac{2}{\Mpl^2}
\left(r\varphi_0'(r)\right)'.
\label{eq:general_tangent_operator}
\end{equation}
Eq.~\eqref{eq:CmuPhi_zero} therefore implies that the scalar component
of the canonical tangent is a homogeneous zero mode of the unsliced reduced
operator:
\begin{equation}
L\varphi_\varphi=0.
\label{eq:canonical_tangent_operator}
\end{equation}

For completeness, the second adapted tangent direction satisfies
\begin{equation}
C_{\mu,M_{{\rm H},0}}
=
e^{\delta_{0{\rm H}}}
r_{{\rm H},0}
\mathcal F_0'(r_{{\rm H},0}).
\label{eq:Cmu_mass_direction}
\end{equation}
This quantity is strictly positive for a non-extremal future outer static
horizon. Since the static solution manifold is two-dimensional, \(C_\mu\)
has rank one and
\begin{equation}
\ker C_\mu
=
\operatorname{span}
\left\{
\left.
\partial_\varphi
\right|_{M_{{\rm H},0}}
\right\}.
\label{eq:Cmu_kernel_canonical}
\end{equation}
Thus the fixed-horizon-mass modulus direction is the unique tangent direction
whose residual mass-integration constant vanishes. Its scalar component
\(\varphi_\varphi\) consequently spans the tangent zero mode selected by the
canonical geometry of the static solution manifold.

The Canonical Tangent Theorem establishes this distinguished direction and
the homogeneous equation
\(L\varphi_\varphi=0\), but it does not determine the sign or nodal structure
of \(\varphi_\varphi\). The stronger property
\begin{equation}
\varphi_\varphi(r)>0
\end{equation}
throughout the charged non-extremal exterior is specific to the magnetic GHS
family and will be proved explicitly in
Sec.~\ref{subsec:explicit_GHS_tangent}. In particular, its absence of zeros
allows \(\varphi_\varphi\) to be used globally in the reduction-of-order
constructions of Secs.~\ref{subsec:displaced_generator} and~\ref{subsec:explicit_forced_response}.

The results summarized in this subsection depend only on the geometry of the
static solution manifold and on the frozen-background linearized
constraints. Their explicit realization for the magnetic GHS family is
carried out in Sec.~\ref{subsec:explicit_GHS_tangent}.

\subsection{Local slice}
\label{subsec:local_slice_recap}

The first-order decomposition introduced in
Sec.~\ref{subsec:adiabatic_expansion} is understood componentwise:
\begin{align}
\frac{\varphi-\varphi_{\rm inst}-\eta}{\Mpl}
&=
\mathcal O_{\rm ad}(\epsilon_\varphi^2),
\\
\delta-\delta_{\rm inst}-\xi
&=
\mathcal O_{\rm ad}(\epsilon_\varphi^2),
\\
\frac{m-m_{\rm inst}-\mu}{M_{\rm H}^{\rm inst}}
&=
\mathcal O_{\rm ad}(\epsilon_\varphi^2),
\end{align}
with
\begin{equation}
\frac{\eta}{\Mpl}
=
\mathcal O_{\rm ad}(\epsilon_\varphi),
\qquad
\xi
=
\mathcal O_{\rm ad}(\epsilon_\varphi),
\qquad
\frac{\mu}{M_{\rm H}^{\rm inst}}
=
\mathcal O_{\rm ad}(\epsilon_\varphi).
\end{equation}
This decomposition is not unique. Consider a nearby curve of instantaneous
static representatives,
\begin{equation}
\lambda^A(v)
\longrightarrow
\widetilde{\lambda}^A(v)
=
\lambda^A(v)+\alpha^A(v).
\label{eq:representative_shift}
\end{equation}
An admissible first-order shift is restricted componentwise by
\begin{equation}
\frac{\alpha^\varphi}{\Mpl}
=
\mathcal O_{\rm ad}(\epsilon_\varphi),
\qquad
\frac{\alpha^{M_{{\rm H},0}}}{M_{\rm H}^{\rm inst}}
=
\mathcal O_{\rm ad}(\epsilon_\varphi),
\label{eq:representative_shift_amplitude}
\end{equation}
and
\begin{equation}
\frac{r_{\rm H}^{\rm inst}}{\Mpl}\,
\dot\alpha^\varphi
=
\mathcal O_{\rm ad}(\epsilon_\varphi^2),
\qquad
\frac{\dot\alpha^{M_{{\rm H},0}}}{\Mpl^2}
=
\mathcal O_{\rm ad}(\epsilon_\varphi^2).
\label{eq:representative_shift_velocity}
\end{equation}
Expanding the static family to first order gives
\begin{equation}
\Psi_0\!\left(r;\widetilde{\lambda}(v)\right)
=
\Psi_0\!\left(r;\lambda(v)\right)
+
\alpha^A(v)\,\partial_A\Psi_0
+
\mathcal O_{\rm ad}(\epsilon_\varphi^2),
\end{equation}
where the remainder is again understood componentwise with the
normalizations specified above. The same physical configuration is
therefore represented if the lag field is transformed simultaneously as
\begin{equation}
\Psi_{\rm lag}
\longrightarrow
\widetilde{\Psi}_{\rm lag}
=
\Psi_{\rm lag}
-
\alpha^A(v)\,\partial_A\Psi_0.
\label{eq:tangent_redundancy}
\end{equation}
Thus the lag field is defined only modulo a slowly varying tangent vector to
the static solution manifold. The velocity counting in
Eq.~\eqref{eq:representative_shift_velocity} ensures that an admissible
change of representative alters the collective velocity only at second
adiabatic order. It therefore leaves unchanged the first-order source
generated by parameter promotion. In this sense, the leading adiabatic
forcing is independent of the representative chosen within the equivalence
class.

Recall that the static family is parametrized by
\begin{equation}
\lambda^A
=
\left(
\varphi,M_{{\rm H},0}
\right).
\end{equation}
As in Paper~I, we first remove the freedom in the static horizon-mass
direction by imposing the representation convention
\begin{equation}
\alpha^{M_{{\rm H},0}}(v)
\equiv
0.
\label{eq:mass_direction_convention}
\end{equation}
This convention fixes how the exact horizon radius is decomposed into the
horizon radius of the instantaneous static representative and the
lag-induced displacement. It does not impose any condition on the exact
dynamical horizon itself.

The only remaining tangential ambiguity is therefore the canonical modulus
direction,
\begin{equation}
\Psi_{\rm lag}
\sim
\Psi_{\rm lag}
-
\alpha^\varphi(v)
\left.
\partial_\varphi\Psi_0
\right|_{M_{{\rm H},0}},
\qquad
\alpha^\varphi
=
\mathcal O_{\rm ad}(\epsilon_\varphi),
\qquad
\dot\alpha^\varphi
=
\mathcal O_{\rm ad}(\epsilon_\varphi^2).
\label{eq:remaining_tangent_redundancy}
\end{equation}
A single scalar condition transverse to this orbit is therefore sufficient
to select a unique instantaneous representative.

The static family and the lag fields are written throughout in the matching
normalization
\begin{equation}
\delta_0(r_{\rm m};\lambda)=0,
\qquad
\xi(v,r_{\rm m})=0.
\label{eq:matching_normalization}
\end{equation}
Because the first relation holds identically on the entire static solution
manifold,
\begin{equation}
\partial_A\delta_0(r_{\rm m};\lambda)=0
\end{equation}
for every tangent direction \(A\). In particular,
\begin{equation}
\delta_\varphi(r_{\rm m})=0.
\end{equation}
Under the surviving change of representative,
\begin{equation}
\xi(v,r_{\rm m})
\longrightarrow
\xi(v,r_{\rm m})
-
\alpha^\varphi(v)\delta_\varphi(r_{\rm m})
=
\xi(v,r_{\rm m}).
\end{equation}
Hence
\begin{equation}
\xi(v,r_{\rm m})=0
\label{eq:matching_gauge_lag}
\end{equation}
is invariant along the residual tangent orbit. It fixes the normalization
of the advanced-time coordinate but cannot select a representative on the
static solution manifold and therefore does not define a slice.

Following Paper~I, we remove the remaining tangential ambiguity by imposing
the local mass slice
\begin{equation}
\mu(v,r_{\rm m})=0.
\label{eq:local_slice_condition}
\end{equation}
Under a displacement along the canonical tangent, the mass lag transforms
as
\begin{equation}
\mu(v,r)
\longrightarrow
\mu(v,r)
-
\alpha^\varphi(v)m_\varphi(r),
\end{equation}
and in particular
\begin{equation}
\mu(v,r_{\rm m})
\longrightarrow
\mu(v,r_{\rm m})
-
\alpha^\varphi(v)m_\varphi(r_{\rm m}),
\label{eq:mu_slice_transformation}
\end{equation}
where
\begin{equation}
m_\varphi(r)
\equiv
\left.
\frac{\partial m_0(r;\lambda)}{\partial\varphi}
\right|_{M_{{\rm H},0},\,r}.
\end{equation}
The condition given in Eq.~\eqref{eq:local_slice_condition} therefore determines
\(\alpha^\varphi(v)\) uniquely if and only if
\begin{equation}
m_\varphi(r_{\rm m})\neq0.
\label{eq:slice_transversality}
\end{equation}
This is the transversality condition between the chosen mass slice and the
canonical tangent direction. It states that evaluation of the static mass
profile at \(r_{\rm m}\) is sensitive to motion along the residual modulus
orbit. It is a property of the chosen slicing functional, not a physical
restriction on the underlying dynamical solution. If it were to fail for a
particular slice, that slice would cease to distinguish neighboring
representatives and would have to be replaced by another transverse
functional. For the charged non-extremal GHS family, the condition is
verified explicitly on every nontrivial finite matching interval in
Sec.~\ref{subsec:displaced_generator}.

The linearized radial mass constraint gives
\begin{equation}
\mu(v,r)
=
f_\varphi(r)\eta(v,r)
+
C_\mu^{\rm lag}(v)e^{-\delta_0(r)},
\qquad
f_\varphi(r)
\equiv
\frac{r^2}{2}\,
\mathcal F_0(r)\varphi_0'(r),
\label{eq:lag_mass_general}
\end{equation}
where \(C_\mu^{\rm lag}(v)\) is the radial integration datum left
undetermined by the bulk constraint. Evaluating this relation at the
matching radius and using
\begin{equation}
\delta_0(r_{\rm m})=0,
\qquad
\mu(v,r_{\rm m})=0,
\end{equation}
gives
\begin{equation}
C_\mu^{\rm lag}(v)
=
-f_\varphi(r_{\rm m})\eta(v,r_{\rm m}).
\label{eq:Cmu_lag_slice}
\end{equation}

The Canonical Tangent Theorem gives
\begin{equation}
C_{\mu,\varphi}=0,
\end{equation}
so that the tangent mass decomposition reduces to
\begin{equation}
m_\varphi(r)
=
f_\varphi(r)\varphi_\varphi(r).
\label{eq:mPhi_slice_identity}
\end{equation}
At the matching radius,
\begin{equation}
m_\varphi(r_{\rm m})
=
f_\varphi(r_{\rm m})\varphi_\varphi(r_{\rm m}).
\label{eq:transversality_identity}
\end{equation}
The transversality condition can therefore be written equivalently as
\begin{equation}
f_\varphi(r_{\rm m})\varphi_\varphi(r_{\rm m})
\neq0.
\end{equation}
Thus the local mass evaluation distinguishes neighboring representatives
precisely when it acts nontrivially on the canonical tangent.

It is important to distinguish the residual tangent parameter
\(\alpha^\varphi\) from the radial integration datum
\(C_\mu^{\rm lag}\). Under a general change of representative,
\begin{equation}
C_\mu^{\rm lag}
\longrightarrow
C_\mu^{\rm lag}
-
\alpha^A C_{\mu,A}.
\end{equation}
After the horizon-mass convention
\(\alpha^{M_{{\rm H},0}}=0\) has been imposed, the only remaining change of
representative is the canonical modulus-tangent shift. Since
\(C_{\mu,\varphi}=0\), this transformation leaves the radial integration
datum invariant:
\begin{equation}
C_\mu^{\rm lag}
\longrightarrow
C_\mu^{\rm lag}.
\end{equation}
Thus \(C_\mu^{\rm lag}\) is unchanged under the residual tangential
ambiguity. The role of the slice is not to modify this datum, but to select
the unique scalar-lag representative for which it is expressed through the
boundary relation
\begin{equation}
C_\mu^{\rm lag}(v)
=
-f_\varphi(r_{\rm m})\eta(v,r_{\rm m}),
\end{equation}
as given in Eq.~\eqref{eq:Cmu_lag_slice}.

Substitution of Eq.~\eqref{eq:Cmu_lag_slice} into the radial mass constraint
reconstructs the mass lag entirely from the selected scalar lag:
\begin{equation}
\mu(v,r)
=
f_\varphi(r)\eta(v,r)
-
f_\varphi(r_{\rm m})
\eta(v,r_{\rm m})
e^{-\delta_0(r)}.
\label{eq:mu_reconstructed_slice}
\end{equation}
The lapse lag is similarly reconstructed from
\begin{equation}
\partial_r\xi
=
\frac{r}{\Mpl^2}\,
\varphi_0'(r)\partial_r\eta,
\qquad
\xi(v,r_{\rm m})=0.
\label{eq:lag_lapse_constraint}
\end{equation}
Once the slice has been imposed, the first-order metric lags are therefore
determined by the single scalar lag field \(\eta\).

The local slice must not be confused with the outer matching condition. The
slice fixes the decomposition of a given physical configuration into an
instantaneous static representative and a lag field. The outer matching
condition instead supplies physical data from the exterior and participates
in determining the physical near-zone solution itself. The slice is defined
entirely within the finite radial problem and does not rely on the
large-radius approximation later used to derive the particular massless
overlap functional.

\subsection{Sliced operator}
\label{subsec:sliced_operator}

Before imposing the local slice, the reduced scalar lag equation contains
both the prescribed exterior forcing and the residual integration datum
\(C_\mu^{\rm lag}(v)\) inherited from the radial mass constraint:
\begin{equation}
2r^2\partial_v\partial_r\eta
+
2r\,\partial_v\eta
+
L\eta
=
C_\mu^{\rm lag}(v)K(r)
+
S_{\rm ext}(v,r),
\label{eq:unsliced_reduced_scalar_equation}
\end{equation}
where
\begin{equation}
K(r)
\equiv
\frac{2}{\Mpl^2}
\left(
r\varphi_0'(r)
\right)',
\label{eq:K_definition_sliced}
\end{equation}
and
\begin{equation}
S_{\rm ext}(v,r)
=
-2\dot\varphi_{\rm ext}(v)
\left[
r^2\varphi_\varphi'(r)
+
r\varphi_\varphi(r)
\right]
\label{eq:exterior_source}
\end{equation}
is the leading inhomogeneous source generated by the prescribed exterior
rolling. All background quantities are evaluated on the instantaneous static
representative \(\Psi_0(r;\lambda(v))\) and are treated, at each fixed
advanced time, as frozen radial coefficients.

Once the local mass slice has been imposed, the radial integration datum is
related to the boundary value of the selected scalar lag representative by
Eq.~\eqref{eq:Cmu_lag_slice},
\begin{equation}
C_\mu^{\rm lag}(v)
=
-f_\varphi(r_{\rm m})\,
\eta(v,r_{\rm m}),
\label{eq:Cmu_lag_slice_again}
\end{equation}
where
\begin{equation}
f_\varphi(r)
\equiv
\frac{r^2}{2}\,
\mathcal F_0(r)\varphi_0'(r).
\label{eq:fvarphi_definition_sliced}
\end{equation}
Substituting Eq.~\eqref{eq:Cmu_lag_slice_again} into
Eq.~\eqref{eq:unsliced_reduced_scalar_equation} gives
\begin{equation}
2r^2\partial_v\partial_r\eta
+
2r\,\partial_v\eta
+
L\eta
+
f_\varphi(r_{\rm m})K(r)\eta(v,r_{\rm m})
=
S_{\rm ext}(v,r).
\label{eq:sliced_equation_expanded}
\end{equation}
This motivates the definition of the sliced radial operator
\begin{equation}
L_{\rm sl}\eta
\equiv
L\eta
+
f_\varphi(r_{\rm m})K(r)\eta(r_{\rm m}),
\label{eq:sliced_operator}
\end{equation}
so that the sliced evolution equation becomes
\begin{equation}
2r^2\partial_v\partial_r\eta
+
2r\,\partial_v\eta
+
L_{\rm sl}\eta
=
S_{\rm ext}.
\label{eq:reduced_scalar_equation}
\end{equation}

The unsliced frozen-background operator has the Sturm--Liouville differential
form
\begin{equation}
L\eta
\equiv
\left(
p(r)\eta'(r)
\right)'
-
V_{\rm eff}(r)\eta(r),
\qquad
p(r)
\equiv
e^{\delta_0(r)}r^2\mathcal F_0(r),
\label{eq:unsliced_operator_definition}
\end{equation}
where \(V_{\rm eff}\) is the effective radial potential obtained after
eliminating the metric perturbations through the linearized radial
constraints. Its general expression was derived in Paper~I, and its explicit
GHS specialization will be evaluated below.

The difference between \(L_{\rm sl}\) and \(L\) originates entirely from the
elimination of \(C_\mu^{\rm lag}\) through the local slice. In operator
notation,
\begin{equation}
L_{\rm sl}
=
L
+
f_\varphi(r_{\rm m})
\,|K\rangle
\langle{\rm ev}_{r_{\rm m}}|,
\qquad
{\rm ev}_{r_{\rm m}}[\eta]
\equiv
\eta(r_{\rm m}).
\label{eq:sliced_rank_one}
\end{equation}
On the finite horizon-regular function space introduced in Paper~I,
\({\rm ev}_{r_{\rm m}}\) is a bounded linear functional. The slice-induced
term is therefore a bounded operator of rank at most one. Its range is
contained in
\begin{equation}
\operatorname{span}\{K\}.
\end{equation}
The differential part of \(L_{\rm sl}\) remains local, while the finite-rank
term depends on the value of the scalar lag at the matching radius.

The update has rank exactly one provided
\begin{equation}
f_\varphi(r_{\rm m})K
\not\equiv
0.
\label{eq:rank_one_nontriviality}
\end{equation}
Under the slice-transversality condition
\begin{equation}
m_\varphi(r_{\rm m})\neq0,
\label{eq:slice_transversality_again}
\end{equation}
the canonical tangent identity
\begin{equation}
m_\varphi(r_{\rm m})
=
f_\varphi(r_{\rm m})
\varphi_\varphi(r_{\rm m})
\label{eq:transversality_identity_again}
\end{equation}
implies separately that
\begin{equation}
f_\varphi(r_{\rm m})\neq0,
\qquad
\varphi_\varphi(r_{\rm m})\neq0.
\end{equation}
In the transverse branch, the nontriviality condition given in Eq.~\eqref{eq:rank_one_nontriviality} is therefore equivalent to
\begin{equation}
K\not\equiv0.
\end{equation}

The canonical tangent satisfies
\begin{equation}
L\varphi_\varphi=0.
\label{eq:L_phiPhi_zero_again}
\end{equation}
The sliced operator instead acts on it according to
\begin{align}
L_{\rm sl}\varphi_\varphi
&=
f_\varphi(r_{\rm m})
K(r)\varphi_\varphi(r_{\rm m})
\nonumber\\
&=
m_\varphi(r_{\rm m})K(r).
\label{eq:sliced_tangent_identity}
\end{align}
Consequently, if
\begin{equation}
m_\varphi(r_{\rm m})\neq0,
\qquad
K\not\equiv0,
\label{eq:nontrivial_rank_one_conditions}
\end{equation}
then
\begin{equation}
L_{\rm sl}\varphi_\varphi\neq0,
\qquad
\varphi_\varphi\notin\ker L_{\rm sl}.
\label{eq:canonical_tangent_displaced}
\end{equation}
Thus the local slice displaces the canonical zero mode of the unsliced
operator whenever the slice is transverse and the induced finite-rank update
is nontrivial. These conditions are logically distinct. Transversality
ensures that the slice selects a unique representative, whereas
\(K\not\equiv0\) ensures that this choice modifies the reduced radial
operator.

The displacement of \(\varphi_\varphi\) does not imply that
\(L_{\rm sl}\) has trivial kernel. Since \(L_{\rm sl}-L\) is finite rank, the
Fredholm index is preserved. Under the analytic hypotheses of Paper~I, the
sliced operator remains surjective with a one-dimensional kernel, generated
by a generally different radial profile. The explicit construction of this
displaced generator for the magnetic GHS family is carried out in
Sec.~\ref{subsec:displaced_generator}.

Eq.~\eqref{eq:reduced_scalar_equation} is the full first-order sliced
evolution equation. On the slowly forced adiabatic branch, the scalar lag has
first-order amplitude but inherits its time dependence only from the slowly
varying exterior data and collective coordinates. Hence
\begin{equation}
\partial_v\eta
=
\mathcal O_{\rm ad}(\epsilon_\varphi^2),
\qquad
\partial_v\partial_r\eta
=
\mathcal O_{\rm ad}(\epsilon_\varphi^2),
\label{eq:adiabatic_lag_time_derivatives}
\end{equation}
and the transport terms satisfy
\begin{equation}
2r^2\partial_v\partial_r\eta
+
2r\,\partial_v\eta
=
\mathcal O_{\rm ad}(\epsilon_\varphi^2).
\end{equation}
The leading first-order quasistatic equation is therefore
\begin{equation}
L_{\rm sl}\eta
=
S_{\rm ext}.
\label{eq:leading_sliced_radial_problem}
\end{equation}
This restriction applies only to the slowly forced branch. It cannot be
applied to a generic freely excited homogeneous perturbation with
black hole scale time dependence, for which the transport terms remain at
the same perturbative order as the radial operator.

The local slice and the outer matching condition play distinct roles. The
slice fixes the decomposition of the physical configuration into an
instantaneous static representative and a lag field, and thereby determines
the sliced operator \(L_{\rm sl}\). Its construction uses only the exact
finite-interval radial problem together with the local condition that removes
the remaining tangential freedom; it is therefore independent of any
large-radius approximation or specific exterior model.

The outer matching functional is supplied separately by the exterior
problem. It provides the physical boundary data required to complete the
radial boundary-value problem. It is this completed problem, rather than the
sliced bulk equation alone, whose existence and uniqueness are controlled by
the Fredholm matching condition. A particular form of the matching functional
will be introduced only after the corresponding massless large-radius
overlap model has been specified.

\subsection{Displaced kernel generator}
\label{subsec:displaced_kernel_generator}

The local slice replaces the unsliced radial operator \(L\) by the
finite-rank perturbation
\begin{equation}
L_{\rm sl}
=
L
+
f_\varphi(r_{\rm m})
\,|K\rangle
\langle{\rm ev}_{r_{\rm m}}|.
\label{eq:Lsl_recap}
\end{equation}
The results of Paper~I apply provided the frozen radial problem satisfies the hypotheses stated there: regular coefficients on a finite radial interval, a smooth non-extremal horizon endpoint, and a well-defined finite matching surface. More precisely, the leading coefficient
\begin{equation}
p(r)
=
e^{\delta_0(r)}r^2\mathcal F_0(r)
\end{equation}
must have a simple zero at the static horizon and remain strictly positive
on the open exterior interval, while the remaining coefficients must extend
regularly to the horizon. Under these conditions, horizon regularity selects a one-dimensional space of homogeneous solutions. All of these properties are verified explicitly for the magnetic GHS background in
Sec.~\ref{subsec:explicit_GHS_kernel}.

Under these hypotheses,
\begin{equation}
L:X\longrightarrow Y
\end{equation}
is a surjective Fredholm operator of index one. Its horizon-regular kernel is
therefore one-dimensional. To identify that kernel with the canonical
tangent, one must additionally verify that
\(\varphi_\varphi\) is a nonzero element of the operator domain \(X\).
Since the Canonical Tangent Theorem gives
\begin{equation}
L\varphi_\varphi=0,
\end{equation}
these properties imply
\begin{equation}
\ker L
=
\operatorname{span}\{\varphi_\varphi\}.
\label{eq:ker_L_canonical}
\end{equation}
The required admissibility and nonvanishing of the canonical tangent are
established explicitly for the magnetic GHS family in
Secs.~\ref{subsec:explicit_GHS_tangent} and
\ref{subsec:explicit_GHS_kernel}.

Assume now that the local mass slice is transverse to the canonical tangent,
\begin{equation}
m_\varphi(r_{\rm m})\neq0.
\label{eq:generic_slice_conditions}
\end{equation}
Lemma~1 of Paper~I then implies that
\begin{equation}
L_{\rm sl}:X\longrightarrow Y
\end{equation}
is also surjective and Fredholm of index one. Consequently,
\begin{equation}
\dim\ker L_{\rm sl}=1.
\end{equation}
We denote by \(\psi_\star\) any nonzero generator of this kernel,
\begin{equation}
\ker L_{\rm sl}
=
\operatorname{span}\{\psi_\star\},
\label{eq:psi_star_kernel}
\end{equation}
where \(\psi_\star\) is defined only up to multiplication by a nonzero
constant.

The canonical tangent itself obeys
\begin{equation}
L_{\rm sl}\varphi_\varphi
=
m_\varphi(r_{\rm m})K(r),
\label{eq:sliced_tangent_identity_recap}
\end{equation}
as shown in Sec.~\ref{subsec:sliced_operator}. Hence, if the finite-rank
update is nontrivial,
\begin{equation}
K\not\equiv0,
\label{eq:nontrivial_kernel_update}
\end{equation}
then transversality implies
\begin{equation}
L_{\rm sl}\varphi_\varphi\neq0,
\qquad
\varphi_\varphi\notin\ker L_{\rm sl}.
\end{equation}
Since the sliced kernel remains one-dimensional, its generator must then
satisfy
\begin{equation}
\psi_\star\not\propto\varphi_\varphi.
\label{eq:displaced_generator_not_tangent}
\end{equation}
Thus, the slice displaces the canonical zero mode from the kernel without
removing the one-dimensional homogeneous freedom required by the Fredholm
index.

The displaced generator can be constructed directly. Let \(\chi_0\) be any
horizon-regular particular solution of
\begin{equation}
L\chi_0=K.
\label{eq:particular_kernel_equation}
\end{equation}
A trial combination
\begin{equation}
\psi
=
\varphi_\varphi-c\,\chi_0
\end{equation}
satisfies
\begin{align}
L_{\rm sl}\psi
&=
K(r)
\left[
-c
+
f_\varphi(r_{\rm m})
\left(
\varphi_\varphi(r_{\rm m})
-
c\,\chi_0(r_{\rm m})
\right)
\right].
\label{eq:Lsl_trial_generator}
\end{align}
Whenever
\begin{equation}
1+
f_\varphi(r_{\rm m})\chi_0(r_{\rm m})
\neq0,
\label{eq:conventional_generator_denominator}
\end{equation}
one may therefore choose
\begin{equation}
c
=
\frac{
f_\varphi(r_{\rm m})\varphi_\varphi(r_{\rm m})
}{
1+
f_\varphi(r_{\rm m})\chi_0(r_{\rm m})
}
=
\frac{
m_\varphi(r_{\rm m})
}{
1+
f_\varphi(r_{\rm m})\chi_0(r_{\rm m})
},
\label{eq:conventional_generator_coefficient}
\end{equation}
which gives the conventionally normalized generator
\begin{equation}
\psi_\star
=
\varphi_\varphi
-
\frac{
m_\varphi(r_{\rm m})
}{
1+
f_\varphi(r_{\rm m})\chi_0(r_{\rm m})
}
\chi_0.
\label{eq:conventional_displaced_generator}
\end{equation}

The denominator in Eq.~\eqref{eq:conventional_displaced_generator} belongs
only to this choice of normalization. A generator that remains algebraically
well defined even where that denominator vanishes is obtained by clearing it:
\begin{equation}
\widehat\psi_\star(r)
\equiv
\left[
1+
f_\varphi(r_{\rm m})\chi_0(r_{\rm m})
\right]
\varphi_\varphi(r)
-
m_\varphi(r_{\rm m})\chi_0(r).
\label{eq:globally_regular_generator_general}
\end{equation}
Using
\begin{equation}
L\varphi_\varphi=0,
\qquad
L\chi_0=K,
\qquad
m_\varphi(r_{\rm m})
=
f_\varphi(r_{\rm m})\varphi_\varphi(r_{\rm m}),
\end{equation}
one finds directly
\begin{equation}
L_{\rm sl}\widehat\psi_\star=0.
\label{eq:widehat_generator_kernel}
\end{equation}
Moreover, its value at the matching radius is
\begin{equation}
\widehat\psi_\star(r_{\rm m})
=
\varphi_\varphi(r_{\rm m}),
\label{eq:widehat_generator_matching_value}
\end{equation}
which is nonzero under slice transversality. Thus
\(\widehat\psi_\star\) is a nonzero generator of
\(\ker L_{\rm sl}\) throughout the transverse branch, including at points
where the conventional normalization given in Eq.~\eqref{eq:conventional_displaced_generator} becomes singular.

The construction is also independent of the freedom in the choice of
particular solution. Indeed, the replacement
\begin{equation}
\chi_0
\longrightarrow
\chi_0+c_0\varphi_\varphi,
\qquad
c_0\in\mathbb R,
\end{equation}
leaves \(\widehat\psi_\star\) in
Eq.~\eqref{eq:globally_regular_generator_general} unchanged. The regular
generator therefore depends only on the sliced operator and not on the
particular horizon-regular solution chosen for \(L\chi_0=K\).

If instead
\begin{equation}
K\equiv0,
\end{equation}
then the finite-rank update vanishes,
\begin{equation}
L_{\rm sl}=L,
\end{equation}
and no kernel displacement occurs. In that degenerate branch, one may simply
choose
\begin{equation}
\psi_\star=\varphi_\varphi.
\end{equation}
The condition \(K\not\equiv0\) is therefore needed only to conclude that the
sliced-kernel generator differs from the canonical tangent; it is not
required for the existence of a one-dimensional sliced kernel.

For the magnetic GHS family, the functions
\(\varphi_\varphi\), \(K\), and a horizon-regular particular solution
\(\chi_0\) can all be obtained in closed analytical form. Substitution into
Eq.~\eqref{eq:globally_regular_generator_general} yields the explicit
globally regular representative used below. This specialization is carried
out in Sec.~\ref{subsec:displaced_generator}.

Neither \(\psi_\star\) nor \(\widehat\psi_\star\) is itself the physical
forced deformation. They generate the unique homogeneous radial direction
left by the sliced bulk equation. The local slice has fixed the ambiguity in
the decomposition into an instantaneous representative and a lag field, but
it has not fixed the amplitude of this remaining homogeneous solution.

That amplitude is determined only after the near-zone problem is completed
by an outer matching condition. The relevant question is therefore whether
the matching functional acts nontrivially on the sliced-kernel generator.
This leads to the Fredholm solvability criterion discussed in the following
subsection.

\subsection{Fredholm solvability criterion}
\label{subsec:fredholm}

On the slowly forced adiabatic branch,
\begin{equation}
\frac{\eta}{\Mpl}
=
\mathcal O_{\rm ad}(\epsilon_\varphi),
\qquad
\frac{r_{\rm H}^{\rm inst}}{\Mpl}\,
\partial_v\eta
=
\mathcal O_{\rm ad}(\epsilon_\varphi^2).
\end{equation}
so that the transport terms in the sliced evolution equation, Eq.~\eqref{eq:reduced_scalar_equation}, begin only at second adiabatic order. At
each fixed advanced time, the leading quasistatic response therefore obeys
the frozen radial equation
\begin{equation}
L_{\rm sl}\eta
=
S_{\rm ext},
\label{eq:leading_sliced_radial_problem_repeated}
\end{equation}
on the finite interval
\begin{equation}
r_{\rm H}^{\rm inst}(v)
\leq
r
\leq
r_{\rm m}.
\end{equation}
The inner condition is regularity at the future horizon of the
instantaneous static representative. The outer condition is supplied by the
exterior matching problem and is represented abstractly as
\begin{equation}
\mathcal B_{\rm m}\eta
=
J_{\rm ext}(v).
\label{eq:outer_matching_condition}
\end{equation}
Here
\begin{equation}
\mathcal B_{\rm m}:X\longrightarrow\mathbb R
\end{equation}
is a bounded linear functional, while \(J_{\rm ext}(v)\) is the
corresponding inhomogeneous datum. At this stage,
\(\mathcal B_{\rm m}\) is kept abstract: no particular exterior model,
overlap expansion, or large-radius approximation has yet been imposed.

All operators and radial profiles in this subsection are evaluated on the
instantaneous static representative at the fixed value of \(v\) under
consideration. Their parametric dependence on \(v\) will generally be left
implicit.

As reviewed in
Sec.~\ref{subsec:displaced_kernel_generator}, the sliced operator
\begin{equation}
L_{\rm sl}:X\longrightarrow Y
\end{equation}
is surjective and Fredholm of index one, with a one-dimensional kernel,
\begin{equation}
\ker L_{\rm sl}
=
\operatorname{span}\{\psi_\star\},
\label{eq:sliced_kernel_recap}
\end{equation}
where \(\psi_\star\) is any nonzero generator and is defined only up to
multiplication by a nonzero constant.

Let \(\eta_0\in X\) be any horizon-regular particular solution of
\begin{equation}
L_{\rm sl}\eta_0
=
S_{\rm ext}.
\label{eq:particular_sliced_solution}
\end{equation}
Such a solution exists because \(L_{\rm sl}\) is surjective. Every
horizon-regular solution of the radial equation then has the form
\begin{equation}
\eta(v,r)
=
\eta_0(v,r)
+
t(v)\psi_\star(r),
\label{eq:general_sliced_solution}
\end{equation}
where \(t(v)\) is the remaining homogeneous amplitude. Imposing the outer
matching condition gives the scalar equation
\begin{equation}
t(v)\,
\mathcal B_{\rm m}\psi_\star
=
J_{\rm ext}(v)
-
\mathcal B_{\rm m}\eta_0.
\label{eq:matching_amplitude_equation}
\end{equation}

If
\begin{equation}
\mathcal B_{\rm m}\psi_\star
\neq
0,
\label{eq:fredholm_condition}
\end{equation}
the homogeneous amplitude is fixed uniquely:
\begin{equation}
t(v)
=
\frac{
J_{\rm ext}(v)-\mathcal B_{\rm m}\eta_0
}{
\mathcal B_{\rm m}\psi_\star
}.
\label{eq:matching_amplitude_solution}
\end{equation}
The completed leading-order boundary-value problem therefore admits a
unique solution for every admissible pair
\begin{equation}
\bigl(
S_{\rm ext},
J_{\rm ext}
\bigr)
\in
Y\times\mathbb R
\end{equation}
if and only if Eq.~\eqref{eq:fredholm_condition} holds.

Equivalently, the generator of the one-dimensional kernel of
\(L_{\rm sl}\) must not satisfy the homogeneous outer matching condition.
Because \(\psi_\star\) is defined only up to multiplication by a nonzero
constant, the numerical value of
\(\mathcal B_{\rm m}\psi_\star\) depends on the normalization of the
generator, but its vanishing or nonvanishing does not.

The complementary case is also instructive. If
\begin{equation}
\mathcal B_{\rm m}\psi_\star=0,
\end{equation}
then the outer condition is insensitive to the remaining homogeneous
direction. Eq.~\eqref{eq:matching_amplitude_equation} has a solution
only if the data satisfy the compatibility condition
\begin{equation}
J_{\rm ext}(v)
=
\mathcal B_{\rm m}\eta_0.
\label{eq:degenerate_matching_compatibility}
\end{equation}
When this condition fails, no solution exists. When it holds, the amplitude
\(t(v)\) remains arbitrary, and the solution is not unique. Thus
\(\mathcal B_{\rm m}\psi_\star\neq0\) is precisely the condition that
guarantees both existence for arbitrary admissible data and uniqueness.

This is the Slice and Solvability Theorem of Paper~I. In operator form, the
completed radial boundary-value problem is
\begin{equation}
\mathcal A\eta
\equiv
\bigl(
L_{\rm sl}\eta,
\mathcal B_{\rm m}\eta
\bigr)
=
\bigl(
S_{\rm ext},
J_{\rm ext}
\bigr),
\qquad
\mathcal A:
X
\longrightarrow
Y\times\mathbb R.
\label{eq:completed_operator_problem}
\end{equation}
Adjoining one bounded scalar functional to the codomain lowers the Fredholm
index by one, so that
\begin{equation}
\operatorname{ind}\mathcal A
=
\operatorname{ind}L_{\rm sl}-1
=
0.
\end{equation}
Moreover,
\begin{equation}
\ker\mathcal A
=
\ker L_{\rm sl}
\cap
\ker\mathcal B_{\rm m}.
\end{equation}
Using Eq.~\eqref{eq:sliced_kernel_recap}, one finds
\begin{equation}
\ker\mathcal A
=
\{0\}
\quad\Longleftrightarrow\quad
\mathcal B_{\rm m}\psi_\star\neq0.
\end{equation}
Since \(\mathcal A\) is Fredholm of index zero, triviality of its kernel is
equivalent to triviality of its cokernel. Hence
\begin{equation}
\mathcal A
\ \text{is bijective}
\quad\Longleftrightarrow\quad
\mathcal B_{\rm m}\psi_\star\neq0.
\end{equation}

The local slice and the outer matching condition play logically distinct
roles. The slice fixes the decomposition into an instantaneous static
representative and a lag field and thereby determines \(L_{\rm sl}\) and
its homogeneous kernel. The independently constructed matching functional
\(\mathcal B_{\rm m}\) then removes the remaining homogeneous radial freedom
precisely when it acts nontrivially on the kernel generator.

In Sec.~\ref{sec:GHS_application}, the canonical tangent, the source profile
\(K(r)\), and a globally regular generator of
\(\ker L_{\rm sl}\) are evaluated explicitly for the magnetic GHS family.
A specific matching functional is introduced only after its associated
large-radius overlap regime has been specified. Its action on the globally
regular kernel generator is then evaluated explicitly, thereby testing the
non-degeneracy condition given in Eq.~\eqref{eq:fredholm_condition} within that matching
model.


\section{Application to the magnetic GHS family}
\label{sec:GHS_application}

We now specialize the general construction of Sec.~\ref{sec:canonical_adiabatic_deformation}
to the magnetic GHS family. All static background quantities are written in
the matching normalization
\begin{equation}
\delta_0(r_{\rm m};\lambda)=0,
\label{eq:GHS_matching_normalization_section4}
\end{equation}
obtained in Sec.~\ref{subsec:static_limit}. Consequently,
\begin{equation}
\delta_\varphi(r_{\rm m})=0
\end{equation}
throughout the static family. At each fixed advanced time, the formulas below
are evaluated on the instantaneous static representative
\(\lambda(v)\), although this time dependence will usually be left implicit.

\subsection{Explicit canonical tangent}
\label{subsec:explicit_GHS_tangent}

We now evaluate explicitly the canonical tangent introduced in
Sec.~\ref{subsec:canonical_tangent_recap}. The magnetic charge \(Q_m\) is
held fixed throughout, and all parameter derivatives are taken at fixed exact
areal radius \(r\).

On the static solution manifold, recall from Eq.~\eqref{eq:canonical_tangent_definition} that the canonical tangent is
\begin{equation}
\varphi_\varphi(r)
=
\left.
\frac{\partial\varphi_0(r)}{\partial\varphi}
\right|_{M_{{\rm H},0},\,r},
\qquad
\delta_\varphi(r)
=
\left.
\frac{\partial\delta_0(r)}{\partial\varphi}
\right|_{M_{{\rm H},0},\,r},
\qquad
m_\varphi(r)
=
\left.
\frac{\partial m_0(r)}{\partial\varphi}
\right|_{M_{{\rm H},0},\,r}.
\label{eq:GHS_tangent_definition}
\end{equation}

The static GHS family may be parametrized by \((\varphi,a)\), where
\begin{equation}
a
\equiv
\frac{r_+-r_-}{r_+},
\qquad
r_{{\rm H},0}
=
r_+\sqrt a,
\qquad
M_{{\rm H},0}
=
\frac{\Mpl^2}{2}\,
r_{{\rm H},0}.
\label{eq:GHS_parameters_again}
\end{equation}
In the fixed-charge sector,
\begin{equation}
M_{{\rm H},0}
=
\frac{\Mpl Q_m}{2}
e^{-\alpha\varphi/\Mpl}
\sqrt{\frac{a}{1-a}},
\label{eq:GHS_MH_again}
\end{equation}
and hence
\begin{equation}
\left.
\frac{\partial a}{\partial\varphi}
\right|_{M_{{\rm H},0}}
=
\frac{2\alpha}{\Mpl}\,
a(1-a).
\label{eq:a_Phi_fixed_MH}
\end{equation}
For any function \(X(r;\varphi,a)\) on the static solution manifold, the
canonical derivative at fixed areal radius therefore takes the form
\begin{equation}
\left.
\frac{\partial X}{\partial\varphi}
\right|_{M_{{\rm H},0},\,r}
=
\left.
\frac{\partial X}{\partial\varphi}
\right|_{a,\,r}
+
\frac{2\alpha}{\Mpl}\,
a(1-a)
\left.
\frac{\partial X}{\partial a}
\right|_{\varphi,\,r}.
\label{eq:canonical_derivative_GHS}
\end{equation}

Since fixing \(M_{{\rm H},0}\) also fixes \(r_{{\rm H},0}\),
\begin{equation}
r_+
=
\frac{r_{{\rm H},0}}{\sqrt a},
\qquad
r_-
=
\frac{r_{{\rm H},0}(1-a)}{\sqrt a},
\label{eq:rpm_fixed_rH}
\end{equation}
and therefore
\begin{equation}
\left.
\frac{\partial r_+}{\partial\varphi}
\right|_{M_{{\rm H},0}}
=
-\frac{\alpha}{\Mpl}(1-a)r_+,
\qquad
\left.
\frac{\partial r_-}{\partial\varphi}
\right|_{M_{{\rm H},0}}
=
-\frac{\alpha}{\Mpl}(1+a)r_-.
\label{eq:rpm_Phi_fixed_MH}
\end{equation}

It is useful to define
\begin{equation}
D(r)
\equiv
2\rho(r)-r_-
=
\sqrt{r_-^2+4r^2}.
\label{eq:D_definition}
\end{equation}
The areal-radius relation
\begin{equation}
r^2
=
\rho(\rho-r_-)
\label{eq:areal_relation_again}
\end{equation}
implies, at fixed \(r\),
\begin{equation}
\left.
\frac{\partial\rho}{\partial r_-}
\right|_r
=
\frac{\rho}{D}.
\label{eq:rho_rminus_fixed_r}
\end{equation}

The static scalar profile of Eq.~\eqref{eq:GHS_scalar_again} is given by
\begin{equation}
\varphi_0(r)
=
\varphi
+
\frac{\Mpl}{2\alpha}
\ln\!\left[
\frac{\rho(r)}{\rho(r)-r_-}
\right].
\label{eq:GHS_scalar_again_2}
\end{equation}
Using
\begin{equation}
\left.
\frac{\partial}{\partial r_-}
\ln\!\left[
\frac{\rho}{\rho-r_-}
\right]
\right|_r
=
\frac{2}{D},
\label{eq:log_derivative_identity}
\end{equation}
together with Eq.~\eqref{eq:rpm_Phi_fixed_MH}, one obtains
\begin{equation}
\varphi_\varphi(r)
=
1-
\frac{(1+a)r_-}{D(r)}.
\label{eq:explicit_varphi_Phi}
\end{equation}
Its asymptotic and static-horizon values are
\begin{equation}
\varphi_\varphi(\infty)=1,
\qquad
\varphi_\varphi(r_{{\rm H},0})=a.
\label{eq:varphi_Phi_limits}
\end{equation}
The second identity follows from
\begin{equation}
D(r_{{\rm H},0})
=
r_+(1+a).
\label{eq:D_horizon_again}
\end{equation}

The sign of the canonical tangent has a direct interpretation in the present
family. In throat variables,
\begin{equation}
\varphi_\varphi(b)
=
\frac{a(1+a)+2b}{1+a+2b},
\label{eq:varphiPhi_throat_positive}
\end{equation}
and therefore
\begin{equation}
a
\leq
\varphi_\varphi(b)
<
1,
\qquad
0<a<1,
\qquad
0\leq b<\infty.
\label{eq:varphiPhi_positive_bounds}
\end{equation}
Moreover,
\begin{equation}
\frac{d\varphi_\varphi}{db}
=
\frac{2(1-a^2)}
{(1+a+2b)^2}
>0.
\label{eq:varphiPhi_monotonicity}
\end{equation}
Thus, at fixed static horizon-mass parameter, increasing the modulus label
\(\varphi\) increases the static scalar profile at every point of the exterior.
The canonical zero mode is strictly positive and monotonic, with no zeros on
the exterior interval: it increases from
\(\varphi_\varphi(r_{{\rm H},0})=a\) at the horizon to
\(\varphi_\varphi(\infty)=1\).

This positivity is a special property of the magnetic GHS family and is not
required by the general Canonical Tangent Theorem of Paper~I. In the present
application, it ensures that division by \(\varphi_\varphi\) is globally well
defined throughout the charged non-extremal exterior. It therefore permits
the reduction-of-order constructions used below without encountering an
interior zero of the canonical mode.

Since \(\varphi_\varphi\) is obtained by varying the static modulus label at
fixed \(M_{{\rm H},0}\), the Canonical Tangent Theorem gives
\begin{equation}
L\varphi_\varphi=0.
\label{eq:GHS_canonical_zero_mode}
\end{equation}

We next evaluate the lapse component. In the asymptotic normalization,
\begin{equation}
e^{\delta_0^{(\infty)}(r)}
=
\frac{2r}{D(r)},
\qquad
\delta_0^{(\infty)}(r)
=
\ln\!\left[
\frac{2r}{D(r)}
\right].
\label{eq:delta_asymptotic_again}
\end{equation}
Differentiation at fixed \(M_{{\rm H},0}\) and fixed \(r\) gives
\begin{equation}
\delta_\varphi^{(\infty)}(r)
=
\frac{\alpha}{\Mpl}
(1+a)
\frac{r_-^2}{r_-^2+4r^2}.
\label{eq:delta_Phi_asymptotic}
\end{equation}
In particular,
\begin{equation}
\delta_\varphi^{(\infty)}(\infty)=0,
\qquad
\delta_\varphi^{(\infty)}(r_{{\rm H},0})
=
\frac{\alpha}{\Mpl}
\frac{(1-a)^2}{1+a}.
\label{eq:delta_Phi_asymptotic_limits}
\end{equation}

The matching-normalized lapse was obtained in
Sec.~\ref{subsec:static_limit} from
\begin{equation}
\delta_0(r)
=
\delta_0^{(\infty)}(r)
-
\delta_0^{(\infty)}(r_{\rm m}).
\label{eq:delta_matching_from_asymptotic}
\end{equation}
Differentiating this relation along the canonical direction gives
\begin{equation}
\delta_\varphi(r)
=
\delta_\varphi^{(\infty)}(r)
-
\delta_\varphi^{(\infty)}(r_{\rm m}),
\label{eq:delta_Phi_matching_relation}
\end{equation}
or explicitly
\begin{equation}
\delta_\varphi(r)
=
\frac{\alpha}{\Mpl}
(1+a)r_-^2
\left[
\frac{1}{r_-^2+4r^2}
-
\frac{1}{r_-^2+4r_{\rm m}^2}
\right].
\label{eq:explicit_delta_Phi}
\end{equation}
By construction,
\begin{equation}
\delta_\varphi(r_{\rm m})=0.
\label{eq:delta_Phi_matching_zero}
\end{equation}

The mass component follows from the canonical tangent decomposition derived
in Sec.~\ref{subsec:canonical_tangent_recap}. For the magnetic GHS
background,
\begin{align}
f_\varphi(r)
&\equiv
\frac{r^2}{2}\,
\mathcal F_0(r)\varphi_0'(r)
\nonumber\\
&=
-\frac{\Mpl r_-D(r)}{8\alpha r}
\left(
1-\frac{r_+}{\rho(r)}
\right).
\label{eq:explicit_fvarphi_GHS}
\end{align}
Since \(C_{\mu,\varphi}=0\), one has
\begin{equation}
m_\varphi(r)
=
f_\varphi(r)\varphi_\varphi(r).
\label{eq:GHS_tangent_constraint_check}
\end{equation}
Using Eq.~\eqref{eq:explicit_varphi_Phi}, this becomes
\begin{equation}
m_\varphi(r)
=
-\frac{\Mpl r_-}{8\alpha r}
\left(
1-\frac{r_+}{\rho(r)}
\right)
\left[
D(r)-(1+a)r_-
\right].
\label{eq:explicit_m_Phi}
\end{equation}
Direct differentiation of the Misner--Sharp mass function given by Eq.~\eqref{eq:GHS_mass_areal} at fixed \(M_{{\rm H},0}\) and fixed \(r\)
reproduces the same expression.

At the static horizon,
\begin{equation}
\rho(r_{{\rm H},0})=r_+,
\end{equation}
and therefore
\begin{equation}
f_\varphi(r_{{\rm H},0})=0,
\qquad
m_\varphi(r_{{\rm H},0})=0.
\label{eq:GHS_tangent_horizon_check}
\end{equation}
For every point in the open charged exterior,
\(r>r_{{\rm H},0}\), one has
\begin{equation}
1-\frac{r_+}{\rho(r)}>0,
\qquad
\varphi_\varphi(r)>0.
\end{equation}
Since \(r_->0\) and \(\alpha>0\), Eq.~\eqref{eq:explicit_fvarphi_GHS}
therefore gives
\begin{equation}
f_\varphi(r)<0,
\qquad
m_\varphi(r)
=
f_\varphi(r)\varphi_\varphi(r)
<0,
\qquad
r>r_{{\rm H},0}.
\label{eq:GHS_mass_tangent_sign}
\end{equation}
In particular, for every nontrivial finite matching interval,
\begin{equation}
r_{\rm m}>r_{{\rm H},0}
\qquad\Longrightarrow\qquad
m_\varphi(r_{\rm m})<0.
\label{eq:GHS_slice_transversality_explicit}
\end{equation}
The local mass slice is therefore transverse throughout the open charged
non-extremal exterior. This conclusion is intrinsic to the finite-interval
radial construction and does not require the large-radius overlap
approximation introduced later.

The horizon relations above concern the horizon of the static reference
solution. After promotion, they hold at the horizon
\(r_{\rm H}^{\rm inst}(v)\) of the instantaneous static representative, not
at the exact dynamical horizon \(r_{\rm H}^{\rm exact}(v)\).

The magnetic GHS family therefore realizes all defining properties of the
canonical tangent in closed form. Its scalar component is a strictly positive
homogeneous zero mode of \(L\), while its lapse and mass components are
determined by the linearized radial constraints. The explicit sign of the
mass tangent also verifies the transversality of the local mass slice on
every nontrivial finite matching interval.

\subsection{Explicit source kernel}
\label{subsec:explicit_GHS_kernel}

The finite-rank modification of the reduced operator is governed by the
background source profile introduced in
Sec.~\ref{subsec:sliced_operator},
\begin{equation}
K(r)
=
\frac{2}{\Mpl^2}
\left(
r\varphi_0'(r)
\right)'.
\label{eq:K_definition_again}
\end{equation}
Since \(K(r)\) depends only on the static scalar profile, it is independent
of the constant lapse renormalization used to pass from the asymptotic
normalization to the finite-radius matching normalization.

For the magnetic GHS background,
\begin{equation}
\varphi_0'(r)
=
-\frac{\Mpl r_-}
{\alpha rD(r)},
\qquad
D(r)
=
\sqrt{r_-^2+4r^2},
\label{eq:GHS_phi_prime_again}
\end{equation}
and therefore
\begin{equation}
r\varphi_0'(r)
=
-\frac{\Mpl r_-}{\alpha D(r)}.
\label{eq:r_phi_prime}
\end{equation}
Using
\begin{equation}
D'(r)
=
\frac{4r}{D(r)},
\label{eq:Dprime}
\end{equation}
one finds
\begin{equation}
\left(
r\varphi_0'
\right)'
=
\frac{4\Mpl r_-r}
{\alpha D(r)^3}.
\label{eq:r_phi_prime_prime}
\end{equation}
Hence
\begin{equation}
K(r)
=
\frac{8rr_-}
{\alpha\Mpl D(r)^3}
=
\frac{8rr_-}
{\alpha\Mpl
\left(r_-^2+4r^2\right)^{3/2}}.
\label{eq:explicit_K_GHS}
\end{equation}

For the charged non-extremal GHS branch,
\begin{equation}
0<a<1,
\label{eq:nonextremal_range_again}
\end{equation}
one has \(r_->0\). Since \(\alpha>0\),
Eq.~\eqref{eq:explicit_K_GHS} shows that \(K(r)\) is smooth and strictly
positive throughout the static exterior,
\begin{equation}
K(r)>0,
\qquad
r_{{\rm H},0}\leq r<\infty.
\end{equation}

At the static horizon,
\begin{equation}
D(r_{{\rm H},0})
=
r_+(1+a),
\end{equation}
and Eq.~\eqref{eq:explicit_K_GHS} gives
\begin{equation}
K(r_{{\rm H},0})
=
\frac{8a(1-a)}
{\alpha\Mpl r_{{\rm H},0}(1+a)^3}
=
\frac{4\Mpl}
{\alpha M_{{\rm H},0}}
\frac{a(1-a)}
{(1+a)^3}
>0.
\label{eq:K_horizon}
\end{equation}
The formal endpoint limits are
\begin{equation}
K(r_{{\rm H},0})
\longrightarrow0
\qquad
(a\to0),
\qquad
K(r_{{\rm H},0})
\longrightarrow0
\qquad
(a\to1).
\label{eq:K_horizon_limits}
\end{equation}
The first limit corresponds to the extremal GHS endpoint, which lies outside
the uniformly non-degenerate horizon framework assumed in the adiabatic
construction. The second is the Schwarzschild endpoint, where the scalar
profile becomes constant.

For later use in the overlap analysis, the formal large-radius expansion of
the source profile is
\begin{equation}
K(r)
=
\frac{r_-}
{\alpha\Mpl r^2}
\left[
1
-\frac{3r_-^2}{8r^2}
+
\mathcal O(r^{-4})
\right],
\label{eq:K_asymptotic}
\end{equation}
and hence
\begin{equation}
K(r)
=
\mathcal O(r^{-2}),
\qquad
r\to\infty.
\label{eq:K_decay}
\end{equation}
This asymptotic expansion is used only to characterize the large-radius end
of the overlap region. The intrinsic near-zone boundary-value problem
remains defined on the finite interval
\([r_{\rm H}^{\rm inst}(v),r_{\rm m}]\).

In the formal Schwarzschild limit \(r_-\to0\), the static scalar profile, Eq.~\eqref{eq:GHS_scalar_again_2},
becomes constant,
\begin{equation}
\varphi_0(r)\equiv\varphi,
\label{eq:phi_constant_schwarzschild}
\end{equation}
and therefore
\begin{equation}
K(r)\equiv0.
\label{eq:K_schwarzschild_zero}
\end{equation}
The finite-rank contribution to \(L_{\rm sl}\) then vanishes. At the same
time,
\begin{equation}
m_\varphi(r_{\rm m})
=
f_\varphi(r_{\rm m})\varphi_\varphi(r_{\rm m})
=
0,
\end{equation}
so the local mass slice ceases to be transverse. The strict Schwarzschild
endpoint is therefore a degenerate consistency limit rather than a member of
the transverse charged branch analyzed below.

For a general static magnetic EMD background, the
effective radial potential derived in Paper~I is
\begin{equation}
V_{\rm eff}(r)
=
e^{\delta_0(r)}r^2
\left[
V''\!\left(\varphi_0(r)\right)
+
\frac{Q_m^2}{2r^4}
B''\!\left(\varphi_0(r)\right)
\right]
+
\frac{d}{dr}
\left[
e^{\delta_0(r)}
\frac{
r^3\mathcal F_0(r)\varphi_0'(r)^2
}{
\Mpl^2
}
\right].
\label{eq:Veff_general_GHS_section}
\end{equation}
Recall that in the massless magnetic GHS theory,
\begin{equation}
V(\varphi)=0,
\qquad
B(\varphi)=e^{-2\alpha\varphi/\Mpl},
\end{equation}
and substitution of the exact static profiles gives
\begin{equation}
V_{\rm eff}(r)
=
\frac{2D_{\rm m}}{r_{\rm m}}\,
\frac{
r\,r_-D_{\rm H}
}{
D(r)^3
},
\qquad
D_{\rm H}
\equiv
D(r_{{\rm H},0})
=
r_+(1+a).
\label{eq:Veff_GHS_explicit_r}
\end{equation}
Equivalently, in the throat coordinates,
\begin{equation}
V_{\rm eff}(b)
=
\frac{2D_{\rm m}}{r_{\rm m}}\,
\frac{
(1-a^2)\sqrt{(1+b)(a+b)}
}{
(1+a+2b)^3
}.
\label{eq:Veff_GHS_explicit_b}
\end{equation}
The same result follows directly from the canonical zero-mode identity,
\begin{equation}
V_{\rm eff}(r)
=
\frac{
\left[
p(r)\varphi_\varphi'(r)
\right]'
}{
\varphi_\varphi(r)
}.
\label{eq:Veff_from_canonical_zero_mode}
\end{equation}
This expression is globally well defined throughout the charged
non-extremal exterior because \(\varphi_\varphi(r)>0\). In particular,
\(V_{\rm eff}\) is smooth and strictly positive there. Its formal
large-radius behavior is
\begin{equation}
V_{\rm eff}(r)
=
\frac{D_{\rm m}r_-D_{\rm H}}
{4r_{\rm m}r^2}
+
\mathcal O(r^{-4}).
\label{eq:Veff_large_r}
\end{equation}

\paragraph{Verification of the Fredholm hypotheses.}
We now verify that the charged non-extremal GHS background satisfies the
concrete hypotheses required to apply the Fredholm results of Paper~I on a
finite, nontrivial near-zone interval
\begin{equation}
r_{{\rm H},0}
\leq
r
\leq
r_{\rm m},
\qquad
r_{\rm m}>r_{{\rm H},0}.
\label{eq:GHS_finite_interval_Fredholm}
\end{equation}
At each fixed advanced time, all coefficients are evaluated on the
corresponding instantaneous static representative, with their dependence on
\(\lambda(v)\) left implicit.

The leading Sturm--Liouville coefficient introduced in Eq.~\eqref{eq:unsliced_operator_definition} is
\begin{equation}
p(r)
=
e^{\delta_0(r)}r^2\mathcal F_0(r).
\end{equation}
In the throat coordinates, this expression takes the explicit form
\begin{equation}
p(b)
=
\frac{D_{\rm m}}{4r_{\rm m}}\,
\frac{r(b)D(b)\,b}{1+b}.
\label{eq:p_GHS_Fredholm_check}
\end{equation}
For \(0<a<1\), the factors \(r(b)\), \(D(b)\), and \(1+b\) are smooth and
strictly positive for \(b\geq0\). Hence
\begin{equation}
p(b)>0,
\qquad
b>0,
\end{equation}
whereas at the static horizon
\begin{equation}
p(0)=0,
\qquad
\left.
\frac{dp}{db}
\right|_{b=0}
=
\frac{D_{\rm m}}{4r_{\rm m}}\,
r_{{\rm H},0}D_{\rm H}
>0.
\label{eq:p_simple_horizon_zero}
\end{equation}
Since \(dr/db\) is finite and strictly positive at \(b=0\) for
\(0<a<1\), \(p(r)\) has a simple zero at the non-extremal horizon and no
further zero on the open exterior interval.

The remaining coefficients of the reduced operator are regular. In
particular, Eqs.~\eqref{eq:Veff_GHS_explicit_r} and
\eqref{eq:explicit_K_GHS} show that \(V_{\rm eff}\) and \(K\) are smooth and
bounded on \([r_{{\rm H},0},r_{\rm m}]\). Moreover,
\begin{equation}
V_{\rm eff}(r)>0,
\qquad
K(r)>0,
\qquad
r_{{\rm H},0}\leq r\leq r_{\rm m}.
\label{eq:Veff_K_positive_finite_interval}
\end{equation}

The horizon is therefore a regular singular endpoint of the homogeneous
equation,
\begin{equation}
Lu=0.
\end{equation}
Since \(p\) has a simple zero while \(V_{\rm eff}\) remains finite there,
the local Frobenius analysis yields one horizon-regular solution and a
second independent branch with logarithmic behavior. More explicitly, near
\(r=r_{{\rm H},0}\), the regular branch admits the expansion,
\begin{equation}
u(r)
=
u_{\rm H}
+
\mathcal O(r-r_{{\rm H},0}),
\end{equation}
whereas the second independent branch behaves as
\begin{equation}
u_{\rm sing}(r)
=
u_{\rm reg}(r)
\ln\!\left[r-r_{{\rm H},0}\right]
+
\mathcal O(1).
\end{equation}
The logarithmic branch is excluded from the horizon-regular domain.
Horizon regularity therefore selects a one-dimensional local homogeneous
solution space.

The canonical tangent provides a globally regular, nonzero representative
of this regular branch:
\begin{equation}
L\varphi_\varphi=0,
\qquad
\varphi_\varphi(r)
=
1-\frac{(1+a)r_-}{D(r)}.
\end{equation}
As shown in Eqs.~\eqref{eq:varphiPhi_positive_bounds} and
\eqref{eq:varphiPhi_monotonicity}, \(\varphi_\varphi\) is strictly positive
and regular throughout the charged non-extremal exterior. It therefore
belongs to the horizon-regular domain used in Paper~I. Together with the
one-dimensionality of the regular homogeneous solution space, this gives
\begin{equation}
\ker L
=
\operatorname{span}\{\varphi_\varphi\}
\end{equation}
on that domain.

Finally, evaluation at the finite matching radius,
\begin{equation}
{\rm ev}_{r_{\rm m}}[u]
=
u(r_{\rm m}),
\end{equation}
is a bounded linear functional on the regular Banach space used in Paper~I.
Consequently,
\begin{equation}
L_{\rm sl}
=
L
+
f_\varphi(r_{\rm m})
\,|K\rangle
\langle{\rm ev}_{r_{\rm m}}|
\end{equation}
is a bounded finite-rank perturbation of the unsliced operator.

These properties place the charged non-extremal magnetic GHS problem within
the Fredholm class analyzed in Paper~I. On the horizon-regular domain,
\(L\) is surjective and Fredholm of index one. For every nontrivial finite matching interval,
\begin{equation}
r_{\rm m}>r_{{\rm H},0},
\end{equation}
the explicit expressions derived in the previous subsection give
\begin{equation}
m_\varphi(r_{\rm m})<0.
\end{equation}
Hence the local mass slice is transverse. The sliced operator \(L_{\rm sl}\) is
therefore likewise surjective and Fredholm of index one, with a
one-dimensional kernel.

The endpoint limits \(a\to0\) and \(a\to1\) are excluded because uniform
horizon non-degeneracy and slice transversality, respectively, are then
lost. All statements in this verification concern the exact finite-interval
radial problem and do not rely on the large-radius overlap approximation
introduced later to derive the particular massless matching functional.

 
\subsection{Explicit displaced homogeneous generator}
\label{subsec:displaced_generator}

Paper~I shows that, in the nontrivial finite-rank branch, the
one-dimensional kernel of the sliced operator is generated by a homogeneous
mode that is generally displaced from the canonical tangent. We now
construct this generator explicitly for the magnetic GHS background.

We begin by constructing a horizon-regular particular solution of
\begin{equation}
L\chi_0
=
K,
\label{eq:chi0_inhomogeneous_equation}
\end{equation}
where \(K(r)\) is the source profile derived in
Sec.~\ref{subsec:explicit_GHS_kernel}.

Recall that the reduced operator has the Sturm--Liouville form
\begin{equation}
Lu
=
\left(
p(r)u'(r)
\right)'
-
V_{\rm eff}(r)u(r),
\qquad
p(r)
=
e^{\delta_0(r)}r^2\mathcal F_0(r).
\label{eq:L_Sturm_Liouville_GHS}
\end{equation}

The canonical tangent satisfies
\begin{equation}
L\varphi_\varphi
=
0.
\label{eq:varphiPhi_homogeneous_again}
\end{equation}
Moreover, as shown in
Eqs.~\eqref{eq:varphiPhi_positive_bounds} and
\eqref{eq:varphiPhi_monotonicity},
\begin{equation}
\varphi_\varphi(r)>0,
\qquad
0<a<1,
\qquad
b\geq0,
\end{equation}
throughout the charged non-extremal static exterior. This strict positivity
is special to the magnetic GHS family and is not implied by the general
Canonical Tangent Theorem of Paper~I. It ensures that division by
\(\varphi_\varphi\) is globally well defined and permits the reduction-of-order
ansatz
\begin{equation}
\chi_0(r)
=
\varphi_\varphi(r)I(r).
\label{eq:chi0_reduction_ansatz}
\end{equation}
Using \(L\varphi_\varphi=0\), one obtains
\begin{equation}
L\!\left(\varphi_\varphi I\right)
=
\frac{1}{\varphi_\varphi}
\left(
p\varphi_\varphi^2 I'
\right)'.
\label{eq:reduction_order_identity}
\end{equation}
Eq.~\eqref{eq:chi0_inhomogeneous_equation} consequently reduces to
\begin{equation}
\left(
p\varphi_\varphi^2 I'
\right)'
=
\varphi_\varphi K.
\label{eq:I_equation}
\end{equation}

Let us define the Volterra primitive,
\begin{equation}
Q(r)
\equiv
\int_{r_{{\rm H},0}}^{\,r}
\varphi_\varphi(s)K(s)\,ds.
\label{eq:Q_definition}
\end{equation}
Since
\begin{equation}
Q(r_{{\rm H},0})=0,
\end{equation}
the horizon-regular first integral is
\begin{equation}
p(r)\varphi_\varphi(r)^2 I'(r)
=
Q(r).
\label{eq:I_first_integral}
\end{equation}
For the charged non-extremal GHS background,
\(\varphi_\varphi>0\) and \(K>0\), so
\begin{equation}
Q(r)>0,
\qquad
r>r_{{\rm H},0}.
\end{equation}

Direct integration gives
\begin{equation}
Q(r)
=
\frac{2r_-}{\alpha\Mpl}
\left[
\frac{1}{2r_+}
-
\frac{1}{D(r)}
+
\frac{(1+a)r_-}{2D(r)^2}
\right],
\label{eq:explicit_Q_GHS}
\end{equation}
where recall that
\begin{equation}
D(r)
=
2\rho(r)-r_-
=
\sqrt{r_-^2+4r^2},
\qquad
D_{\rm H}
\equiv
D(r_{{\rm H},0})
=
r_+(1+a),
\qquad
D_{\rm m}
\equiv
D(r_{\rm m}).
\label{eq:DH_definition}
\end{equation}

It is useful to rewrite the first integral in terms of the throat
coordinate. Using
\begin{equation}
r
=
r_+\sqrt{(1+b)(a+b)},
\qquad
D
=
r_+(1+a+2b),
\qquad
\frac{dr}{db}
=
\frac{
r_+(1+a+2b)
}{
2\sqrt{(1+b)(a+b)}
},
\label{eq:b_relations_generator}
\end{equation}
one finds
\begin{equation}
Q(b)
=
\frac{
4(1-a)b(a+b)
}{
\alpha\Mpl(1+a+2b)^2
}.
\label{eq:Q_GHS_b}
\end{equation}

In the finite-radius matching normalization,
\begin{equation}
p(r)
=
e^{\delta_0(r)}r^2\mathcal F_0(r)
=
\frac{D_{\rm m}}{4r_{\rm m}}\,
\frac{rD\,b}{1+b}.
\label{eq:p_GHS_matching}
\end{equation}
Eq.~\eqref{eq:I_first_integral} therefore becomes
\begin{equation}
\frac{dI}{db}
=
\frac{
8(1-a)r_{\rm m}
}{
\alpha\Mpl r_+D_{\rm m}
}
\frac{1}{
\left[a(1+a)+2b\right]^2
}.
\label{eq:I_b_equation}
\end{equation}
Imposing
\begin{equation}
I(0)=0,
\label{eq:I_horizon_normalization}
\end{equation}
which is equivalent to
\(\chi_0(r_{{\rm H},0})=0\) because
\(\varphi_\varphi(r_{{\rm H},0})=a>0\), gives
\begin{equation}
I(b)
=
\frac{
8(1-a)r_{\rm m}
}{
\alpha\Mpl r_+D_{\rm m}
}
\frac{
b
}{
a(1+a)\left[a(1+a)+2b\right]
}.
\label{eq:I_GHS_b}
\end{equation}

Equivalently, returning to the areal-radius variables,
Eq.~\eqref{eq:I_GHS_b} may be written as
\begin{equation}
I(r)
=
-\frac{4r_-r_{\rm m}}
{\alpha\Mpl D_{\rm m}}
\left[
\frac{1}
{r_+D(r)-r_-D_{\rm H}}
-
\frac{1}
{D_{\rm H}(r_+-r_-)}
\right].
\label{eq:explicit_I_GHS}
\end{equation}

The closed form of \(\chi_0\) follows most transparently in the throat
coordinate. The identities
\begin{equation}
r_+D(r)-r_-D_{\rm H}
=
r_+^2
\left[
a(1+a)+2b
\right],
\qquad
\varphi_\varphi(r)
=
\frac{
a(1+a)+2b
}{
1+a+2b
}
\label{eq:generator_cancellation_identities}
\end{equation}
show that the factor
\(a(1+a)+2b\) cancels between \(I\) and \(\varphi_\varphi\). Thus
\begin{equation}
\chi_0(r)
=
\frac{
8(1-a)r_{\rm m}
}{
\alpha\Mpl D_{\rm m}r_+a(1+a)
}
\frac{b}{1+a+2b}.
\label{eq:chi0_closed_form}
\end{equation}
This solution satisfies
\begin{equation}
L\chi_0
=
K,
\qquad
\chi_0(r_{{\rm H},0})=0,
\label{eq:chi0_properties}
\end{equation}
and is regular throughout the charged non-extremal static exterior.

We now construct a generator of the sliced kernel. Let us define
\begin{equation}
\mathcal N_{\rm m}
\equiv
1+
f_\varphi(r_{\rm m})\chi_0(r_{\rm m}),
\label{eq:Nm_definition}
\end{equation}
where
\begin{equation}
f_\varphi(r)
=
\frac{r^2}{2}\,
\mathcal F_0(r)\varphi_0'(r).
\label{eq:fvarphi_again_generator}
\end{equation}
A convenient globally regular representative is
\begin{equation}
\widehat\psi_\star(r)
=
\mathcal N_{\rm m}\varphi_\varphi(r)
-
f_\varphi(r_{\rm m})
\varphi_\varphi(r_{\rm m})
\chi_0(r).
\label{eq:psihat_definition}
\end{equation}
Using the definition of \(\mathcal N_{\rm m}\), one immediately finds
\begin{equation}
\widehat\psi_\star(r_{\rm m})
=
\varphi_\varphi(r_{\rm m})
>0.
\label{eq:psihat_matching_value}
\end{equation}
Thus \(\widehat\psi_\star\) is nonzero independently of the value, or sign,
of \(\mathcal N_{\rm m}\).

Since
\begin{equation}
L\varphi_\varphi=0,
\qquad
L\chi_0=K,
\end{equation}
the unsliced operator acts on \(\widehat\psi_\star\) as
\begin{equation}
L\widehat\psi_\star
=
-
f_\varphi(r_{\rm m})
\varphi_\varphi(r_{\rm m})K.
\label{eq:L_psihat}
\end{equation}
The sliced operator therefore gives
\begin{align}
L_{\rm sl}\widehat\psi_\star
&=
L\widehat\psi_\star
+
f_\varphi(r_{\rm m})
K(r)\widehat\psi_\star(r_{\rm m})
\nonumber\\
&=
0.
\label{eq:Lsl_psihat_zero}
\end{align}
Hence
\begin{equation}
\widehat\psi_\star
\in
\ker L_{\rm sl}.
\end{equation}
Since the sliced kernel is one-dimensional,
\(\widehat\psi_\star\) generates it:
\begin{equation}
\ker L_{\rm sl}
=
\operatorname{span}\{\widehat\psi_\star\}.
\label{eq:kernel_psihat}
\end{equation}

We express the throat-coordinate location of the matching surface using
\begin{equation}
b_{\rm m}
\equiv
\frac{\rho(r_{\rm m})-r_+}{r_+}.
\label{eq:bm_definition_generator}
\end{equation}
A nontrivial finite radial interval satisfies
\begin{equation}
r_{\rm m}>r_{{\rm H},0}
\qquad\Longleftrightarrow\qquad
b_{\rm m}>0.
\end{equation}
For every such interval, the explicit GHS profiles satisfy
\begin{equation}
f_\varphi(r_{\rm m})<0,
\qquad
\varphi_\varphi(r_{\rm m})>0,
\end{equation}
and therefore
\begin{equation}
m_\varphi(r_{\rm m})
=
f_\varphi(r_{\rm m})\varphi_\varphi(r_{\rm m})
<0.
\label{eq:explicit_slice_transversality_generator}
\end{equation}
The local mass slice is therefore transverse on every nontrivial finite
interval of the charged non-extremal exterior. The formal endpoint
\(b_{\rm m}=0\), at which the matching surface coincides with the static
horizon, does not define a nontrivial near-zone boundary-value problem and
is excluded.

Whenever
\begin{equation}
\mathcal N_{\rm m}\neq0,
\label{eq:Nm_nonzero}
\end{equation}
one may equivalently introduce the normalized representative
\begin{equation}
\psi_\star(r)
=
\frac{
\widehat\psi_\star(r)
}{
\mathcal N_{\rm m}
},
\label{eq:normalized_psi_star}
\end{equation}
whose coefficient along the canonical tangent is unity. The possible
vanishing of \(\mathcal N_{\rm m}\), and the regularity of
\(\widehat\psi_\star\) across that locus, are analyzed in the following
subsection.

All constructions in the present subsection are exact on the finite radial
interval and depend only on the instantaneous GHS background and the local
mass slice. They do not use the large-radius approximation introduced later
to derive the particular exterior matching functional
\(\mathcal B_{\rm m}=r_{\rm m}\partial_r+1\).

\subsection{Regularity and normalization of the displaced generator}
\label{subsec:regularity_normalization}

The previous subsection provides closed-form expressions for the particular
solution \(\chi_0\) and for the kernel representative
\(\widehat\psi_\star\). We now establish their regularity throughout the
charged non-extremal static exterior and analyze the normalization factor
\(\mathcal N_{\rm m}\).

From Eq.~\eqref{eq:chi0_closed_form},
\begin{equation}
\chi_0(r)
=
\frac{
8(1-a)r_{\rm m}
}{
\alpha\Mpl D_{\rm m}r_+a(1+a)
}
\frac{b}{1+a+2b},
\qquad
b
=
\frac{\rho-r_+}{r_+}
\geq0.
\label{eq:chi0_regularity_form}
\end{equation}
For
\begin{equation}
0<a<1,
\end{equation}
one has
\begin{equation}
\frac{d\chi_0}{db}
=
\frac{
8(1-a)r_{\rm m}
}{
\alpha\Mpl D_{\rm m}r_+a
}
\frac{1}{(1+a+2b)^2}
>0.
\label{eq:chi0_monotonicity}
\end{equation}
Thus \(\chi_0\) is strictly increasing throughout the static exterior. Its
endpoint values are
\begin{equation}
\chi_0(r_{{\rm H},0})=0,
\label{eq:chi0_horizon_value}
\end{equation}
and
\begin{align}
\chi_0(\infty)
&=
\frac{
4(1-a)r_{\rm m}
}{
\alpha\Mpl D_{\rm m}r_+a(1+a)
}
\nonumber\\
&=
\frac{
4r_-r_{\rm m}
}{
\alpha\Mpl D_{\rm m}D_{\rm H}(r_+-r_-)
},
\label{eq:chi0_infinity_value}
\end{align}
where \(D_{\rm H}=r_+(1+a)\). Hence \(\chi_0\) is smooth,
non-negative, monotonically increasing, and bounded on the complete static
exterior,
\begin{equation}
r_{{\rm H},0}\leq r<\infty.
\end{equation}
The dynamical near-zone problem uses only its restriction to the finite,
nontrivial interval
\([r_{\rm H}^{\rm inst}(v),r_{\rm m}]\), with
\(r_{\rm m}>r_{\rm H}^{\rm inst}(v)\).

The displaced homogeneous generator constructed in
Eq.~\eqref{eq:psihat_definition} is
\begin{equation}
\widehat\psi_\star(r)
=
\mathcal N_{\rm m}\varphi_\varphi(r)
-
f_\varphi(r_{\rm m})
\varphi_\varphi(r_{\rm m})
\chi_0(r).
\label{eq:psihat_again}
\end{equation}
Both \(\varphi_\varphi\) and \(\chi_0\) are regular throughout the charged
non-extremal static exterior, and therefore so is
\(\widehat\psi_\star\). Moreover,
\begin{equation}
\widehat\psi_\star(r_{\rm m})
=
\varphi_\varphi(r_{\rm m}),
\label{eq:psihat_matching_normalization}
\end{equation}
independently of the value of \(\mathcal N_{\rm m}\). Since
\(\varphi_\varphi(r)>0\) for \(0<a<1\) and \(b\geq0\), this matching value is
strictly positive:
\begin{equation}
\widehat\psi_\star(r_{\rm m})
=
\varphi_\varphi(r_{\rm m})
>0.
\label{eq:psihat_nonzero_matching}
\end{equation}
Thus \(\widehat\psi_\star\) is a globally regular, nonzero generator of
\(\ker L_{\rm sl}\) for every nontrivial finite matching interval on the
charged non-extremal branch.

The normalization factor is
\begin{equation}
\mathcal N_{\rm m}
=
1+
f_\varphi(r_{\rm m})\chi_0(r_{\rm m}).
\label{eq:Nm_again}
\end{equation}
The explicit GHS expressions give
\begin{equation}
f_\varphi(r_{\rm m})\chi_0(r_{\rm m})
=
-
\frac{
(1-a)^2b_{\rm m}^2
}{
\alpha^2a(1+a)(1+b_{\rm m})
(1+a+2b_{\rm m})
},
\label{eq:fvarphi_chi0_product}
\end{equation}
where \(b_{\rm m} \) is given in Eq.~\eqref{eq:bm_definition_generator}, and hence
\begin{equation}
\mathcal N_{\rm m}
=
1-
\frac{
(1-a)^2b_{\rm m}^2
}{
\alpha^2a(1+a)(1+b_{\rm m})
(1+a+2b_{\rm m})
}.
\label{eq:explicit_Nm}
\end{equation}

As simple consistency checks, the exact expression satisfies
\begin{equation}
\mathcal N_{\rm m}
\longrightarrow
1,
\qquad
b_{\rm m}\longrightarrow0,
\label{eq:Nm_horizon_limit}
\end{equation}
corresponding formally to a matching surface approaching the static horizon.
Likewise,
\begin{equation}
\mathcal N_{\rm m}
\longrightarrow
1,
\qquad
a\longrightarrow1,
\label{eq:Nm_Schwarzschild_limit}
\end{equation}
reflecting the disappearance of the finite-rank correction in the
Schwarzschild limit.

For fixed \(b_{\rm m}\), the locus
\(\mathcal N_{\rm m}=0\), when present, is determined by
Eq.~\eqref{eq:explicit_Nm}. In the formal large-matching-radius limit
\begin{equation}
b_{\rm m}\longrightarrow\infty,
\end{equation}
one obtains
\begin{equation}
\mathcal N_{\rm m}
\longrightarrow
1-
\frac{(1-a)^2}
{2\alpha^2a(1+a)}.
\label{eq:Nm_asymptotic_limit}
\end{equation}
For the canonical GHS coupling
\begin{equation}
\alpha=\frac1{\sqrt2},
\end{equation}
this becomes
\begin{equation}
\mathcal N_{\rm m}
\longrightarrow
\frac{3a-1}{a(1+a)}.
\label{eq:Nm_canonical_limit}
\end{equation}
The limiting normalization therefore vanishes at
\begin{equation}
a_{\rm crit}^{(\infty)}
=
\frac13.
\label{eq:acrit}
\end{equation}
The superscript emphasizes that
\(a_{\rm crit}^{(\infty)}\) refers to the formal limit
\(b_{\rm m}\to\infty\). At finite matching radius, the exact zero of
\(\mathcal N_{\rm m}\), if it exists, is instead determined by
Eq.~\eqref{eq:explicit_Nm}. Equivalently, for fixed \(a\), its location in
\(b_{\rm m}\) and its position relative to the controlled large-radius
overlap regime are analyzed in
Sec.~\ref{subsec:explicit_solvability}.

The locus \(\mathcal N_{\rm m}=0\) is not a singularity of the sliced
kernel. It invalidates only the conventionally normalized representative
\begin{equation}
\psi_\star
=
\frac{\widehat\psi_\star}{\mathcal N_{\rm m}},
\label{eq:normalized_generator_again}
\end{equation}
whose coefficient along the canonical tangent is fixed to unity. The
globally regular representative \(\widehat\psi_\star\) remains finite and
satisfies
\begin{equation}
L_{\rm sl}\widehat\psi_\star=0
\label{eq:psihat_kernel_again}
\end{equation}
throughout the charged non-extremal branch, including on
\(\mathcal N_{\rm m}=0\).

Indeed, on this locus,
\begin{equation}
\widehat\psi_\star(r)
=
-
f_\varphi(r_{\rm m})
\varphi_\varphi(r_{\rm m})
\chi_0(r)
=
-
m_\varphi(r_{\rm m})\chi_0(r).
\label{eq:psihat_on_Nm_zero}
\end{equation}
For every nontrivial finite matching interval, one has
\(r_{\rm m}>r_{{\rm H},0}\), or equivalently \(b_{\rm m}>0\). The explicit
GHS profiles then give
\begin{equation}
f_\varphi(r_{\rm m})<0,
\qquad
\varphi_\varphi(r_{\rm m})>0,
\end{equation}
and therefore
\begin{equation}
m_\varphi(r_{\rm m})
=
f_\varphi(r_{\rm m})\varphi_\varphi(r_{\rm m})
<0.
\label{eq:transversality_again_regular}
\end{equation}
Since \(\chi_0(r)\geq0\), with strict inequality for
\(r>r_{{\rm H},0}\), Eq.~\eqref{eq:psihat_on_Nm_zero} is regular and
non-negative throughout the exterior and strictly positive away from the
horizon. It is therefore a nontrivial kernel generator. Its non-vanishing
also follows directly from
\(\widehat\psi_\star(r_{\rm m})=\varphi_\varphi(r_{\rm m})>0\).

The vanishing of \(\mathcal N_{\rm m}\) is therefore a normalization
degeneracy only. The relevant geometric object is the one-dimensional
subspace
\begin{equation}
\ker L_{\rm sl},
\end{equation}
rather than any particular normalized generator or decomposition of that
generator into \(\varphi_\varphi\) and \(\chi_0\). In particular, the value and
sign of \(\mathcal N_{\rm m}\) refer to the chosen convention in which the
canonical-tangent coefficient of \(\psi_\star\) is fixed to unity; they have
no intrinsic geometric or physical meaning.

As \(\mathcal N_{\rm m}\) crosses zero, the normalized representative
\(\psi_\star=\widehat\psi_\star/\mathcal N_{\rm m}\) ceases to be defined at
the crossing and changes sign relative to the globally regular
representative on its two sides. The kernel direction itself remains
continuous and is represented globally by \(\widehat\psi_\star\), which
stays finite and nonzero across \(\mathcal N_{\rm m}=0\). Thus the locus
produces no discontinuity or bifurcation of the sliced kernel and does not
generate a new physical branch of solutions.

The regularity statements in this subsection are intrinsic to the exact
finite-interval GHS radial problem. They do not rely on the large-radius
overlap approximation used later to derive the particular matching
functional
\(\mathcal B_{\rm m}=r_{\rm m}\partial_r+1\).


\subsection{Leading overlap matching}
\label{subsec:far_field_matching}

Having constructed a globally regular generator of the one-dimensional
kernel of the sliced operator, we now specify the outer matching functional
used in the remainder of this paper. The finite near-zone problem does not
determine this functional by itself: as emphasized in Paper~I, both the
matching operator and its inhomogeneous datum must be supplied by the
exterior problem. Here we adopt the leading matching law associated with a
massless, slowly varying exterior in a controlled large-radius overlap
regime.

The near-zone and exterior expansions must overlap on a radial interval
satisfying
\begin{equation}
r_{\rm H}^{\rm inst}
\ll
r
\lesssim
r_{\rm m}
\ll
L_{\rm cos},
\label{eq:matching_overlap_hierarchy}
\end{equation}
where recall that \(L_{\rm cos}\) denotes the shortest characteristic length scale of the
exterior solution. For the explicit GHS large-radius expansion used below,
this scale hierarchy must be supplemented by the requirement that the
matching surface lie in the asymptotic end of the static GHS exterior. In
terms of the conventional GHS coordinate, we reiterate that
\begin{equation}
\frac{r_+}{\rho_{\rm m}}
=
\frac{1}{1+b_{\rm m}}
\ll1,
\qquad
b_{\rm m}
\equiv
\frac{\rho_{\rm m}-r_+}{r_+},
\label{eq:GHS_large_radius_overlap_condition}
\end{equation}
or equivalently
\begin{equation}
b_{\rm m}\gg1.
\label{eq:bm_large_overlap}
\end{equation}
This condition is stronger than merely requiring the matching surface to be
many areal-horizon radii away. Indeed,
\begin{equation}
\frac{r_{\rm m}}{r_{{\rm H},0}}
=
\sqrt{
\frac{(1+b_{\rm m})(a+b_{\rm m})}{a}
},
\label{eq:rm_over_rH_bm}
\end{equation}
so that \(r_{\rm m}/r_{{\rm H},0}\) may become large near extremality even
when \(b_{\rm m}={\mathcal O}(1)\). The condition
\(b_{\rm m}\gg1\), rather than the ratio
\(r_{\rm m}/r_{{\rm H},0}\) alone, controls the asymptotic expansion about the
flat-space end of the GHS exterior.

In this controlled overlap region, the coefficients of the instantaneous
static GHS background approach their flat-space values. In the matching
normalization,
\begin{equation}
\mathcal F_0(r)
=
1+{\mathcal O}(r^{-1}),
\qquad
e^{\delta_0(r)}
=
e^{\delta_\infty}
\left[
1+{\mathcal O}(r^{-2})
\right],
\label{eq:GHS_far_coefficients}
\end{equation}
where
\begin{equation}
e^{\delta_\infty}
=
\frac{D_{\rm m}}{2r_{\rm m}}
\neq0.
\label{eq:delta_infinity_matching}
\end{equation}
Consequently,
\begin{equation}
p(r)
=
e^{\delta_0(r)}r^2\mathcal F_0(r)
=
e^{\delta_\infty}r^2
\left[
1+{\mathcal O}(r^{-1})
\right].
\label{eq:p_far_asymptotic}
\end{equation}

For the massless GHS background, the effective radial potential and the
finite-rank source profile satisfy
\begin{equation}
V_{\rm eff}(r)
=
{\mathcal O}(r^{-2}),
\qquad
K(r)
=
{\mathcal O}(r^{-2}).
\label{eq:far_potential_decay}
\end{equation}
The homogeneous sliced equation is
\begin{equation}
\left(
p\,\eta_{\rm hom}'
\right)'
-
V_{\rm eff}\eta_{\rm hom}
+
f_\varphi(r_{\rm m})K(r)\eta_{\rm hom}(r_{\rm m})
=
0.
\label{eq:homogeneous_sliced_overlap}
\end{equation}
Using Eqs.~\eqref{eq:p_far_asymptotic} and
\eqref{eq:far_potential_decay}, its two independent large-radius behaviors
are
\begin{equation}
\eta_{\rm hom}(r)
=
A
+
\frac{B}{r}
+
{\mathcal O}(r^{-2}).
\label{eq:far_homogeneous_expansion}
\end{equation}

The coefficient \(A\) represents a constant homogeneous displacement of the
scalar field. The near-zone equations alone do not determine whether this
mode belongs to the lag field or is absorbed into the collective modulus
coordinate. We adopt the matching convention in which the prescribed
exterior modulus fixes the constant scalar mode of the total configuration.
A homogeneous lag variation at fixed exterior modulus therefore has
\begin{equation}
A=0.
\label{eq:constant_mode_removed}
\end{equation}
The remaining homogeneous behavior is
\begin{equation}
\eta_{\rm hom}(r)
=
\frac{B}{r}
+
{\mathcal O}(r^{-2}),
\label{eq:decaying_homogeneous_overlap}
\end{equation}
and hence satisfies, at leading order in the overlap expansion,
\begin{equation}
r\,\eta_{\rm hom}'(r)
+
\eta_{\rm hom}(r)
=
{\mathcal O}(r^{-2}).
\label{eq:Robin_homogeneous}
\end{equation}

We therefore adopt the leading overlap matching functional
\begin{equation}
\mathcal B_{\rm m}\eta
\equiv
r_{\rm m}\eta'(r_{\rm m})
+
\eta(r_{\rm m})
\label{eq:Bm_GHS}
\end{equation}
for the remainder of the analysis. The completed outer condition is
\begin{equation}
\mathcal B_{\rm m}\eta
=
J_{\rm ext}(v),
\label{eq:complete_GHS_matching_condition}
\end{equation}
where \(J_{\rm ext}(v)\) is determined by the particular exterior solution.

Eq.~\eqref{eq:Bm_GHS} constrains the homogeneous radial freedom. The
non-decaying behavior generated by the forced response need not satisfy the
homogeneous Robin condition; its contribution at the matching surface is
included in the inhomogeneous datum \(J_{\rm ext}\).

Subleading corrections in the large-radius overlap expansion would modify
both the functional \(\mathcal B_{\rm m}\) and the datum \(J_{\rm ext}\).
The same is true for a different exterior background or a different
convention relating the collective modulus to the exterior scalar mode.
Thus Eq.~\eqref{eq:Bm_GHS} is not a universal boundary condition implied by
the GHS interior alone. It is the leading Robin functional associated with
the specified massless overlap model, with
\begin{equation}
b_{\rm m}\gg1,
\qquad
r_{\rm m}\ll L_{\rm cos},
\label{eq:controlled_massless_overlap_domain}
\end{equation}
and with the convention that the prescribed exterior modulus fixes the
constant scalar mode.

Once the model functional
\(\mathcal B_{\rm m}=r_{\rm m}\partial_r+1\) has been fixed, its action on the
exact GHS radial profiles can be evaluated algebraically for arbitrary
\(b_{\rm m}\geq0\). Such an algebraic continuation outside
Eq.~\eqref{eq:controlled_massless_overlap_domain} does not, however, imply
that \(\mathcal B_{\rm m}\) represents the physical matching operator of a
massless exterior there. Its interpretation as an exterior matching
condition is restricted to the controlled overlap regime in which it was
derived.

Within this matching model, the Fredholm criterion is controlled by the
action of \(\mathcal B_{\rm m}\) on the globally regular sliced-kernel
generator \(\widehat\psi_\star\). This action is evaluated explicitly in the
following subsection.

 
\subsection{Explicit test of the solvability condition}
\label{subsec:explicit_solvability}

For the leading massless overlap matching functional adopted in
Sec.~\ref{subsec:far_field_matching}, the Slice and Solvability Theorem
reviewed in Sec.~\ref{subsec:fredholm} reduces existence and uniqueness of
the completed radial problem to the non-vanishing of
\begin{equation}
\mathcal B_{\rm m}\widehat\psi_\star.
\label{eq:Fredholm_denominator_GHS}
\end{equation}
We now evaluate this quantity explicitly for the globally regular generator
constructed in Sec.~\ref{subsec:displaced_generator}. Once the model
functional
\(\mathcal B_{\rm m}=r_{\rm m}\partial_r+1\)
has been fixed, the calculation below is algebraically exact for arbitrary
\(b_{\rm m}\geq0\). Its interpretation as a physical exterior matching
criterion is restricted, however, to the controlled large-radius overlap
regime specified in Sec.~\ref{subsec:far_field_matching}.

The conversion between areal-radius and throat-coordinate derivatives is
\begin{equation}
r\frac{d}{dr}
=
\frac{
2(1+b)(a+b)
}{
1+a+2b
}
\frac{d}{db}.
\label{eq:r_derivative_in_b}
\end{equation}
By linearity of the matching functional,
\begin{equation}
\mathcal B_{\rm m}\widehat\psi_\star
=
\mathcal N_{\rm m}
\mathcal B_{\rm m}\varphi_\varphi
-
f_\varphi(r_{\rm m})
\varphi_\varphi(r_{\rm m})
\mathcal B_{\rm m}\chi_0.
\label{eq:Bm_psihat_linear}
\end{equation}

For \(b_{\rm m} \) given in Eq.~\eqref{eq:bm_definition_generator}, we introduce
\begin{equation}
s_{\rm m}
\equiv
1+a+2b_{\rm m}.
\label{eq:bm_sm_definitions}
\end{equation}
Using the explicit canonical tangent given in Eq.~\eqref{eq:explicit_varphi_Phi} and
\begin{equation}
\mathcal B_{\rm m}
=
r_{\rm m}\partial_r+1,
\end{equation}
one obtains
\begin{equation}
\mathcal B_{\rm m}\varphi_\varphi
=
\frac{
a(1+a)+2b_{\rm m}
}{
s_{\rm m}
}
+
\frac{
4(1-a^2)(1+b_{\rm m})(a+b_{\rm m})
}{
s_{\rm m}^3
}.
\label{eq:Bm_varphiPhi}
\end{equation}

Similarly, writing
\begin{equation}
\chi_0(r)
=
C_{\rm m}
\frac{b}{1+a+2b},
\end{equation}
with
\begin{equation}
C_{\rm m}
\equiv
\frac{
8(1-a)r_{\rm m}
}{
\alpha\Mpl D_{\rm m}r_+a(1+a)
},
\label{eq:Cm_definition}
\end{equation}
gives
\begin{equation}
\mathcal B_{\rm m}\chi_0
=
C_{\rm m}
\left[
\frac{b_{\rm m}}{s_{\rm m}}
+
\frac{
2(1+a)(1+b_{\rm m})(a+b_{\rm m})
}{
s_{\rm m}^3
}
\right].
\label{eq:Bm_chi0}
\end{equation}

The cancellation of the normalization-dependent quantities can be exhibited
without expanding them separately. Using
\begin{equation}
\mathcal N_{\rm m}
=
1+
f_\varphi(r_{\rm m})\chi_0(r_{\rm m}),
\end{equation}
Eq.~\eqref{eq:Bm_psihat_linear} becomes
\begin{align}
\mathcal B_{\rm m}\widehat\psi_\star
&=
\mathcal B_{\rm m}\varphi_\varphi
+
f_\varphi(r_{\rm m})r_{\rm m}
\left[
\chi_0\varphi_\varphi'
-
\varphi_\varphi\chi_0'
\right]_{r_{\rm m}}
\nonumber\\
&=
\mathcal B_{\rm m}\varphi_\varphi
-
f_\varphi(r_{\rm m})r_{\rm m}
\varphi_\varphi(r_{\rm m})^2
I'(r_{\rm m}),
\label{eq:Bm_psihat_Wronskian}
\end{align}
where \(\chi_0=\varphi_\varphi I\) has been used in the second line.

Since
\begin{equation}
p\varphi_\varphi^2I'=Q,
\end{equation}
and, in the matching normalization,
\begin{equation}
p(r_{\rm m})
=
r_{\rm m}^2\mathcal F_0(r_{\rm m}),
\end{equation}
while
\begin{equation}
f_\varphi(r_{\rm m})
=
\frac{r_{\rm m}^2}{2}
\mathcal F_0(r_{\rm m})
\varphi_0'(r_{\rm m}),
\end{equation}
one obtains the reduced expression
\begin{equation}
\mathcal B_{\rm m}\widehat\psi_\star
=
\mathcal B_{\rm m}\varphi_\varphi
-
\frac{
r_{\rm m}\varphi_0'(r_{\rm m})
}{2}
Q(r_{\rm m}).
\label{eq:Bm_psihat_reduced}
\end{equation}

The explicit GHS profiles give
\begin{equation}
r_{\rm m}\varphi_0'(r_{\rm m})
=
-\frac{
\Mpl(1-a)
}{
\alpha s_{\rm m}
},
\qquad
Q(r_{\rm m})
=
\frac{
4(1-a)b_{\rm m}(a+b_{\rm m})
}{
\alpha\Mpl s_{\rm m}^2
}.
\label{eq:matching_varphi0_Q}
\end{equation}
Substituting these expressions together with
Eq.~\eqref{eq:Bm_varphiPhi} into
Eq.~\eqref{eq:Bm_psihat_reduced} yields
\begin{equation}
\mathcal B_{\rm m}\widehat\psi_\star
=
\frac{
a(1+a)+2b_{\rm m}
}{
s_{\rm m}
}
+
\frac{
4(1-a^2)(1+b_{\rm m})(a+b_{\rm m})
}{
s_{\rm m}^3
}
+
\frac{
2b_{\rm m}(1-a)^2(a+b_{\rm m})
}{
\alpha^2s_{\rm m}^3
}.
\label{eq:Bm_psihat_final}
\end{equation}

For the charged non-extremal GHS family,
\begin{equation}
0<a<1,
\qquad
b_{\rm m}\geq0,
\end{equation}
all three terms on the right-hand side of
Eq.~\eqref{eq:Bm_psihat_final} are non-negative. The first term is strictly
positive, since
\begin{equation}
a(1+a)+2b_{\rm m}>0,
\qquad
s_{\rm m}>0.
\end{equation}
Consequently,
\begin{equation}
\mathcal B_{\rm m}\widehat\psi_\star>0 \;.
\label{eq:Bm_positive}
\end{equation}

The large-radius limit provides an additional consistency check. As
\begin{equation}
b_{\rm m}\longrightarrow\infty,
\end{equation}
Eq.~\eqref{eq:Bm_psihat_final} gives
\begin{equation}
\mathfrak D
=
\mathcal B_{\rm m}\widehat\psi_\star
=
1
+
\mathcal O(b_{\rm m}^{-1}).
\label{eq:D_large_radius_limit}
\end{equation}
This agrees with the asymptotic behavior
\(\varphi_\varphi\rightarrow1\) and
\(r\varphi_\varphi'\rightarrow0\), for which the leading Robin functional
acts on the canonical tangent as
\(\mathcal B_{\rm m}\varphi_\varphi\rightarrow1\).

Eq.~\eqref{eq:Bm_positive} is an exact algebraic statement about the
model functional
\(\mathcal B_{\rm m}=r_{\rm m}\partial_r+1\)
acting on the exact GHS sliced-kernel generator. Eq.
\eqref{eq:Bm_psihat_final} is independent of all dimensionful scales,
including \(r_+\), \(r_{\rm m}\), and \(D_{\rm m}\), and depends only on the
dimensionless parameters \((a,b_{\rm m},\alpha)\).

In particular, it remains finite and strictly positive on the locus
\begin{equation}
\mathcal N_{\rm m}=0,
\end{equation}
where the conventionally normalized representative
\begin{equation}
\psi_\star
=
\frac{
\widehat\psi_\star
}{
\mathcal N_{\rm m}
}
\end{equation}
ceases to be defined. Formulating the non-degeneracy condition in terms of
the globally regular generator \(\widehat\psi_\star\) therefore removes the
spurious singularity associated with this normalization convention.

For the canonical GHS coupling
\(\alpha=1/\sqrt{2}\), the finite-radius normalization-degeneracy locus is
present only for
\begin{equation}
0<a<\frac13,
\end{equation}
and is located at
\begin{equation}
b_{\rm crit}(a)
=
\frac{
(1+a)
\left[
a(a+3)
+
\sqrt{a}\,(1-a)\sqrt{a+8}
\right]
}{
4(1-3a)
}.
\label{eq:bcrit_finite_matching}
\end{equation}
Its position relative to the controlled overlap regime is not uniform along
the charged branch. Near extremality,
\begin{equation}
b_{\rm crit}(a)
=
\sqrt{\frac{a}{2}}
+
\frac{3a}{4}
+
\mathcal O(a^{3/2}),
\qquad
a\longrightarrow0,
\label{eq:bcrit_extremal_expansion}
\end{equation}
so the degeneracy lies well outside the large-radius overlap domain. By
contrast,
\begin{equation}
b_{\rm crit}(a)
\longrightarrow\infty,
\qquad
a\longrightarrow\frac13^-,
\label{eq:bcrit_one_third_limit}
\end{equation}
and the locus then enters the controlled overlap region. Thus the positivity
of
\(\mathcal B_{\rm m}\widehat\psi_\star\)
on \(\mathcal N_{\rm m}=0\) is an exact algebraic property of the model
functional at every point where that locus exists. Its interpretation as a
physical massless-overlap solvability statement applies only when
\begin{equation}
b_{\rm crit}(a)\gg1,
\qquad
r_{\rm m}\ll L_{\rm cos}.
\end{equation}

The formal Schwarzschild limit provides a consistency check. As
\begin{equation}
a\longrightarrow1,
\end{equation}
one has
\begin{equation}
\chi_0\longrightarrow0,
\qquad
\mathcal N_{\rm m}\longrightarrow1,
\qquad
\widehat\psi_\star\longrightarrow\varphi_\varphi,
\end{equation}
and Eq.~\eqref{eq:Bm_psihat_final} reduces to
\begin{equation}
\mathcal B_{\rm m}\widehat\psi_\star
\longrightarrow1.
\label{eq:Bm_Schwarzschild_limit}
\end{equation}
As discussed in Sec.~\ref{subsec:explicit_GHS_kernel}, the strict
Schwarzschild endpoint is a degenerate limit of the mass slice rather than a
member of the transverse charged branch.

We conclude that, for the model functional
\begin{equation}
\mathcal B_{\rm m}
=
r_{\rm m}\partial_r+1,
\end{equation}
the Fredholm denominator is algebraically strictly positive for all
\(0<a<1\) and \(b_{\rm m}\geq0\), including every point of the locus
\(\mathcal N_{\rm m}=0\) at which that locus exists. Its interpretation as
the non-degeneracy condition of a massless exterior matching problem is
restricted to the controlled overlap regime
\begin{equation}
b_{\rm m}\gg1,
\qquad
r_{\rm m}\ll L_{\rm cos}.
\end{equation}
Within that regime, the completed leading-order near-zone problem admits a
unique solution for every admissible inhomogeneous matching datum
\(J_{\rm ext}(v)\).



\section{Explicit leading-order near-zone response}
\label{sec:physical_consequences}

The previous section completed the Fredholm construction for the magnetic GHS
family and established the non-degeneracy of the leading massless overlap
problem. We now assemble these results into the explicit leading-order
adiabatic response.

The scalar lag is determined analytically throughout the finite near zone.
Its forced radial profile is obtained in closed form, while the amplitude of
the remaining sliced homogeneous mode is fixed uniquely by the exterior
matching datum. The metric lag fields then follow from the radial constraints:
the mass lag is algebraic in the scalar lag, and the lapse lag is obtained by
one explicit radial quadrature.

The resulting solution is exact in its radial dependence at first adiabatic
order. It provides the complete leading-order near-zone response for every
massless exterior compatible with the overlap assumptions adopted in
Sec.~\ref{subsec:far_field_matching}. The dependence on the far zone enters
through the prescribed rolling rate \(\dot\varphi_{\rm ext}(v)\) and the
inhomogeneous matching datum \(J_{\rm ext}(v)\).

\subsection{Fredholm reduction of the leading-order problem}
\label{subsec:Fredholm_reduction}

The leading-order radial equation derived in
Sec.~\ref{subsec:fredholm} is
\begin{equation}
L_{\rm sl}\eta
=
S_{\rm ext},
\label{eq:leading_eta_again}
\end{equation}
where
\begin{equation}
S_{\rm ext}(v,r)
=
-2\dot\varphi_{\rm ext}(v)
\left[
r^2\varphi_\varphi'(r)
+
r\varphi_\varphi(r)
\right].
\label{eq:Sext_again}
\end{equation}
At each fixed advanced time, this equation is posed on
\begin{equation}
r_{\rm H}^{\rm inst}(v)
\leq
r
\leq
r_{\rm m},
\label{eq:leading_problem_domain}
\end{equation}
with regularity at the future horizon of the instantaneous static
representative and with the overlap matching condition
\begin{equation}
\mathcal B_{\rm m}\eta
=
J_{\rm ext}(v).
\label{eq:matching_again_physical}
\end{equation}
For the leading massless overlap model adopted in
Sec.~\ref{subsec:far_field_matching},
\begin{equation}
\mathcal B_{\rm m}
=
r_{\rm m}\partial_r+1.
\label{eq:Bm_again}
\end{equation}
Its interpretation as the physical matching functional of a massless
exterior is restricted to the controlled overlap regime given in Eq.~\eqref{eq:controlled_massless_overlap_domain}.

The explicit GHS construction determines the one-dimensional kernel of the
sliced operator,
\begin{equation}
\ker L_{\rm sl}
=
\operatorname{span}\{\widehat\psi_\star\},
\label{eq:kernel_again}
\end{equation}
together with the corresponding Fredholm denominator
\begin{equation}
\mathfrak D
\equiv
\mathcal B_{\rm m}\widehat\psi_\star.
\label{eq:Fredholm_denominator_definition}
\end{equation}
Section~\ref{subsec:explicit_solvability} established the exact algebraic
inequality
\begin{equation}
\mathfrak D>0,
\qquad
0<a<1,
\qquad
b_{\rm m}\geq0,
\label{eq:Fredholm_denominator_positive}
\end{equation}
including every point of the locus
\(\mathcal N_{\rm m}=0\) at which that locus exists. Outside the controlled
large-radius overlap regime, Eq.~\eqref{eq:Fredholm_denominator_positive}
remains a valid property of the model functional
\(\mathcal B_{\rm m}=r_{\rm m}\partial_r+1\), but it should not be interpreted
as a solvability statement for a physical massless exterior matching
problem.

Let \(\eta_0\) be any horizon-regular particular solution of
Eq.~\eqref{eq:leading_eta_again}. The general horizon-regular solution is
then
\begin{equation}
\eta(v,r)
=
\eta_0(v,r)
+
t(v)\widehat\psi_\star(r),
\label{eq:eta_general_physical}
\end{equation}
where the remaining homogeneous amplitude is fixed by the outer matching
condition:
\begin{equation}
t(v)\mathfrak D
=
J_{\rm ext}(v)
-
\mathcal B_{\rm m}\eta_0.
\label{eq:fredholm_matching_final}
\end{equation}
Since \(\mathfrak D\) is strictly positive, \(t(v)\) is uniquely determined
for every admissible matching datum \(J_{\rm ext}(v)\). Consequently, within
the controlled massless overlap regime
\eqref{eq:controlled_massless_overlap_domain}, the completed leading-order
near-zone boundary-value problem has a unique solution.

For the magnetic GHS family, the particular solution \(\eta_0\) can itself
be constructed in closed analytical form. We now determine its radial
profile explicitly.


 
\subsection{Explicit forced scalar response}
\label{subsec:explicit_forced_response}

At fixed advanced time, the source
\(S_{\rm ext}\) is proportional to the prescribed exterior rolling rate.
We therefore write
\begin{equation}
\eta_0(v,r)
=
\dot\varphi_{\rm ext}(v)\,q_0(r),
\label{eq:eta0_q0_definition}
\end{equation}
where the radial response function \(q_0\) satisfies
\begin{equation}
L_{\rm sl}q_0
=
-2
\left(
r^2\varphi_\varphi'
+
r\varphi_\varphi
\right).
\label{eq:q0_sliced_equation}
\end{equation}

We first construct a particular solution of the corresponding unsliced
equation. We first set
\begin{equation}
q_{\rm H}(r)
=
\varphi_\varphi(r)J(r).
\label{eq:qH_reduction_ansatz}
\end{equation}
Since
\begin{equation}
L\varphi_\varphi=0,
\end{equation}
the reduction-of-order identity gives
\begin{equation}
L\!\left(\varphi_\varphi J\right)
=
\frac{1}{\varphi_\varphi}
\left(
p\varphi_\varphi^2J'
\right)',
\qquad
p(r)
=
e^{\delta_0(r)}r^2\mathcal F_0(r).
\label{eq:forced_reduction_identity}
\end{equation}
Moreover,
\begin{equation}
2\varphi_\varphi
\left(
r^2\varphi_\varphi'
+
r\varphi_\varphi
\right)
=
\left(
r^2\varphi_\varphi^2
\right)'.
\label{eq:forced_source_total_derivative}
\end{equation}
The unsliced forced equation therefore reduces to
\begin{equation}
\left(
p\varphi_\varphi^2J'
\right)'
=
-
\left(
r^2\varphi_\varphi^2
\right)'.
\label{eq:J_first_derivative_equation}
\end{equation}

The horizon-regular first integral is
\begin{equation}
p(r)\varphi_\varphi(r)^2J'(r)
=
r_{{\rm H},0}^{\,2}
\varphi_\varphi(r_{{\rm H},0})^2
-
r^2\varphi_\varphi(r)^2.
\label{eq:J_first_integral}
\end{equation}
The integration constant has been chosen so that the right-hand side
vanishes at the static horizon, consistently with
\(p(r_{{\rm H},0})=0\). This selects the horizon-regular branch of the
first-order equation.

Let us introduce
\begin{equation}
s(b)
\equiv
1+a+2b,
\qquad
A(b)
\equiv
a(1+a)+2b.
\label{eq:s_A_definitions}
\end{equation}
The exact GHS relations are
\begin{equation}
r
=
r_+\sqrt{(1+b)(a+b)},
\qquad
D
=
r_+s(b),
\qquad
\varphi_\varphi
=
\frac{A(b)}{s(b)}.
\label{eq:GHS_b_relations_forced}
\end{equation}
In the finite-radius matching normalization,
\begin{equation}
p(r)
=
\frac{D_{\rm m}}{4r_{\rm m}}\,
\frac{rD\,b}{1+b}.
\label{eq:p_GHS_matching_forced}
\end{equation}
Eq.~\eqref{eq:J_first_integral} then becomes
\begin{equation}
\frac{dJ}{db}
=
\frac{2r_{\rm m}r_+}{D_{\rm m}}
\left[
-1
-\frac{1}{a+b}
+
\frac{4a(a^2-1)}
{\left[a(1+a)+2b\right]^2}
\right].
\label{eq:J_b_derivative}
\end{equation}

Choosing
\begin{equation}
J(0)=0,
\label{eq:J_horizon_normalization}
\end{equation}
one obtains
\begin{equation}
J(b)
=
\frac{2r_{\rm m}r_+}{D_{\rm m}}\,
\mathcal G(b),
\label{eq:J_closed_form}
\end{equation}
where
\begin{equation}
\mathcal G(b)
\equiv
-b
-
\ln\!\left[
\frac{a+b}{a}
\right]
+
\frac{4b(a-1)}
{a(1+a)+2b}.
\label{eq:G_function_definition}
\end{equation}
In particular, \(\mathcal G(0)=0\). Thus
\begin{equation}
q_{\rm H}(r)
=
\frac{2r_{\rm m}r_+}{D_{\rm m}}\,
\varphi_\varphi(r)
\mathcal G\!\left(b(r)\right)
\label{eq:qH_closed_form}
\end{equation}
is a horizon-regular particular solution of the unsliced forced equation.

To obtain a convenient particular solution of the sliced equation, we use
the freedom to add the unsliced homogeneous mode \(\varphi_\varphi\) and impose
the normalization
\begin{equation}
q_0(r_{\rm m})=0.
\label{eq:q0_matching_value}
\end{equation}
This fixes the normalization of the particular solution; it is not an
additional physical outer matching condition. One finds
\begin{equation}
q_0(r)
=
\frac{2r_{\rm m}r_+}{D_{\rm m}}\,
\varphi_\varphi(r)
\left[
\mathcal G\!\left(b(r)\right)
-
\mathcal G(b_{\rm m})
\right].
\label{eq:q0_closed_form}
\end{equation}
Since
\begin{equation}
Lq_0
=
-2
\left(
r^2\varphi_\varphi'
+
r\varphi_\varphi
\right),
\qquad
q_0(r_{\rm m})=0,
\end{equation}
the finite-rank term in \(L_{\rm sl}\) vanishes on \(q_0\), and therefore
\begin{equation}
L_{\rm sl}q_0
=
-2
\left(
r^2\varphi_\varphi'
+
r\varphi_\varphi
\right).
\label{eq:q0_verification}
\end{equation}

The explicit forced scalar lag is consequently
\begin{equation}
\eta_0(v,r)
=
\dot\varphi_{\rm ext}(v)\,q_0(r)
=
\frac{
2r_{\rm m}r_+\dot\varphi_{\rm ext}(v)
}{
D_{\rm m}
}
\varphi_\varphi(r)
\left[
\mathcal G\!\left(b(r)\right)
-
\mathcal G(b_{\rm m})
\right].
\label{eq:eta0_explicit}
\end{equation}
The construction of \(q_0\) and \(\eta_0\) up to this point is exact on
every nontrivial finite radial interval and does not use the large-radius
overlap approximation.

For use in the matching equation, we now evaluate
\(\mathcal B_{\rm m}\eta_0\) for the model functional adopted in
Sec.~\ref{subsec:far_field_matching}. The radial and throat-coordinate
derivatives are related by
\begin{equation}
r\frac{d}{dr}
=
\frac{
2(1+b)(a+b)
}{
1+a+2b
}
\frac{d}{db}.
\label{eq:r_derivative_in_b_forced}
\end{equation}
Since \(q_0(r_{\rm m})=0\),
\begin{equation}
\mathcal B_{\rm m}q_0
=
r_{\rm m}q_0'(r_{\rm m}).
\end{equation}
We define
\begin{equation}
A_{\rm m}
\equiv
a(1+a)+2b_{\rm m},
\qquad
s_{\rm m}
\equiv
1+a+2b_{\rm m}
\label{eq:Am_sm_forced}
\end{equation}
by evaluating Eq.~\eqref{eq:s_A_definitions} at the matching surface \(r=r_{\rm m}\). We then obtain
\begin{equation}
\mathcal R_{\rm m}
\equiv
\mathcal B_{\rm m}q_0
=
\frac{
4r_{\rm m}
(1+b_{\rm m})(a+b_{\rm m})A_{\rm m}
}{
s_{\rm m}^3
}
\mathcal G'(b_{\rm m}),
\label{eq:R_m_explicit}
\end{equation}
where
\begin{equation}
\mathcal G'(b)
=
-1
-\frac{1}{a+b}
+
\frac{
4a(a^2-1)
}{
\left[a(1+a)+2b\right]^2
}.
\label{eq:G_prime}
\end{equation}
Consequently,
\begin{equation}
\mathcal B_{\rm m}\eta_0
=
\dot\varphi_{\rm ext}(v)\mathcal R_{\rm m}.
\label{eq:Bm_eta0_explicit}
\end{equation}

For the fixed model functional
\(\mathcal B_{\rm m}=r_{\rm m}\partial_r+1\),
Eqs.~\eqref{eq:R_m_explicit} and
\eqref{eq:Bm_eta0_explicit} are exact algebraic evaluations for arbitrary
\(b_{\rm m}\geq0\). Their physical interpretation as overlap-matching data
is understood within the controlled domain given in Eq.~\eqref{eq:controlled_massless_overlap_domain}.

\subsection{Complete scalar and metric response}
\label{subsec:Explicit_near-zone_solution}

Substituting Eq.~\eqref{eq:Bm_eta0_explicit} into the Fredholm matching
equation gives
\begin{equation}
t(v)
=
\frac{
J_{\rm ext}(v)
-
\dot\varphi_{\rm ext}(v)\mathcal R_{\rm m}
}{
\mathfrak D
},
\qquad
\mathfrak D
\equiv
\mathcal B_{\rm m}\widehat\psi_\star
>0.
\label{eq:t_solution}
\end{equation}
For the model functional
\(\mathcal B_{\rm m}=r_{\rm m}\partial_r+1\), the positivity of
\(\mathfrak D\) is an exact algebraic property of the GHS profiles for
\(0<a<1\) and \(b_{\rm m}\geq0\). Its interpretation as the
non-degeneracy condition of a physical massless exterior matching problem is
understood within the controlled overlap regime given in
Eq.~\eqref{eq:controlled_massless_overlap_domain}.

For admissible slowly varying matching data,
\begin{equation}
\frac{t(v)}{\Mpl}
=
\mathcal O_{\rm ad}(\epsilon_\varphi).
\end{equation}
The complete leading-order scalar lag is therefore
\begin{equation}
\eta(v,r)
=
\dot\varphi_{\rm ext}(v)\,q_0(r)
+
\frac{
J_{\rm ext}(v)
-
\dot\varphi_{\rm ext}(v)\mathcal R_{\rm m}
}{
\mathfrak D
}
\widehat\psi_\star(r).
\label{eq:eta_solution}
\end{equation}
All radial profiles entering Eq.~\eqref{eq:eta_solution} are known
analytically, with their dependence on the instantaneous collective
coordinates \(\lambda(v)\) left implicit. The exterior enters through the
prescribed rolling rate \(\dot\varphi_{\rm ext}(v)\) and the inhomogeneous
matching datum \(J_{\rm ext}(v)\).

Because
\begin{equation}
q_0(r_{\rm m})=0,
\qquad
\widehat\psi_\star(r_{\rm m})
=
\varphi_\varphi(r_{\rm m}),
\end{equation}
the scalar lag at the matching surface is
\begin{equation}
\eta(v,r_{\rm m})
=
t(v)\varphi_\varphi(r_{\rm m}).
\label{eq:eta_at_rm}
\end{equation}

The mass lag follows algebraically from the sliced radial constraint,
\begin{equation}
\mu(v,r)
=
f_\varphi(r)\eta(v,r)
-
f_\varphi(r_{\rm m})
\eta(v,r_{\rm m})
e^{-\delta_0(r)},
\label{eq:mu_reconstruction}
\end{equation}
where
\begin{equation}
f_\varphi(r)
=
\frac{r^2}{2}
\mathcal F_0(r)\varphi_0'(r).
\end{equation}
Using
\begin{equation}
m_\varphi(r_{\rm m})
=
f_\varphi(r_{\rm m})\varphi_\varphi(r_{\rm m}),
\label{eq:mPhi_rm_again_response}
\end{equation}
together with Eqs.~\eqref{eq:eta_solution} and
\eqref{eq:eta_at_rm}, one obtains
\begin{equation}
\mu(v,r)
=
\dot\varphi_{\rm ext}(v)
f_\varphi(r)q_0(r)
+
t(v)
\left[
f_\varphi(r)\widehat\psi_\star(r)
-
m_\varphi(r_{\rm m})e^{-\delta_0(r)}
\right].
\label{eq:mu_complete_explicit}
\end{equation}
The local mass slice is manifestly satisfied:
\begin{equation}
\mu(v,r_{\rm m})=0.
\label{eq:mu_slice_check_complete}
\end{equation}

The lapse lag satisfies the radial constraint
\begin{equation}
\partial_r\xi
=
\frac{r}{\Mpl^2}
\varphi_0'(r)\partial_r\eta,
\qquad
\xi(v,r_{\rm m})=0.
\label{eq:xi_reconstruction}
\end{equation}
It is therefore reconstructed as
\begin{equation}
\xi(v,r)
=
-
\int_r^{r_{\rm m}}
\frac{s}{\Mpl^2}
\varphi_0'(s)
\partial_s\eta(v,s)\,ds.
\label{eq:xi_integrated}
\end{equation}
Equivalently, in the throat coordinate,
\begin{equation}
\xi(v,b)
=
\frac{1-a}{\alpha\Mpl}
\int_b^{b_{\rm m}}
\frac{
\partial_x\eta(v,x)
}{
1+a+2x
}\,dx.
\label{eq:xi_complete_quadrature}
\end{equation}
The integrand is explicit in terms of rational functions of \(x\) and the
logarithm contained in \(\mathcal G(x)\). The lapse lag is therefore
determined by a single elementary radial quadrature.

Eqs~\eqref{eq:eta_solution},
\eqref{eq:mu_complete_explicit}, and
\eqref{eq:xi_complete_quadrature} give the complete leading-order near-zone
configuration relative to the chosen instantaneous static representative.
The scalar and mass lags are obtained directly in closed analytical form,
while the lapse lag follows from one explicit quadrature.

The inner endpoint of this construction is the horizon
\(r_{\rm H}^{\rm inst}(v)\) of the instantaneous static representative. The
exact dynamical horizon \(r_{\rm H}^{\rm exact}(v)\) is displaced from it by
the mass lag and is reconstructed in the following subsection.

\subsection{Displacement of the exact dynamical horizon}
\label{subsec:exact_horizon_displacement}

The inner endpoint of the radial construction is the horizon
\(r_{\rm H}^{\rm inst}(v)\) of the instantaneous static representative. The
exact marginal horizon of the full dynamical solution is instead written as
\begin{equation}
r_{\rm H}^{\rm exact}(v)
=
r_{\rm H}^{\rm inst}(v)
+
\Delta r_{\rm H}(v)
+
\mathcal O_{\rm ad}(\epsilon_\varphi^2).
\label{eq:exact_inst_horizon_split}
\end{equation}
Before the representation convention in the static horizon-mass direction is
fixed, the decomposition of \(r_{\rm H}^{\rm exact}\) into an instantaneous
background contribution and a lag-induced displacement is representation
dependent. In the convention adopted throughout this paper,
\(\Delta r_{\rm H}\) denotes the displacement relative to the selected
instantaneous static representative, whereas the total areal radius
\(r_{\rm H}^{\rm exact}\) is representation independent.

Expanding the exact horizon condition
\begin{equation}
\mathcal F\!\left(
v,r_{\rm H}^{\rm exact}(v)
\right)
=
0
\end{equation}
about \(r=r_{\rm H}^{\rm inst}(v)\) gives
\begin{equation}
\mathcal F_0'(r_{\rm H}^{\rm inst})
\Delta r_{\rm H}
-
\frac{
2\mu(v,r_{\rm H}^{\rm inst})
}{
\Mpl^2r_{\rm H}^{\rm inst}
}
=
0.
\label{eq:linearized_exact_horizon_condition}
\end{equation}
Therefore
\begin{equation}
\Delta r_{\rm H}(v)
=
\frac{
2\mu(v,r_{\rm H}^{\rm inst})
}{
\Mpl^2r_{\rm H}^{\rm inst}
\mathcal F_0'(r_{\rm H}^{\rm inst})
}.
\label{eq:exact_horizon_displacement_general}
\end{equation}

At the instantaneous static horizon,
\begin{equation}
f_\varphi(r_{\rm H}^{\rm inst})=0.
\end{equation}
The sliced mass constraint consequently gives
\begin{equation}
\mu(v,r_{\rm H}^{\rm inst})
=
-f_\varphi(r_{\rm m})
\eta(v,r_{\rm m})
e^{-\delta_0(r_{\rm H}^{\rm inst})}.
\label{eq:mu_at_inst_horizon}
\end{equation}
Using
\begin{equation}
\eta(v,r_{\rm m})
=
t(v)\varphi_\varphi(r_{\rm m}),
\qquad
m_\varphi(r_{\rm m})
=
f_\varphi(r_{\rm m})\varphi_\varphi(r_{\rm m}),
\end{equation}
one obtains
\begin{equation}
\Delta r_{\rm H}(v)
=
-
\frac{
2m_\varphi(r_{\rm m})
e^{-\delta_0(r_{\rm H}^{\rm inst})}
}{
\Mpl^2r_{\rm H}^{\rm inst}
\mathcal F_0'(r_{\rm H}^{\rm inst})
}
\,t(v).
\label{eq:exact_horizon_displacement_t}
\end{equation}
Introducing the surface gravity parameter of the instantaneous static
representative in the matching normalization,
\begin{equation}
\kappa_{\rm H}^{\rm inst}
\equiv
\frac12
e^{\delta_0(r_{\rm H}^{\rm inst})}
\mathcal F_0'(r_{\rm H}^{\rm inst}),
\end{equation}
Eq.~\eqref{eq:exact_horizon_displacement_t} may equivalently be written as
\begin{equation}
\Delta r_{\rm H}(v)
=
-
\frac{
m_\varphi(r_{\rm m})
}{
\Mpl^2r_{\rm H}^{\rm inst}
\kappa_{\rm H}^{\rm inst}
}
\,t(v).
\label{eq:exact_horizon_displacement_kappa}
\end{equation}
For the charged non-extremal GHS branch and every nontrivial finite matching
interval,
\begin{equation}
m_\varphi(r_{\rm m})<0,
\qquad
\kappa_{\rm H}^{\rm inst}>0.
\end{equation}
The horizon displacement therefore has the same sign as the remaining
homogeneous amplitude:
\begin{equation}
\operatorname{sgn}(\Delta r_{\rm H})
=
\operatorname{sgn}(t).
\label{eq:horizon_displacement_sign}
\end{equation}

Substituting the explicit matching amplitude gives
\begin{equation}
\Delta r_{\rm H}(v)
=
-
\frac{
2m_\varphi(r_{\rm m})
e^{-\delta_0(r_{\rm H}^{\rm inst})}
}{
\Mpl^2r_{\rm H}^{\rm inst}
\mathcal F_0'(r_{\rm H}^{\rm inst})
}
\,
\frac{
J_{\rm ext}(v)
-
\dot\varphi_{\rm ext}(v)\mathcal R_{\rm m}
}{
\mathfrak D
}.
\label{eq:exact_horizon_displacement_explicit}
\end{equation}
For the particular massless matching model adopted here, the physical
interpretation of this expression is understood within the controlled
overlap regime given in Eq.~\eqref{eq:controlled_massless_overlap_domain}.

The exact and instantaneous horizon-mass parameters are correspondingly
related by
\begin{equation}
M_{\rm H}^{\rm exact}
-
M_{\rm H}^{\rm inst}
=
\frac{\Mpl^2}{2}\Delta r_{\rm H}
+
\mathcal O_{\rm ad}(\epsilon_\varphi^2).
\label{eq:exact_inst_mass_difference}
\end{equation}
The displacement begins at first adiabatic order, whereas its time derivative
begins at second order:
\begin{equation}
\Delta r_{\rm H}
=
\mathcal O_{\rm ad}(\epsilon_\varphi),
\qquad
\partial_v\Delta r_{\rm H}
=
\mathcal O_{\rm ad}(\epsilon_\varphi^2).
\end{equation}
Since
\(\partial_v r_{\rm H}^{\rm inst}
=\mathcal O_{\rm ad}(\epsilon_\varphi^2)\),
Eq.~\eqref{eq:exact_inst_horizon_split} consequently gives
\begin{equation}
\partial_v r_{\rm H}^{\rm exact}
=
\mathcal O_{\rm ad}(\epsilon_\varphi^2),
\end{equation}
in agreement with the exact quadratic horizon-growth law derived in
Sec.~\ref{subsec:horizon_dynamics}. Thus an instantaneous first-order
separation between the exact and reference horizons is fully compatible with
the absence of first-order physical horizon evolution.

%
\subsection{Role of the exterior}
\label{sec:role_exterior}

The explicit solution obtained above separates the intrinsic radial response
of the black hole from the information supplied by the far zone.

At each fixed advanced time, the instantaneous GHS background determines the
radial functions
\begin{equation}
f_\varphi,
\qquad
K,
\qquad
\varphi_\varphi,
\qquad
\chi_0,
\qquad
\widehat\psi_\star,
\qquad
q_0,
\end{equation}
as well as the operators \(L\) and \(L_{\rm sl}\). Their dependence on the
slowly varying collective coordinates \(\lambda(v)\) is left implicit. Once
an outer matching functional has been specified, the same instantaneous
background determines its action on the sliced-kernel generator. For the
massless model functional adopted here, this gives the Fredholm denominator
\begin{equation}
\mathfrak D
=
\mathcal B_{\rm m}\widehat\psi_\star.
\end{equation}
These radial quantities depend on the static black hole family and on the
instantaneous values of its collective coordinates, but not on the detailed
time history of the exterior solution.

The matching functional
\begin{equation}
\mathcal B_{\rm m}
=
r_{\rm m}\partial_r+1
\end{equation}
is not determined by the GHS interior alone. It is the leading Robin
functional associated with the massless large-radius overlap model of
Sec.~\ref{subsec:far_field_matching}, with the convention that the prescribed
exterior modulus fixes the constant scalar mode in the overlap region. Its
interpretation as the physical matching operator of a massless exterior is
therefore restricted to the controlled domain
\begin{equation}
b_{\rm m}\gg1,
\qquad
r_{\rm m}\ll L_{\rm cos},
\end{equation}
as summarized in
Eq.~\eqref{eq:controlled_massless_overlap_domain}. A different exterior
background, a different matching convention, or subleading overlap
corrections would in general modify both \(\mathcal B_{\rm m}\) and its
inhomogeneous datum.

Once the model functional has been fixed, the exact GHS radial profiles allow
\(\mathfrak D\) and the remaining matching quantities to be evaluated
algebraically for arbitrary \(b_{\rm m}\geq0\). Their continuation outside
the controlled overlap domain remains a valid mathematical property of the
chosen model functional, but it should not be interpreted as a statement
about matching to a physical massless exterior in that regime.

Within the specified overlap model, the exterior enters the near-zone
solution through two functions of advanced time. The first is the prescribed
rolling rate
\begin{equation}
\dot\varphi_{\rm ext}(v),
\end{equation}
which multiplies the universal forced radial profile \(q_0(r)\). The second
is the inhomogeneous matching datum
\begin{equation}
J_{\rm ext}(v),
\end{equation}
which fixes the amplitude of the remaining sliced homogeneous mode through
\begin{equation}
t(v)
=
\frac{
J_{\rm ext}(v)
-
\dot\varphi_{\rm ext}(v)\mathcal R_{\rm m}
}{
\mathfrak D
}.
\end{equation}

Different massless exterior solutions compatible with the same leading
overlap assumptions therefore share the same instantaneous radial response
functions and the same model Fredholm denominator. They differ through the
prescribed functions \(\dot\varphi_{\rm ext}(v)\) and \(J_{\rm ext}(v)\). Within
the controlled overlap regime, the near-zone construction guarantees that
these data determine a unique leading-order response. It does not, however,
determine \(J_{\rm ext}\) without solving the corresponding far-zone problem.

The present analysis does not attempt to characterize the complete class of
global exterior solutions compatible with the overlap construction. Instead,
the exterior enters only through the matching data
\(\dot\varphi_{\rm ext}(v)\) and \(J_{\rm ext}(v)\). Throughout this work,
they are assumed to obey the dimensionless slow-variation counting
\begin{equation}
\frac{r_{\rm H}^{\rm inst}}{\Mpl}\,
\dot\varphi_{\rm ext}
=
\mathcal O_{\rm ad}(\epsilon_\varphi),
\qquad
\frac{J_{\rm ext}}{\Mpl}
=
\mathcal O_{\rm ad}(\epsilon_\varphi),
\qquad
\frac{r_{\rm H}^{\rm inst}}{\Mpl}\,
\partial_v J_{\rm ext}
=
\mathcal O_{\rm ad}(\epsilon_\varphi^2).
\label{eq:admissible_matching_data}
\end{equation}
The corresponding higher time derivatives and the implicit dependence of
the radial profiles on the slowly varying background must likewise remain
compatible with the leading adiabatic truncation.

In the present paper, an admissible exterior datum therefore means a pair
\((\dot\varphi_{\rm ext},J_{\rm ext})\) that satisfies
Eq.~\eqref{eq:admissible_matching_data} and belongs to the leading massless
overlap model in the controlled domain given in  Eq.~\eqref{eq:controlled_massless_overlap_domain}. For every such datum, the
completed near-zone boundary-value problem admits a unique leading-order
solution. The existence of a global exterior spacetime that realizes a
prescribed admissible datum is a separate problem and is not established
here.

The matching convention also clarifies the role of the collective modulus.
The constant scalar mode in the overlap region is assigned to the promoted
static modulus coordinate. The lag field then describes the departure from
the instantaneous static configuration at that prescribed modulus, while
\(J_{\rm ext}\) encodes the remaining information transmitted by the
exterior solution to the matching surface.

This separation should not be interpreted as a claim that the near-zone and
far-zone problems are dynamically independent. The exterior supplies both
the forcing and the outer boundary data, whereas the black hole background
determines the radial operator through which those data are converted into a
local response. The Fredholm construction isolates these two roles and shows
that, for the magnetic GHS family and within the controlled massless overlap
model considered here, no homogeneous obstruction prevents their consistent
combination.

The intrinsic GHS construction applies to the charged non-extremal branch
\begin{equation}
0<a<1.
\end{equation}
The extremal limit lies outside the uniformly non-degenerate horizon
framework assumed in Paper~I. The strict Schwarzschild endpoint is also
degenerate for the present mass slice, since
\begin{equation}
K\equiv0,
\qquad
m_\varphi(r_{\rm m})=0.
\end{equation}
The finite-rank update then disappears and the slice ceases to be transverse.
Both endpoints provide useful formal consistency checks, but neither belongs
to the domain on which the present sliced Fredholm construction is applied.
The additional restriction
\(b_{\rm m}\gg1\), \(r_{\rm m}\ll L_{\rm cos}\) concerns the physical
validity of the particular massless matching model, rather than the intrinsic
existence of the GHS radial operator and its exact closed-form profiles.

\subsection{Discussion and Outlook}
\label{sec:outlook}

Several extensions follow naturally from the present construction.

The first extension concerns weak stabilizing scalar potentials. Let the
exterior potential be characterized locally by
\begin{equation}
m_{\rm eff}^2
=
V''(\varphi_{\rm ext})
>0,
\qquad
L_V
\equiv
m_{\rm eff}^{-1}.
\end{equation}
In a region where \(m_{\rm eff}\) may be treated as approximately constant
and curvature corrections are subleading, the homogeneous exterior solution
selected by decay at large radius behaves as
\begin{equation}
\eta_{\rm hom}(r)
\sim
\frac{B}{r}
e^{-m_{\rm eff}r}.
\label{eq:Yukawa_overlap_mode}
\end{equation}
It obeys
\begin{equation}
r\eta_{\rm hom}'(r)
+
\left(
1+m_{\rm eff}r
\right)
\eta_{\rm hom}(r)
=
0,
\label{eq:Yukawa_Robin_relation}
\end{equation}
and would therefore induce the Yukawa-type matching functional
\begin{equation}
\mathcal B_{\rm m}^{(V)}
=
r_{\rm m}\partial_r
+
1
+
m_{\rm eff}r_{\rm m},
\label{eq:Yukawa_matching_functional}
\end{equation}
up to corrections from the radial variation of the exterior background and
from the overlap geometry.

The near-zone hierarchy inherited from Paper~I includes
\begin{equation}
r_{\rm m}
\ll
L_{\rm cos}
\leq
L_V,
\end{equation}
and hence
\begin{equation}
m_{\rm eff}r_{\rm m}
\ll1.
\label{eq:weak_Yukawa_overlap_hierarchy}
\end{equation}
Within the present construction,
Eq.~\eqref{eq:Yukawa_matching_functional} must therefore be interpreted as a
small mass-induced correction to the massless Robin functional, rather than
as evidence for an order-one exponential suppression within the near zone.
The globally decaying exterior branch may still carry genuine Yukawa
behavior at radii of order \(L_V\) and beyond, but the matching to the near
zone takes place in its small-\(m_{\rm eff}r\) expansion. A regime with
\(m_{\rm eff}r_{\rm m}={\mathcal O}(1)\) would require a modified near-zone/far-zone
split and lies outside the hierarchy assumed in the present papers.

For the potential-deformed problem, the corresponding Fredholm denominator
would be
\begin{equation}
\mathfrak D_V
\equiv
\mathcal B_{\rm m}^{(V)}
\widehat\psi_\star^{(V)},
\label{eq:Yukawa_Fredholm_denominator}
\end{equation}
where \(\widehat\psi_\star^{(V)}\), if it exists, denotes a globally regular
generator of the sliced kernel associated with the potential-deformed static
background. The logical structure of the solvability theorem is unchanged:
existence and uniqueness require the matching functional to act
nontrivially on this kernel generator,
\begin{equation}
\mathfrak D_V\neq0.
\end{equation}
Establishing this condition, however, requires a new analysis. A nonzero
potential may deform the static black hole family, the canonical tangent,
the reduced radial operator, the source kernel, and the matching functional
itself. For a generic potential, neither the deformed static background nor
the corresponding sliced-kernel generator is expected to remain available
in closed form. Their existence, regularity, and non-degeneracy under the
modified matching functional must therefore be established as part of the
completed potential-deformed boundary-value problem. The stabilized problem
cannot be obtained by a direct replacement of \(J_{\rm ext}\) in the
massless formulas.

A second, physically distinct extension concerns genuinely rolling
cosmological backgrounds. In that case no globally static exterior solution
exists, and the matching functional must be derived from the appropriate
time-dependent far-zone expansion. The present construction suggests that,
at leading adiabatic order, the instantaneous static black hole geometry may
continue to determine the intrinsic near-zone operator, while the
cosmological solution supplies the forcing, the matching functional, and its
inhomogeneous datum. Establishing this separation requires an explicit
matched asymptotic analysis of the rolling exterior.

A third direction is the extension to electrically charged and dyonic black
holes. The geometric part of the construction requires a smooth static
solution manifold parametrized by the modulus label and the static
horizon-mass parameter. Whenever such a family exists and the corresponding
reduced radial operator has the Fredholm properties assumed in Paper~I, the
canonical-tangent and local-slice construction can be applied after adapting
the explicit background data.

It would also be useful to extend the present analysis beyond leading
adiabatic order. The transport terms omitted from the frozen radial problem
become definite sources at second order once the leading solution is known.
At that order, the exact horizon-growth law, the evolution of the
instantaneous collective coordinates, and the time dependence of the radial
profiles must be treated consistently.

More broadly, the magnetic GHS family demonstrates that the abstract
Fredholm framework of Paper~I can be realized almost entirely in closed
analytical form. The canonical tangent, the finite-rank source profile, the
sliced-kernel generator, the forced scalar response, the Fredholm
denominator, and the mass lag are obtained explicitly, while the lapse lag
is reconstructed by a single elementary quadrature. This provides a concrete
starting point for studying slowly driven black hole moduli in less
integrable systems.



\section{Conclusions}
\label{sec:conclusions}

In this work, we have carried out the first complete realization of the
Fredholm construction developed in Paper~I for an explicit black hole family.
The magnetic GHS solution provides a rare example
for which every ingredient of the leading adiabatic near-zone problem can be
determined analytically.

Starting from the exact EMD equations in advanced
Eddington--Finkelstein coordinates, we reformulated the dynamical problem as
an adiabatic deformation of an instantaneous static representative while
maintaining a clear distinction between the horizon of that representative
and the exact dynamical marginal horizon. This separation allows the lag
fields to be defined unambiguously and leads naturally to the canonical-slice
construction introduced in Paper~I.

For the magnetic GHS family, we obtained explicit closed-form expressions for
the canonical tangent, the rank-one coefficient, the source profile, the
effective radial potential, a horizon-regular solution of the auxiliary
inhomogeneous equation, and a globally regular representative of the
one-dimensional kernel of the sliced operator. For the model functional
\begin{equation}
\mathcal B_{\rm m}
=
r_{\rm m}\partial_r+1,
\end{equation}
associated with the leading massless large-radius overlap, the Fredholm
denominator was evaluated analytically and shown to satisfy the exact
algebraic inequality
\begin{equation}
\mathfrak D
=
\mathcal B_{\rm m}\widehat\psi_\star
>0,
\qquad
0<a<1,
\qquad
b_{\rm m}\geq0.
\end{equation}
This includes every point of the locus
\(\mathcal N_{\rm m}=0\) at which the conventional normalization of the
kernel representative degenerates. The globally regular representative
\(\widehat\psi_\star\) shows that this locus is a normalization degeneracy
only and does not correspond to a discontinuity or bifurcation of the sliced
kernel.

The strict positivity of \(\mathfrak D\) is stronger than the minimal
Fredholm requirement \(\mathfrak D\neq0\). It shows that the model matching
functional remains sign definite and undergoes no algebraic loss of
solvability or change of matching orientation anywhere on the charged
non-extremal parameter domain. Its interpretation as the non-degeneracy
condition of a genuine massless exterior matching problem is controlled,
however, only in the large-radius overlap regime
\begin{equation}
b_{\rm m}\gg1,
\qquad
r_{\rm m}\ll L_{\rm cos},
\end{equation}
in which \(\mathcal B_{\rm m}\) was derived. The locus
\(\mathcal N_{\rm m}=0\) may lie inside or outside this regime depending on
\(a\); in particular, it lies outside the controlled overlap domain near the
extremal endpoint. Outside the overlap regime, the positivity result remains
an exact algebraic property of the chosen model functional, but it is not
interpreted as a physical exterior-matching statement.

The forced radial problem was likewise solved explicitly. The scalar lag is
obtained as the sum of a universal forced profile and a sliced homogeneous
mode whose amplitude is fixed uniquely by the exterior matching datum. The
mass lag follows algebraically from the radial constraints, while the lapse
lag is reconstructed by a single elementary quadrature. Within the
controlled massless overlap model, these expressions give the complete
leading-order near-zone response for every admissible pair
\((\dot\varphi_{\rm ext},J_{\rm ext})\).

The exact dynamical horizon is then recovered from the lag-corrected
solution, yielding an explicit expression for its displacement relative to
the horizon of the instantaneous static representative. This displacement
is of first adiabatic order, while its time derivative begins at second
order, consistently with the exact quadratic horizon-flux law. The
displacement depends on the convention used to choose the instantaneous
representative, whereas the exact marginal-horizon radius itself is
representation independent.

The present analysis also clarifies the respective roles of the near-zone
and far-zone problems. The intrinsic radial operator, its kernel, and the
local reconstruction of the metric fields depend only on the instantaneous
static black hole background. The exterior supplies the prescribed rolling
rate, the matching functional, and its inhomogeneous datum. For data
belonging to the specified massless overlap model and satisfying the
adiabatic slow-variation conditions, these inputs determine a unique
leading-order near-zone response. The construction of a global exterior
spacetime realizing a prescribed matching datum remains a separate problem
and is not established here.

Taken together, these results transform the abstract geometric and
functional-analytic framework of Paper~I into a completely explicit
analytical construction. The magnetic GHS family therefore provides both a
nontrivial validation of the general theory and a benchmark against which
future investigations of slowly evolving black holes with more general
scalar dynamics and matching prescriptions can be compared.


\clearpage
\appendix

\section{Complete notation dictionary}
\label{app:notation_dictionary}

This appendix summarizes the notation used in Paper~I, our previous static
GHS analysis, and the present work. Paper~I denotes the scalar field and
canonical tangent by \(\phi\) and \(\phi_\Phi\), whereas Paper~II uses
\(\varphi\) and \(\varphi_\varphi\). Both papers distinguish the horizon of
the static or instantaneous representative from the exact dynamical horizon.

Table~\ref{tab:general_notation_dictionary} collects the geometric,
collective-coordinate, and horizon notation.

\begin{table}[H]
\centering
\begingroup
\footnotesize
\setlength{\tabcolsep}{5pt}
\renewcommand{\arraystretch}{1.16}
\begin{tabularx}{\textwidth}{
@{}
>{\raggedright\arraybackslash}p{0.36\textwidth}
>{\raggedright\arraybackslash}X
>{\raggedright\arraybackslash}X
@{}
}
\toprule
Concept
&
Paper~I
&
Paper~II
\\
\midrule

\multicolumn{3}{@{}l}{\textbf{Scalar and collective-coordinate data}}
\\[0.5mm]

Scalar field
&
\(\phi\)
&
\(\varphi\)
\\

Static modulus label
&
\(\Phi\)
&
\(\varphi\)
\\

Prescribed exterior modulus
&
\(\Phi_{\rm ext}(v)\)
&
\(\varphi_{\rm ext}(v)\)
\\

Static collective coordinates
&
\((\Phi,M_{{\rm H},0})\)
&
\((\varphi,M_{{\rm H},0})\)
\\

Promoted collective coordinates
&
\(\lambda^A(v)\)
&
\(\lambda^A(v)\)
\\

Promoted modulus coordinate
&
\(\lambda^\Phi(v)\)
&
\(\lambda^\varphi(v)\)
\\

Adiabatic parameter
&
\(\epsilon_\Phi\)
&
\(\epsilon_\varphi\)
\\

\midrule

\multicolumn{3}{@{}l}{\textbf{Horizon quantities}}
\\[0.5mm]

Static areal horizon radius
&
\(r_{{\rm H},0}\)
&
\(r_{{\rm H},0}=r_+\sqrt a\)
\\

Static horizon-mass parameter
&
\(\displaystyle
M_{{\rm H},0}
=
\frac{\Mpl^2}{2}r_{{\rm H},0}\)
&
\(\displaystyle
M_{{\rm H},0}
=
\frac{\Mpl^2}{2}r_{{\rm H},0}\)
\\

Instantaneous areal horizon radius
&
\(r_{\rm H}^{\rm inst}(v)\)
&
\(r_{\rm H}^{\rm inst}(v)\)
\\

Instantaneous horizon-mass parameter
&
\(M_{\rm H}^{\rm inst}(v)\)
&
\(M_{\rm H}^{\rm inst}(v)\)
\\

Exact dynamical horizon radius
&
\(r_{\rm H}^{\rm exact}(v)\)
&
\(r_{\rm H}^{\rm exact}(v)\)
\\

Exact dynamical horizon mass
&
\(M_{\rm H}^{\rm exact}(v)\)
&
\(M_{\rm H}^{\rm exact}(v)\)
\\

Horizon location in the GHS coordinate
&
\(\text{---}\)
&
\(\rho=r_+\)
\\

Non-extremality parameter
&
\(\text{---}\)
&
\(\displaystyle
a=\frac{r_+-r_-}{r_+}\)
\\

\midrule

\multicolumn{3}{@{}l}{\textbf{Radial coordinates}}
\\[0.5mm]

Exact areal radius
&
\(r\)
&
\(r\)
\\

Conventional GHS coordinate
&
\(\text{---}\)
&
\(\rho\)
\\

Areal-radius dictionary
&
\(\text{---}\)
&
\(\displaystyle
r=R(\rho)=\sqrt{\rho(\rho-r_-)}\)
\\

Dimensionless throat coordinate
&
\(\text{---}\)
&
\(\displaystyle
b=\frac{\rho-r_+}{r_+}\)
\\

Matching radius
&
\(r_{\rm m}\)
&
\(r_{\rm m}\)
\\

\midrule

\multicolumn{3}{@{}l}{\textbf{Static profiles and tangent data}}
\\[0.5mm]

Static scalar profile
&
\(\phi_0\)
&
\(\varphi_0\)
\\

Static GHS scalar profile
&
\(\text{---}\)
&
\(\varphi_{\rm GHS}\)
\\

Static lapse
&
\(\delta_0\)
&
\(\delta_0\)
\\

Static Misner--Sharp mass
&
\(m_0\)
&
\(m_0\)
\\

Canonical scalar tangent
&
\(\phi_\Phi\)
&
\(\varphi_\varphi\)
\\

Canonical lapse tangent
&
\(\delta_\Phi\)
&
\(\delta_\varphi\)
\\

Canonical mass tangent
&
\(m_\Phi\)
&
\(m_\varphi\)
\\

Intrinsic tangent covector
&
\(C_\mu\)
&
\(C_\mu\)
\\

\bottomrule
\end{tabularx}
\endgroup
\caption{Geometric, horizon, and collective-coordinate notation in
Paper~I and Paper~II. Paper~I denotes the scalar field and static modulus
label by \(\phi\) and \(\Phi\), whereas Paper~II uses \(\varphi\) for both
the scalar field and the static modulus label. The radially varying static
profile is distinguished by the subscript \(0\), or by the notation
\(\varphi_{\rm GHS}\) when the explicit GHS solution is displayed.}
\label{tab:general_notation_dictionary}
\end{table}

\clearpage

Table~\ref{tab:operator_notation_dictionary} summarizes the reduced
operators, kernel data, lag fields, and matching quantities. The previous
static GHS analysis did not use this operator-level notation and is therefore
omitted from this table.

\begin{table}[H]
\centering
\begingroup
\footnotesize
\setlength{\tabcolsep}{5pt}
\renewcommand{\arraystretch}{1.12}
\begin{tabularx}{\textwidth}{
@{}
>{\raggedright\arraybackslash}p{0.34\textwidth}
>{\raggedright\arraybackslash}X
>{\raggedright\arraybackslash}X
@{}
}
\toprule
Concept
&
Paper~I
&
Paper~II
\\
\midrule

\multicolumn{3}{@{}l}{\textbf{Reduced radial operator}}
\\[0.5mm]

Leading coefficient
&
\(\displaystyle
p=e^{\delta_0}r^2\mathcal F_0\)
&
\(\displaystyle
p=e^{\delta_0}r^2\mathcal F_0\)
\\

Effective potential
&
\(V_{\rm eff}\)
&
\(V_{\rm eff}\)
\\

Unsliced operator
&
\(L\)
&
\(L\)
\\

Rank-one coefficient
&
\(\displaystyle
f_\phi(r)
=
\frac{r^2}{2}\mathcal F_0\phi_0'(r)\)
&
\(\displaystyle
f_\varphi(r)
=
\frac{r^2}{2}\mathcal F_0\varphi_0'(r)\)
\\

Source kernel
&
\(\displaystyle
K(r)
=
\frac{2}{\Mpl^2}
\bigl(r\phi_0'(r)\bigr)'\)
&
\(\displaystyle
K(r)
=
\frac{2}{\Mpl^2}
\bigl(r\varphi_0'(r)\bigr)'\)
\\

Sliced operator
&
\(L_{\rm sl}\)
&
\(L_{\rm sl}\)
\\

Matching functional
&
\(\mathcal B_{\rm m}\)
&
\(\mathcal B_{\rm m}\)
\\

Completed operator
&
\(\mathcal A=(L_{\rm sl},\mathcal B_{\rm m})\)
&
\(\mathcal A=(L_{\rm sl},\mathcal B_{\rm m})\)
\\

\midrule

\multicolumn{3}{@{}l}{\textbf{Kernel and solvability data}}
\\[0.5mm]

Particular solution of \(L\chi_0=K\)
&
\(\chi_0\)
&
\(\chi_0\)
\\

Abstract sliced-kernel generator
&
\(\psi_\star\)
&
\(\psi_\star\)
\\

Globally regular representative
&
\(\text{---}\)
&
\(\widehat\psi_\star\)
\\

Conventionally normalized representative
&
\(\text{---}\)
&
\(\displaystyle
\psi_\star
=
\frac{\widehat\psi_\star}{\mathcal N_{\rm m}},
\quad
\mathcal N_{\rm m}\neq0\)
\\

Normalization factor
&
\(\text{---}\)
&
\(\mathcal N_{\rm m}\)
\\

Fredholm non-degeneracy quantity
&
\(\mathcal B_{\rm m}\psi_\star\)
&
\(\displaystyle
\mathfrak D
=
\mathcal B_{\rm m}\widehat\psi_\star\)
\\

\midrule

\multicolumn{3}{@{}l}{\textbf{Lag and matching data}}
\\[0.5mm]

Scalar, lapse, and mass lags
&
\(\eta,\ \xi,\ \mu\)
&
\(\eta,\ \xi,\ \mu\)
\\

Residual mass-integration datum
&
\(C_\mu^{\rm lag}(v)\)
&
\(C_\mu^{\rm lag}(v)\)
\\

Exterior source
&
\(S_{\rm ext}\)
&
\(S_{\rm ext}\)
\\

Exterior matching datum
&
\(J_{\rm ext}\)
&
\(J_{\rm ext}\)
\\

Forced particular solution
&
\(\eta_0\)
&
\(\eta_0=\dot\varphi_{\rm ext}q_0\)
\\

Forced radial profile
&
\(\text{---}\)
&
\(q_0\)
\\

Homogeneous amplitude
&
\(t\)
&
\(t(v)\)
\\

\bottomrule
\end{tabularx}
\endgroup
\caption{Reduced-operator, kernel, lag-field, and matching notation.
Paper~I defines the generator of the one-dimensional sliced kernel only up
to a nonzero normalization. In Paper~II,
\(\widehat\psi_\star\) denotes a globally regular representative, whereas
\(\psi_\star=\widehat\psi_\star/\mathcal N_{\rm m}\) denotes the
conventionally normalized representative wherever
\(\mathcal N_{\rm m}\neq0\).}
\label{tab:operator_notation_dictionary}
\end{table}

Our previous static analysis of the magnetic GHS family used the conventional
Schwarzschild-like coordinate \(\rho\), the dimensionless throat coordinate
\begin{equation}
b
=
\frac{\rho-r_+}{r_+},
\end{equation}
and the explicit scalar profile \(\varphi_{\rm GHS}\). The asymptotic
modulus was denoted there by \(\varphi_\infty\) (or, in some formulas, by
\(\Phi\)), and the static family was parametrized by
\((\varphi_\infty,a)\). In the present paper, the same static family is
instead described by the coordinates
\((\varphi,M_{{\rm H},0})\), with \(\varphi\) denoting the static modulus
label and
\begin{equation}
a
=
\frac{r_+-r_-}{r_+},
\qquad
r_{{\rm H},0}
=
r_+\sqrt a,
\qquad
M_{{\rm H},0}
=
\frac{\Mpl^2}{2}r_{{\rm H},0}.
\end{equation}
The two descriptions are related by the smooth change of static-family
coordinates discussed in Sec.~\ref{subsec:static_GHS}. The present
dynamical construction uses the exact areal radius \(r\), while \(\rho\)
and \(b\) are retained only as convenient variables for expressing the
closed-form GHS profiles.

The explicit GHS calculation uses the auxiliary quantities
\begin{align}
D(r)
&\equiv
2\rho(r)-r_-
=
\sqrt{r_-^2+4r^2},
\\
D_{\rm H}
&\equiv
D(r_{{\rm H},0})
=
r_+(1+a),
\\
D_{\rm m}
&\equiv
D(r_{\rm m}),
\\
s(b)
&\equiv
1+a+2b,
\\
A(b)
&\equiv
a(1+a)+2b,
\\
b_{\rm m}
&\equiv
\frac{\rho(r_{\rm m})-r_+}{r_+},
\\
s_{\rm m}
&\equiv
1+a+2b_{\rm m},
\\
A_{\rm m}
&\equiv
a(1+a)+2b_{\rm m}.
\end{align}
The Volterra primitive used in the construction of the displaced kernel is
\begin{equation}
Q(r)
\equiv
\int_{r_{{\rm H},0}}^r
\varphi_\varphi(s)K(s)\,ds.
\end{equation}
The closed-form forced response is written in terms of
\begin{equation}
\mathcal G(b)
\equiv
-b
-
\ln\!\left(
\frac{a+b}{a}
\right)
+
\frac{4b(a-1)}
{a(1+a)+2b}.
\end{equation}
The quantity
\begin{equation}
\mathcal R_{\rm m}
\equiv
\mathcal B_{\rm m}q_0
\end{equation}
is the contribution of the forced particular solution to the overlap
matching condition.

Throughout Secs.~\ref{sec:GHS_application} and
\ref{sec:physical_consequences}, all static radial profiles are understood
to be evaluated on the instantaneous static representative
\begin{equation}
\lambda(v)
=
\left(
\lambda^\varphi(v),
M_{\rm H}^{\rm inst}(v)
\right),
\end{equation}
with their slow dependence on \(v\) generally left implicit. At leading
adiabatic order,
\begin{equation}
\frac{r_{\rm H}^{\rm inst}}{\Mpl}
\left(
\dot\lambda^\varphi
-
\dot\varphi_{\rm ext}
\right)
=
\mathcal O_{\rm ad}(\epsilon_\varphi^2),
\qquad
\frac{\dot M_{\rm H}^{\rm inst}}{\Mpl^2}
=
\mathcal O_{\rm ad}(\epsilon_\varphi^2).
\end{equation}
Thus \(\dot\lambda^\varphi\) and
\(\dot\varphi_{\rm ext}\) agree at leading adiabatic order. The exact
dynamical quantities
\(r_{\rm H}^{\rm exact}(v)\) and
\(M_{\rm H}^{\rm exact}(v)\) are not background labels of the instantaneous
static representative; they are reconstructed from the complete
lag-corrected solution.

\section*{Acknowledgments}
AC acknowledges the support of the Initiative Physique des Infinis (IPI), a research training programme of Idex SUPER at Sorbonne Universit\'e.


\bibliographystyle{JHEP}
\bibliography{GHS_Time_Paper_II}

\end{document}